\documentclass[smallextended]{svjour3}       
\smartqed  
\usepackage{graphicx}
\journalname{J Nonlinear Sci}
\usepackage{subfig}
\usepackage{graphicx}
\renewcommand{\figurename}{\textbf{Fig.}}
\renewcommand{\tablename}{\textbf{Table}}
\usepackage{dcolumn}
\usepackage{bm}
\usepackage{epstopdf}
\usepackage{epsfig}
\usepackage[colorlinks = true,
linkcolor = blue,
urlcolor  = blue,
citecolor = blue,
anchorcolor = blue]{hyperref}
\usepackage{url}
\usepackage{textcomp}
\usepackage{amsmath}
\usepackage{amssymb}
\usepackage{threeparttable,multirow}
\usepackage{amsfonts}
\usepackage{mathrsfs}
\usepackage{graphicx}
\usepackage{hyperref}
\usepackage{xcolor}

\usepackage{bbding}
\begin{document}
	
	\title{Snakes on Lieb Lattice
	}
	%
	
	\author{R.\ Kusdiantara \and
		F.T.\ Akbar \and\\
		N.\ Nuraini \and
		B.E.\ Gunara \and
		H.\ Susanto
	}
	
	
	\institute{Rudy Kusdiantara \at
		Industrial and Financial Mathematics Research Group, Faculty of Mathematics and Natural	Sciences, Institut Teknologi Bandung, Jl.\ Ganesha No.\ 10, Bandung, 40132, Indonesia\\
		Centre of Mathematical Modelling and Simulation, Institut Teknologi Bandung, 1st Floor, Labtek III, Jl.\ Ganesha No.\ 10, Bandung, 40132, Indonesia\\ 
		\email{rudy@math.itb.ac.id}\\
		\\
		Fiki T.\ Akbar \at
		Theoretical High Energy Physics Research Division,Faculty of Mathematics and Natural Sciences, Institut Teknologi Bandung, Jl.\ Ganesha No.\ 10, Bandung, 40132, Indonesia\\  
		\email{ftakbar@fi.itb.ac.id}\\
		\\
		{Nuning Nuraini} \at
		Industrial and Financial Mathematics Research Group, Faculty of Mathematics and Natural	Sciences, Institut Teknologi Bandung, Jl.\ Ganesha No.\ 10, Bandung, 40132, Indonesia\\
		Centre of Mathematical Modelling and Simulation, Institut Teknologi Bandung, 1st Floor, Labtek III, Jl.\ Ganesha No.\ 10, Bandung, 40132, Indonesia\\ 
		\email{nuning@math.itb.ac.id}\\
		\\
		Bobby E.\ Gunara \at
		Theoretical High Energy Physics Research Division,Faculty of Mathematics and Natural Sciences, Institut Teknologi Bandung, Jl.\ Ganesha No.\ 10, Bandung, 40132, Indonesia\\  
		\email{bobby@fi.itb.ac.id}\\
		\\
		Hadi Susanto \at
		Department of Mathematics, College of Arts and Sciences, Khalifa University, PO Box 127788, Abu Dhabi, United Arab Emirates\\
		\email{hadi.susanto@ku.ac.ae}
	}
	
	\date{Received: date / Accepted: date}

	\maketitle
	
	\begin{abstract}
		We consider the discrete Allen--Cahn equation with cubic and quintic nonlinearity on the Lieb lattice. 
		We study localized {nonlinear} solutions of the system that have {linear} multistability and hysteresis in their bifurcation diagram. In this work, we investigate the system's homoclinic snaking, i.e., snaking-like structure of the bifurcation diagram, particularly the effect of the lattice type. Numerical continuation using a pseudo-arclength method is used to obtain localized solutions along the bifurcation diagram. We then develop an active-cell approximation to classify the type of solution at the turning points, which gives good agreement with the numerical results when the sites are weakly coupled. Time-dynamics of localized solutions inside and outside the pinning region is also discussed.
	\end{abstract}
	
	\section{Introduction}
	There has been a great interest in the study of homoclinic snaking \cite{Woods1999}, which is a snaking-like structure in the bifurcation diagram of spatially localized solutions, e.g., homoclinic orbits, 
	that appears in pattern formations in nonlinear systems.
	The Swift-Hohenberg equation with cubic and quintic nonlinearity is the basic model for pattern formation and the commonly studied equation for homoclinic snaking \cite{Burke2007,Burke2007a,Burke2012,Kusdiantara2017,Lloyd2019,Knobloch2019,Uecker2020,Schmidt2020}, which also has been studied previously as a model in, e.g., 
	cellular buckling \cite{Hunt2000}, neuronal model \cite{Laing2001,Avitabile2010}, and optical systems \cite{Firth2007,Yulin2008,Yulin2010,Yulin2011}. 
	Homoclinic snaking has also been observed in different experiments, e.g.,\
	in semiconductor optical systems \cite{Barbay2008}, liquid crystals \cite{Bortolozzo2009,Haudin2011,Bortolozzo2009a}, optical cavities \cite{Tlidi2012}, magnetic fluids \cite{Lloyd2015}, and shell bucklings \cite{Thompson2015}. It is caused by the fronts that are locked to the pattern and causes a pinning effect \cite{Pomeau1986,Bensimon1988}, resulting in a finite regime of bifurcation parameter where multiple localized solutions can co-exist. 
		{Homoclinic snaking 
	also appears in models for vegetation patterns \cite{Cisternas2020}, Schnakenberg system \cite{DeWitt2019}, and Coullet flow \cite{Salewski2019}.}
	
	Homoclinic snaking in higher dimensional systems has been studied as well in \cite{Lloyd2008,Uecker2014,Avitabile2010,Taylor2010}. 
	By using the Swift-Hohenberg equation, several numerical studies show appealing solutions, such as localized spots and hexagon patches, fronts or stripes \cite{Lloyd2008,Coullet2000,Kozyreff2006,Hilali1995,Sakaguchi1996,Tlidi1994,Vladimirov2011}, and localized radial solutions \cite{McCalla2010,Lloyd2009}. 
	Snaking may also be associate with various superpatterns and convectons in three-dimensional doubly diffusive convection that have been studied in \cite{Beaume2011,Beaume2013} and \cite{Dionne1997,Judd2000}, respectively.

	{
	Note that while the aforementioned references reported homoclinic snaking in spatially continuous systems, the snaking is also observed in discrete systems, such as in bistable nonlinear Schr\"odinger lattices \cite{Carretero-Gonzalez2006,Chong2009,Chong2011}, optical cavity solitons \cite{Yulin2008,Yulin2010}, and in small-world networks \cite{McCullen2016}. If in the continuous case the snaking is due to front locking mediated by spatially periodic solutions, in the discrete systems it is due to the imposed lattice, i.e., a discreteness-induced effective potential on the front dynamics, which is characterized by the overlap of the attractive interaction of fronts and the Peierls-Nabarro potential \cite{Braun2004}. Further discussion about discreteness effect that generates a set of bound states also have been studied by Egorov \emph{et al}. \cite{Egorov2013} and Clerc \emph{et al.} \cite{Clerc2017,Clerc2020}. The pinning region in the discrete case was first approximated analytically by Matthews and Susanto \cite{Matthews2011} and Dean \emph{et al.}\ \cite{Dean2015}. 
}

	{Some of the present authors have also studied snaking in higher-dimensional discrete systems \cite{Kusdiantara2019} where details of the bifurcation diagram are rather more involved (see also \cite{Bramburger2020a,Tian2021}). The complexity and width of the snaking diagrams depend on the number of ``patch interfaces'' admitted by the lattice patterns. While in our previous work \cite{Kusdiantara2019}, we considered square, honeycomb, and triangular lattices, in here we study a two-dimensional discrete Allen--Cahn equation with cubic and quintic nonlinearity in the Lieb lattice. }

	{
	The particular lattice is studied because of its physical wide interests and applications, {such as in the design of organic spintronic devices \cite{Cui2020} and quantum materials with tailored properties \cite{Drost2017}. } Two-dimensional materials with a Lieb lattice host exotic electronic band structures, which comes from Hubbard model where the ground state has zero spin angular momentum \cite{Lieb1989}. In nature, Lieb lattice does not exist, and it is difficult to obtain experimentally due to its structural instability \cite{Feng2020}. Even though Lieb lattice is mostly studied theoretically, such as in the Heisenberg model for impurity-tuning of phase transition \cite{Le2019}, a metal-based lattice for photonic zero-energy modes \cite{Chen2019}, and magnetic materials \cite{Cui2020,Oliveira-Lima2020}, it has also been studied experimentally using, e.g., 
	a tin overlayer \cite{Feng2020}, synthesized metal--organic framework \cite {Jiang2019,Jiang2020}, {polariton quantum fluids \cite{Scafirimuto2021}, micropillars \cite{Whittaker2018}, an array of carbon monoxide molecules \cite{Slot2017}, optical waveguide arrays \cite{Mukherjee2015}, and Bose-Einstein condensate \cite{Ozawa2017}.
	} 
	In the present paper, we are interested in the effect of such an exotic lattice to homoclinic snaking in the bistable Allen-Cahn equation. Even though the discrete nonlinear Schr\"odinger equation is a more natural playground to study the lattice from the physical point of view, we chose the Allen-Cahn equation for its simplicity, i.e., it is real-valued, yet it shares the same standing wave (time-independent) solutions with the Schr\"odinger counterpart. The cubic-quintic nonlinearity is also physically relevant as it is quite generic experimentally in optical systems, see, e.g., \cite{Smektala2010,Boudebs2003,Zhan2002} for the experimental observation of optical nonlinearities that may be fitted by a combination of self-focusing cubic and self-defocusing quintic terms.}

	As the main result of the present work, we classify all types of saddle-node bifurcations that form the boundaries of the pinning regions, characterised by the number of `fronts' exhibited by the discrete patterns. We also develop analytical approximations of the {localized nonlinear} solutions and their {linear} stability. 
	When unstable, we analyze their time-dynamics as well as the dynamics of the system inside and outside the pinning region. In addition to its exotic electronic band structures, we find that the Lieb lattice yields complicated structure in the snaking structures, such as many {`switchbacks'}. Nonetheless, we observe that we can approximate the first turning point in the bifurcation diagram using our analysis, which previously failed for square, honeycomb, and triangular lattices \cite{Kusdiantara2019}. 
	
	The paper is outlined as follows. The discrete Allen-Cahn equation and the stability of the uniform solutions are discussed in Sec.\ \ref{sec:unisol}.
	r{red}{We analyze site and bond-centred localized solutions and their snaking in Sec.\ \ref{sec:loc}. }
	Section \ref{sec:saddle} discusses saddle-node bifurcations and their approximation. 
	The critical eigenvalue approximation is also discussed in the section where good agreement is obtained.
	Time-dynamics of localized solutions about the pinning region is discussed in Sec.\ \ref{sec:time}.
	Conclusions are in Sec.\ \ref{sec:conclusion}.
	
	\section{Mathematical model and uniform solution}\label{sec:unisol}
\begin{figure}[h!]
		\centering
		\includegraphics[scale=0.7]{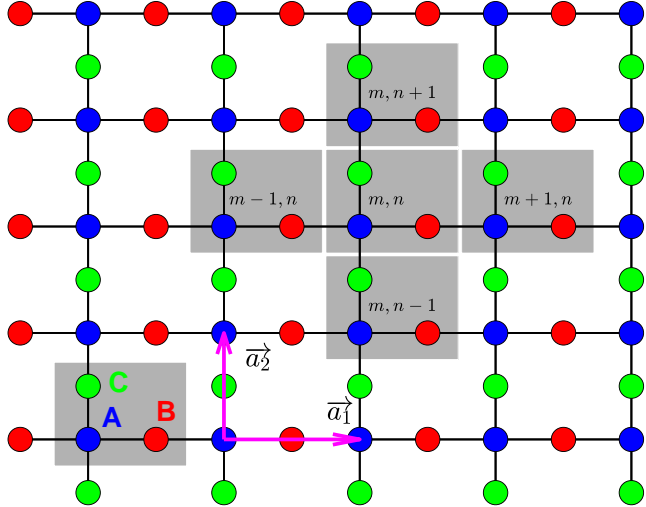}
		\caption{Lieb lattice tight binding structure. 
			The blue, red, and green points represent the stationary field for $A_{m,n}$, $B_{m,n}$, and $C_{m,n}$ respectively. The vector $\protect\overrightarrow{a_1}$ and $\protect\overrightarrow{a_2}$ (magenta) represent perturbation directions (see the text).
		}
		\label{fig:lieb_structure}
	\end{figure}

	In this study, we consider the discrete Allen-Cahn equation with cubic-quintic nonlinearity, {which has linear bistability in the background states \cite{Burke2007,Burke2007a,Taylor2010,Kusdiantara2017,Kusdiantara2019,Chong2009}, i.e.,}
	\begin{equation}
		\frac{d{u}_{m,n}}{dt}=-\mu u_{m,n}+2u_{m,n}^3-u^5_{m,n}+c\Delta u_{m,n}
		\label{eq:ac_ori}
	\end{equation}
	where $\mu$ is a real-valued bifurcation parameter, $c$ is the coupling strength of the nearest-cell, $\Delta$ is discrete Laplacian operator for Lieb lattice on the two-dimensional (2D) integer lattice $\mathbb{Z}^2$, 
	and $u_{m,n}$ is a real-valued stationary field defined on 2D integer lattice.
	Defining $u_{m,n}$ and $\Delta u_{m,n}$ as
	\begin{equation}
		\begin{array}{c}
			u_{m,n}=\left(\begin{array}{c}
				A_{m,n}\\B_{m,n}\\C_{m,n}
			\end{array}\right)\quad \text{and}\quad\\\\
			\Delta u_{m,n}=\left(\begin{array}{c}
				\Delta A_{m,n}\\\Delta B_{m,n}\\\Delta C_{m,n}
			\end{array}\right)=\left(
			\begin{array}{c}
				B_{m,n}+B_{m-1,n}+C_{m,n}+C_{m,n-1}-4A_{m,n}\\
				A_{m,n}+A_{m+1,n}-2B_{m,n}\\
				A_{m,n}+A_{m,n+1}-2C_{m,n}
			\end{array}
			\right),
		\end{array}
	\end{equation}
	we can re-write Eq.\ \eqref{eq:ac_ori} into 
	\begin{equation}
		\begin{array}{rcl}
\dfrac{{dA}_{m,n}}{dt}&=&-\mu A_{m,n}+2A_{m,n}^3-A_{m,n}^5
+c \left(B_{m,n}+B_{m-1,n}+C_{m,n}+C_{m,n-1}\right.\\
&&\left.-4A_{m,n}\right)=\dfrac{\partial \mathcal{H}}{\partial A_{m,n}},\\
\dfrac{{dB}_{m,n}}{dt}&=&-\mu B_{m,n}+2B_{m,n}^3-B_{m,n}^5+c \left(A_{m,n}+A_{m+1,n}-2B_{m,n}\right)=\dfrac{\partial \mathcal{H}}{\partial B_{m,n}},\\
\dfrac{{dC}_{m,n}}{dt}&=&-\mu C_{m,n}+2C_{m,n}^3-C_{m,n}^5+c \left(A_{m,n}+A_{m,n+1}-2C_{m,n}\right)=\dfrac{\partial \mathcal{H}}{\partial C_{m,n}},
		\end{array}
		\label{eq:ac_lieb}
	\end{equation}
where
	\begin{equation}
		\begin{array}{lcl}
			\mathcal{H}\left(A_{m,n},B_{m,n},C_{m,n}\right)=\\
			\displaystyle\sum_{m,n}\left(\frac{\mu}{2}\left(A_{m,n}^2+B_{m,n}^2+C_{m,n}^2\right)
			-\dfrac{1}{2}\left(A_{m,n}^4+B_{m,n}^4+C_{m,n}^4\right)\right.\\
			\left.+\dfrac{1}{6}\left(A_{m,n}^6+B_{m,n}^6+C_{m,n}^6\right)+\dfrac{c}{2}\left(\left(B_{m-1,n}-A_{m,n}\right)^2\right.\right.\\
			\left.\left.+\left(C_{m,n-1}-A_{m,n}\right)^2+\left(A_{m,n}-B_{m,n}\right)^2+\left(A_{m,n}-C_{m,n}\right)^2\right)\right),
		\end{array}
		\label{eq:energy}
	\end{equation}
{is the energy of the system \eqref{eq:ac_ori}}.	A sketch of the field location on the Lieb structure is given in Fig.\ \ref{fig:lieb_structure}. It can then be shown using \eqref{eq:ac_lieb} that {  
		\begin{equation}
			\begin{array}{rcl}
				-\dfrac{d \mathcal{H}}{d t}&=&\left(\dfrac{{dA}_{m,n}}{dt}\right)^2+\left(\dfrac{{dB}_{m,n}}{dt}\right)^2+\left(\dfrac{{dC}_{m,n}}{dt}\right)^2\geq0.
			\end{array}
		\label{dH}
		\end{equation}
	Hence, 
	every solution of \eqref{eq:ac_lieb} flows down along a gradient of the energy \eqref{eq:energy} towards a local minimum that corresponds to a stable time-independent 
	solution. 
	Therefore, oscillatory dynamics is not possible.
}
	
	In particular, we study the time-independent solution of Eq \eqref{eq:ac_ori}, i.e.,\ 
	\begin{equation}
		-\mu {u}_{m,n}+2{u}_{m,n}^3-{u}_{m,n}^5+c\Delta {u}_{m,n}=0.
		\label{eq:ac_ti}
	\end{equation}
	To determine the linear stability of a solution $\tilde{u}_{m,n}=\left(\begin{array}{c}
		\tilde{A}_{m,n}\\\tilde{B}_{m,n}\\ \tilde{C}_{m,n}\\  \end{array}\right)$, we write 
	\begin{equation}
		\left(\begin{array}{c}
			{A}_{m,n}\\{B}_{m,n}\\ {C}_{m,n}\\  \end{array}\right)=\left(\begin{array}{c}
			\tilde{A}_{m,n}\\\tilde{B}_{m,n}\\ \tilde{C}_{m,n}\\  \end{array}\right)+
		\left(\begin{array}{c}
			\hat{A}_{m,n}\\\hat{B}_{m,n}\\\hat{C}_{m,n}\\  \end{array}\right)\epsilon e^{\lambda t}.
		\label{eq:anz1}
	\end{equation}
	By substituting \eqref{eq:anz1} in \eqref{eq:ac_ori} and linearizing about $\epsilon=0$, 
	we obtain the linear equation 
	\begin{equation}
		\lambda \left(\begin{array}{c}
			\hat{A}_{m,n}\\\hat{B}_{m,n}\\\hat{C}_{m,n}\\  \end{array}\right)=\mathcal{L}\left(\begin{array}{c}
			\hat{A}_{m,n}\\\hat{B}_{m,n}\\\hat{C}_{m,n}\\  \end{array}\right),
	\end{equation}
	where 
	\begin{equation}
		\mathcal{L}=
		\left(\begin{array}{ccc}
			\gamma(\mu,\tilde{A}_{m,n})&0&0\\0&\gamma(\mu,\tilde{B}_{m,n})&0\\ 0&0&\gamma(\mu,\tilde{C}_{m,n})\\  \end{array}\right)
		+c\Delta,
	\end{equation}
	and 
	\begin{equation*}
		\gamma\left(\mu,X\right)=-\mu+6X^2-5X^4.
	\end{equation*}
	A solution is said to be linearly stable when all $\lambda\leq0$ and unstable when $\exists\lambda>0$.
	
\begin{figure}[t!]
		\centering
		\includegraphics[scale=0.5]{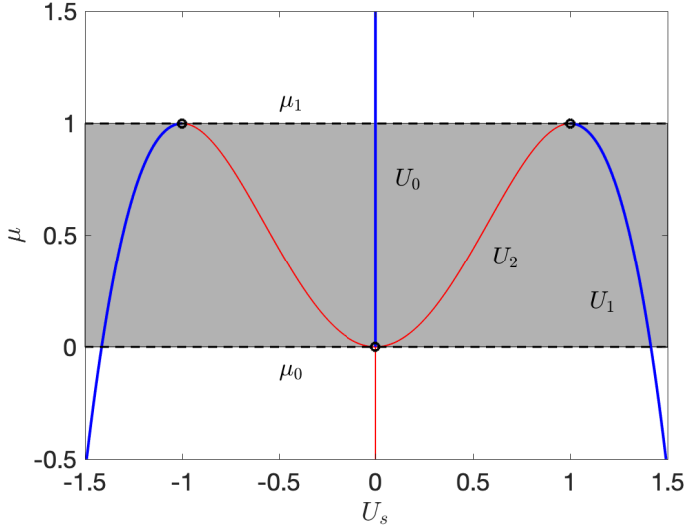}
		\caption{Uniform solution $U_s$ of the discrete Allen-Cahn equation as a function of $\mu$.
			The blue thick and red thin lines indicate stable and unstable solutions, respectively.}
		\label{fig:unisol_lieb}
	\end{figure}

	Generally, the 2D discrete Allen-Cahn equation \eqref{eq:ac_ori} exhibits the same uniform solutions as those in other structures that have been studied in \cite{Taylor2010,Kusdiantara2019}, which are given by 
	\begin{equation}
		-\mu U_s+2U_s^3-U_s^5=0.
	\end{equation}
	It can be solved to yield
	\begin{equation}
		U_0=0\quad \text{and}\quad U_{1,2}^2=1\pm\sqrt{1-\mu}.
	\end{equation}
	We plot the uniform solutions for varying $\mu$ in Fig.\ \ref{fig:unisol_lieb}. 
	{Herein, we define $U_1=\pm\sqrt{1+\sqrt{1-\mu}}$ as the ``upper'', $U_2=\pm\sqrt{1-\sqrt{1-\mu}}$ as the ``lower'', and $U_0=0$ as the ``background'' states.}
	
	\begin{figure}[t!]
		\centering
		{\includegraphics[scale=0.5]{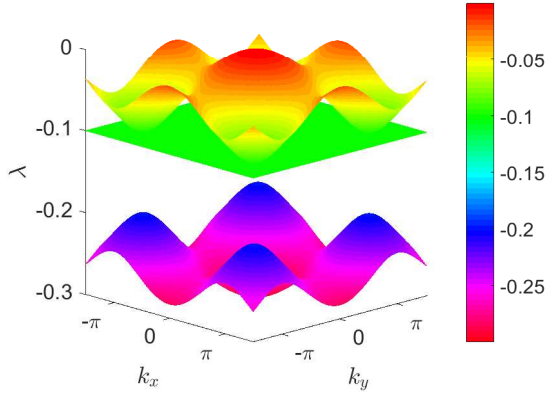}}
		\caption{The dispersion relation of Lieb lattice for zero solution at $\mu=0$ and $c=0.05$.
		{Note the presence of a flat band, which is a special characteristic of the lattice.}
	}
		\label{fig:disrel}
	\end{figure}
	To determine the linear stability of the uniform solutions $\tilde{u}_{m,n}$, i.e.,\ $\tilde{A}_{m,n}=\tilde{B}_{m,n}=\tilde{C}_{m,n}=U_s$, 
	where $s=0,1,2$, one has
	\begin{equation}
		\hat{{u}}_{m,n}=\exp\left({i\left(\left<\overrightarrow{a_1},\textbf{k}\right>m+\left<\overrightarrow{a_2},\textbf{k}\right>n\right)}\right),
	\end{equation}
	where $\textbf{k}=\left(\begin{array}{c}
		k_x\\
		k_y
	\end{array}\right)$ is the wave number of perturbations in the $n$ and $m$ directions and $\overrightarrow{a_1}=\left(\begin{array}{c}
		1\\0\end{array}\right)$ and $\overrightarrow{a_2}=\left(\begin{array}{c}
		0\\1\end{array}\right)$ are ortonormal basis in $\mathbb{R}^2$ as shown in Fig.\ \ref{fig:lieb_structure}. 
	So, the perturbation ansatz would be
	\begin{equation}
		\begin{array}{ccl}
			A_{m,n}&=&U_s+\epsilon e^{\lambda t}\tilde{A}_{m,n},\\
			B_{m,n}&=&U_s+\epsilon e^{\lambda t}\tilde{B}_{m,n},\\
			C_{m,n}&=&U_s+\epsilon e^{\lambda t}\tilde{C}_{m,n}.
		\end{array}
		\label{eq:anz}
	\end{equation}
	By substituting \eqref{eq:anz} into \eqref{eq:ac_lieb} and linearizing about $\epsilon=0$, we obtain the linear equation 
	\begin{equation}
		\lambda \left(
		\begin{array}{c}
			\hat{A}_{m,n}\\
			\hat{B}_{m,n}\\
			\hat{C}_{m,n}
		\end{array}
		\right)=\mathcal{M}
		\left(
		\begin{array}{c}
			\hat{A}_{m,n}\\
			\hat{B}_{m,n}\\
			\hat{C}_{m,n}
		\end{array}
		\right),
		\label{ei}
	\end{equation}
	where
	\begin{equation}
		\mathcal{M}=\left(
		\begin{array}{ccc}
			\gamma\left(\mu,U_s\right)-4c&c\left(1+e^{-ik_x}\right)&c\left(1+e^{-ik_y}\right)\\
			c\left(1+e^{ik_x}\right)&\gamma\left(\mu,U_s\right)-2c&0\\
			c\left(1+e^{ik_y}\right)&0&\gamma\left(\mu,U_s\right)-2c
		\end{array}
		\right).
	\end{equation}
	Hence, we have the dispersion relation of the Lieb lattice, i.e.,\
	\begin{equation}
		\begin{array}{ccl}
			\lambda_1&=&-2c+\gamma\left(\mu,U_s\right),\\
			\lambda_{2,3}\left(k_x,k_y\right)&=&-3c+\gamma\left(\mu,U_s\right)\pm c\sqrt{5+2\left(\cos\left(k_x\right)+\cos\left(k_y\right)\right)}.
		\end{array}
		\label{eq:disrel}
	\end{equation}
	The points $\mu_0=0$ and $\mu_1=1$ in Fig.\ \ref{fig:unisol_lieb} denote the stability change of $U_s$. 
	They correspond to a condition when the maximum of the dispersion relation \eqref{eq:disrel} touches the $k_x,\ k_y$ plane, which is attained at $k_x=2\eta_1\pi,\,k_y=2\eta_2\pi,\, \eta_1,\eta_2\in\mathbb{Z}$. 
	One can note that we have the bistability interval $\mu\in\left[\mu_{1},\mu_0\right]$ for the uniform solutions, see Fig.\ \ref{fig:unisol_lieb}.
	Furthermore, the bifurcation diagram and the stability of the uniform solution in Fig.\ \ref{fig:unisol_lieb} are the same as those in the one-dimensional model \cite{Taylor2010,Chong2009}. Figure\ \ref{fig:disrel} shows that the dispersion relation has a flat band, i.e.,\ $\lambda_1$. This special band structure can bring various exotic electronic properties, see, e.g., \cite{julku,tamura,wang}.

\section{Localized solutions and snaking}\label{sec:loc}
	
	\begin{figure*}[t!]
		\centering
		\subfloat[Site-centred solution]{\includegraphics[scale=0.4]{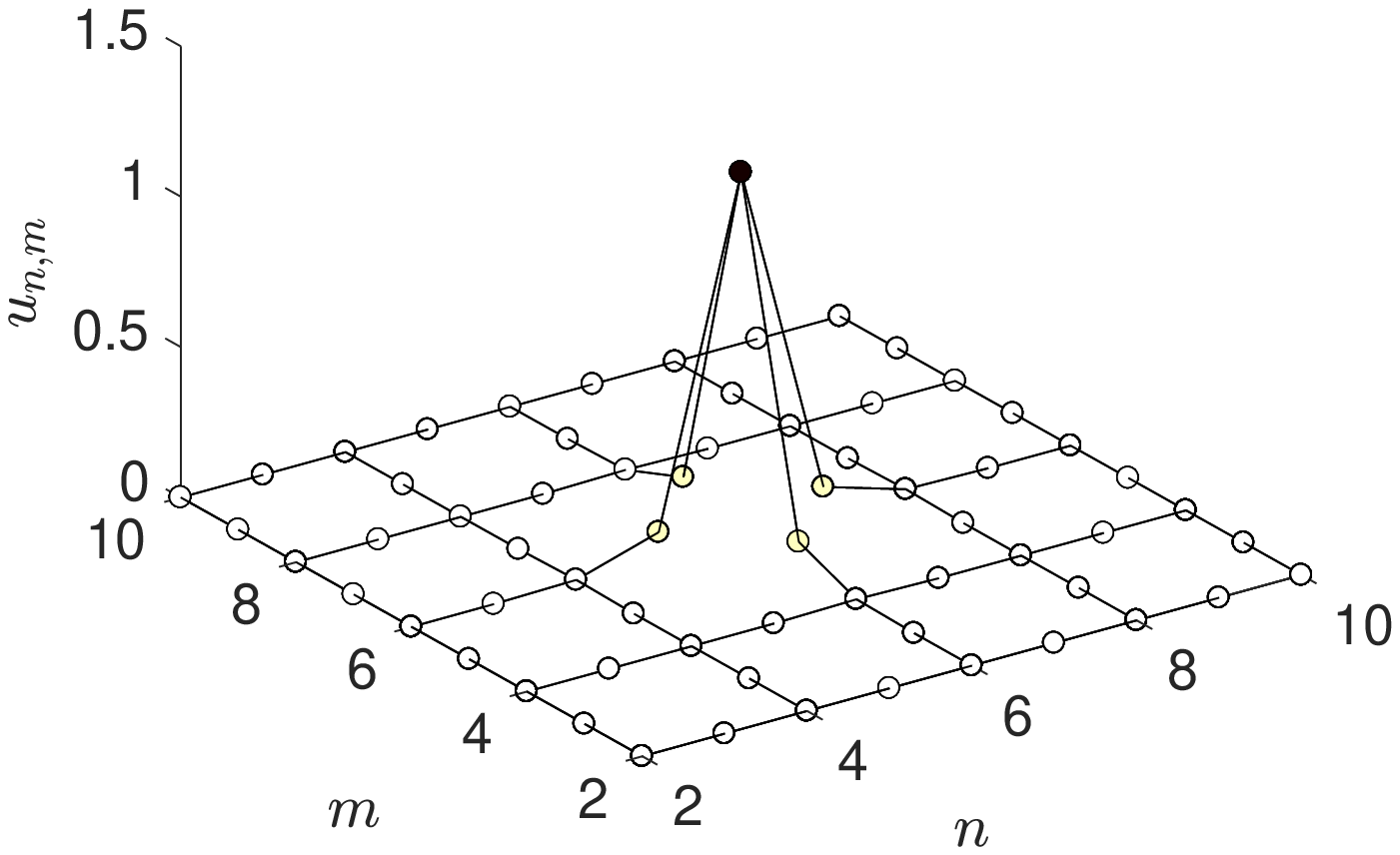}
			\label{subfig:site_ex}}
		\subfloat[Bond-centred solution]{\includegraphics[scale=0.4]{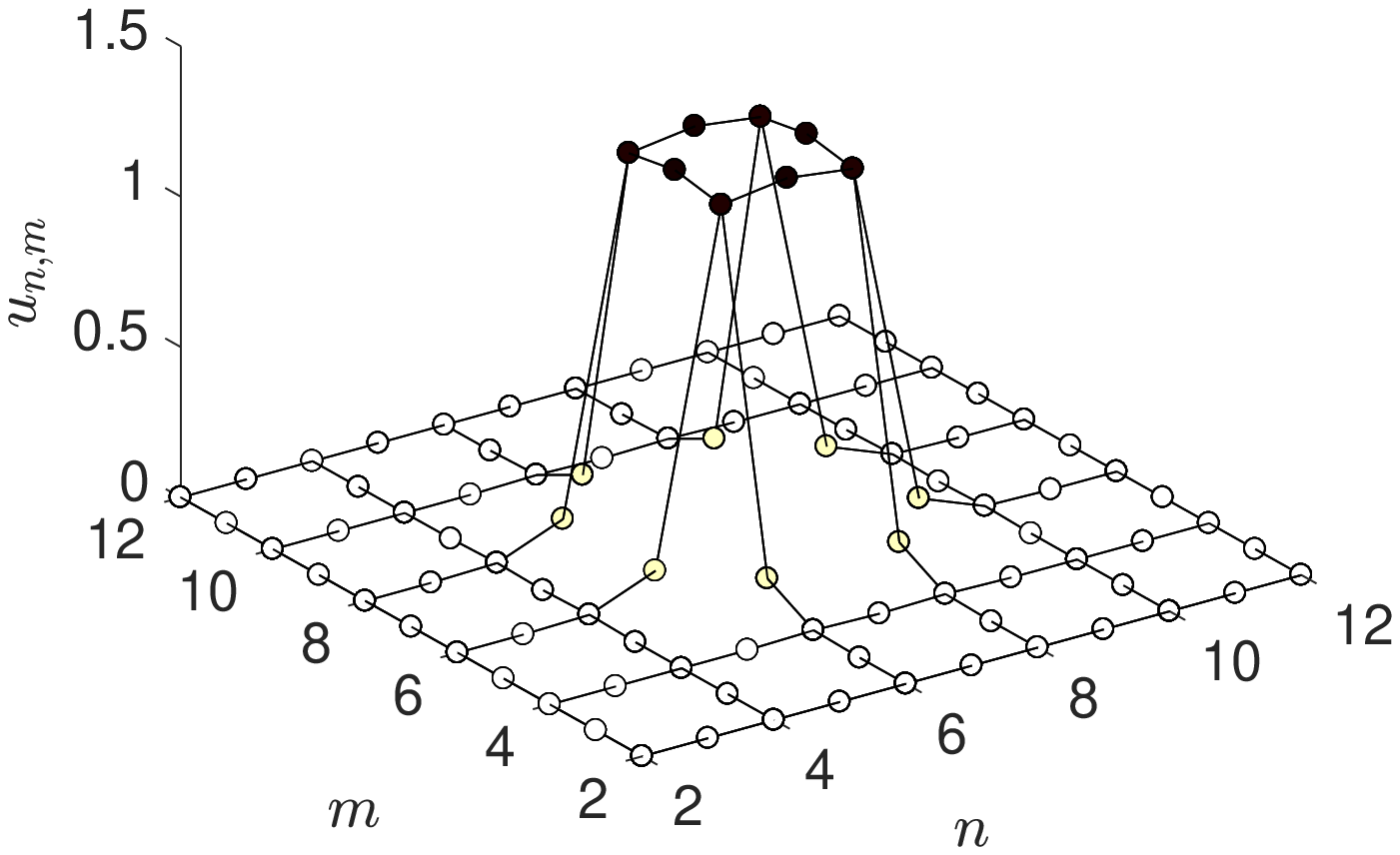}
			\label{subfig:bond_ex}}
		\caption{Structures of fundamental localized solutions for $c=0.05$ and $\mu=0.6$.}
		\label{fig:site_bond_example}
	\end{figure*}

	The discrete Allen-Cahn equation \eqref{eq:ac_ori} admits localized solutions that bifurcate from the zero solution $U_0$ at point $\mu_0$. 
	We are particularly interested in fundamental localized solutions, i.e.,\ site-centred and bond-centred solutions, which are the counter-part of onsite and intersite solutions in the regular one-dimensional lattice case. 
	They are formed by two bistable states from the uniform solutions, i.e.,\ the non-zero solution $U_1$ as the ``upper'' state and the zero solution $U_0$ as the ``background" state.

	\begin{figure*}[htbp!]
		\centering 
		\subfloat[$c=0.05$]{\includegraphics[scale=0.45]{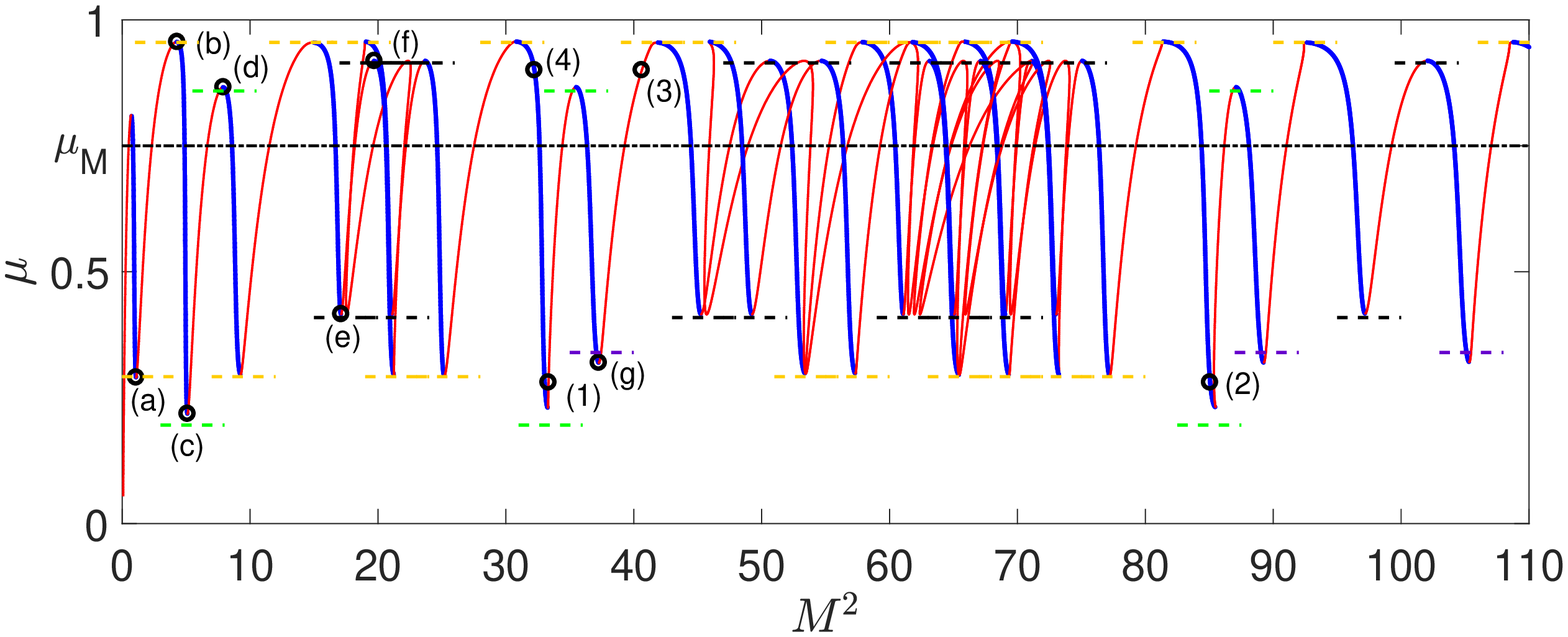}\label{subfig:snake_site_c_0_05}}\\
		\subfloat[$c=0.1$]{\includegraphics[scale=0.45]{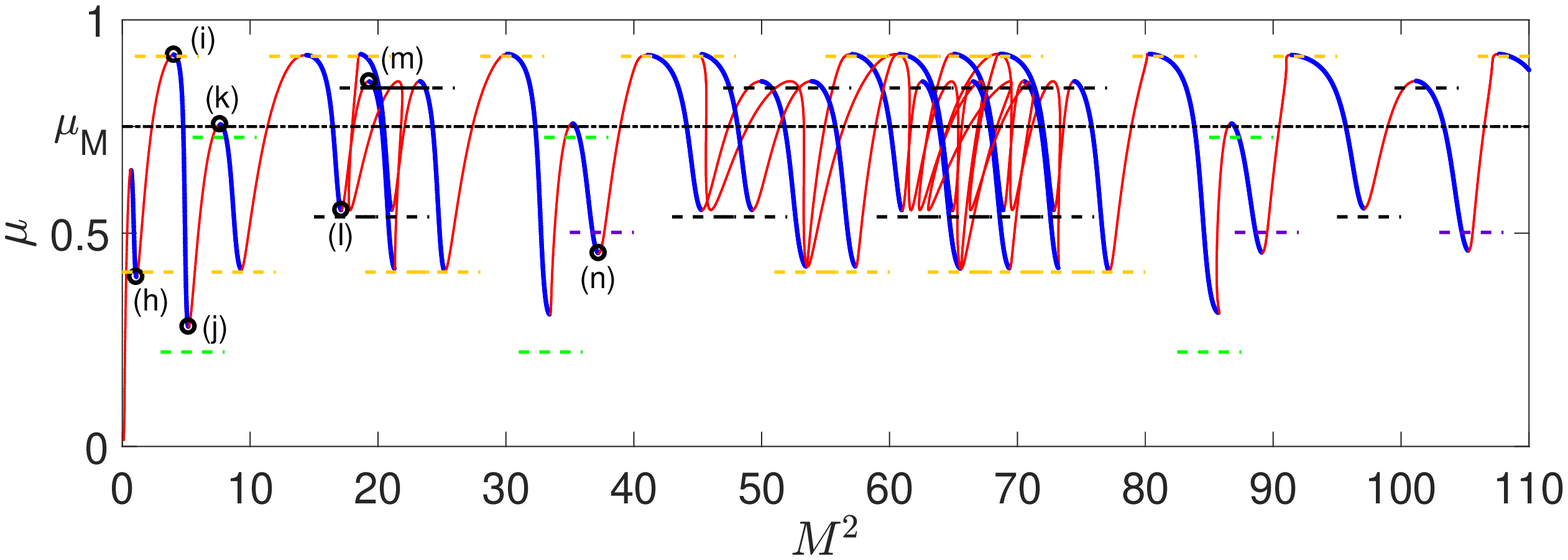}\label{subfig:snake_site_c_0_10}}\\
		\subfloat[$c=0.05$]{\includegraphics[scale=0.35]{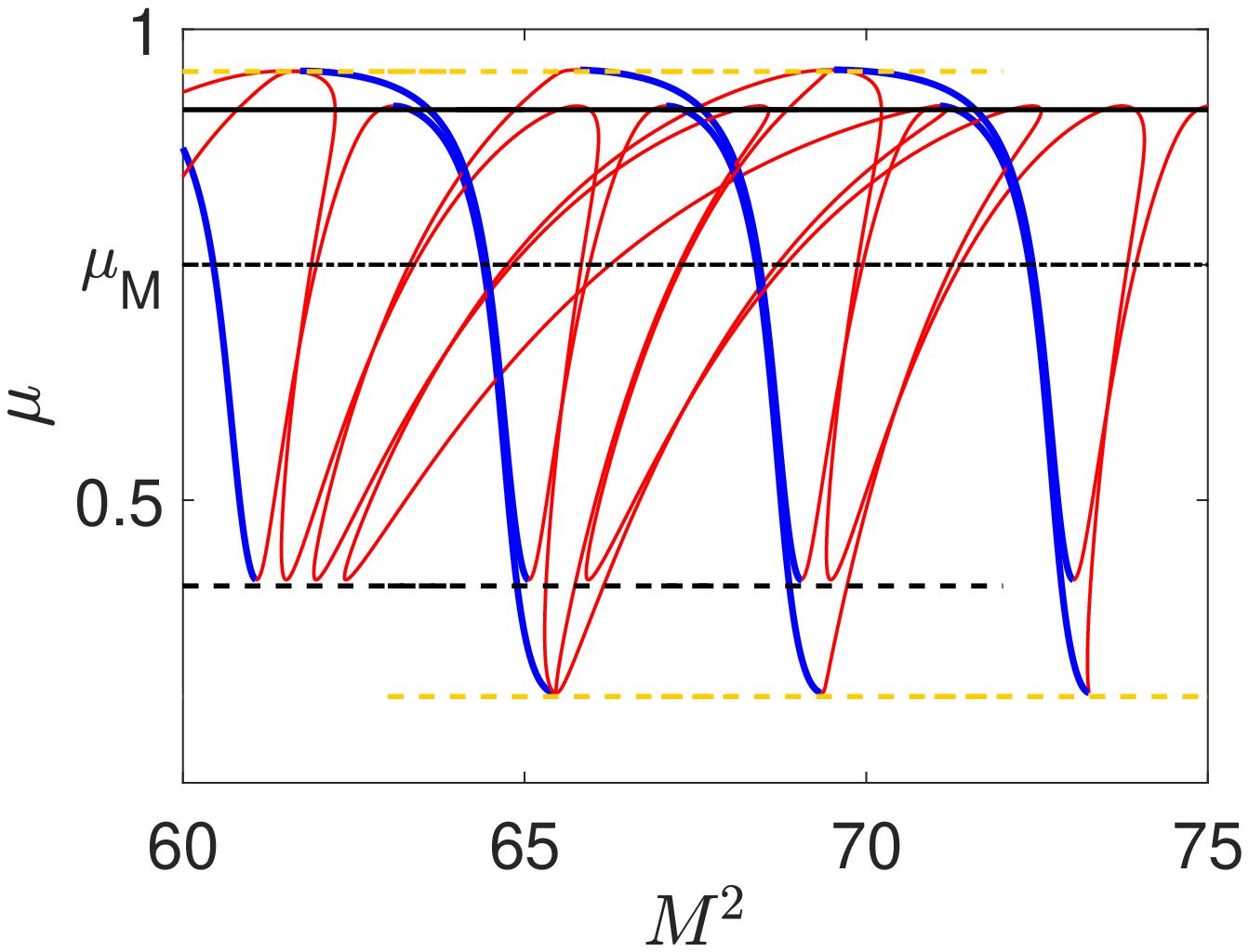}\label{subfig:snake_site_c_0_05_zoom}}
		\subfloat[$c=0.1$]{\includegraphics[scale=0.35]{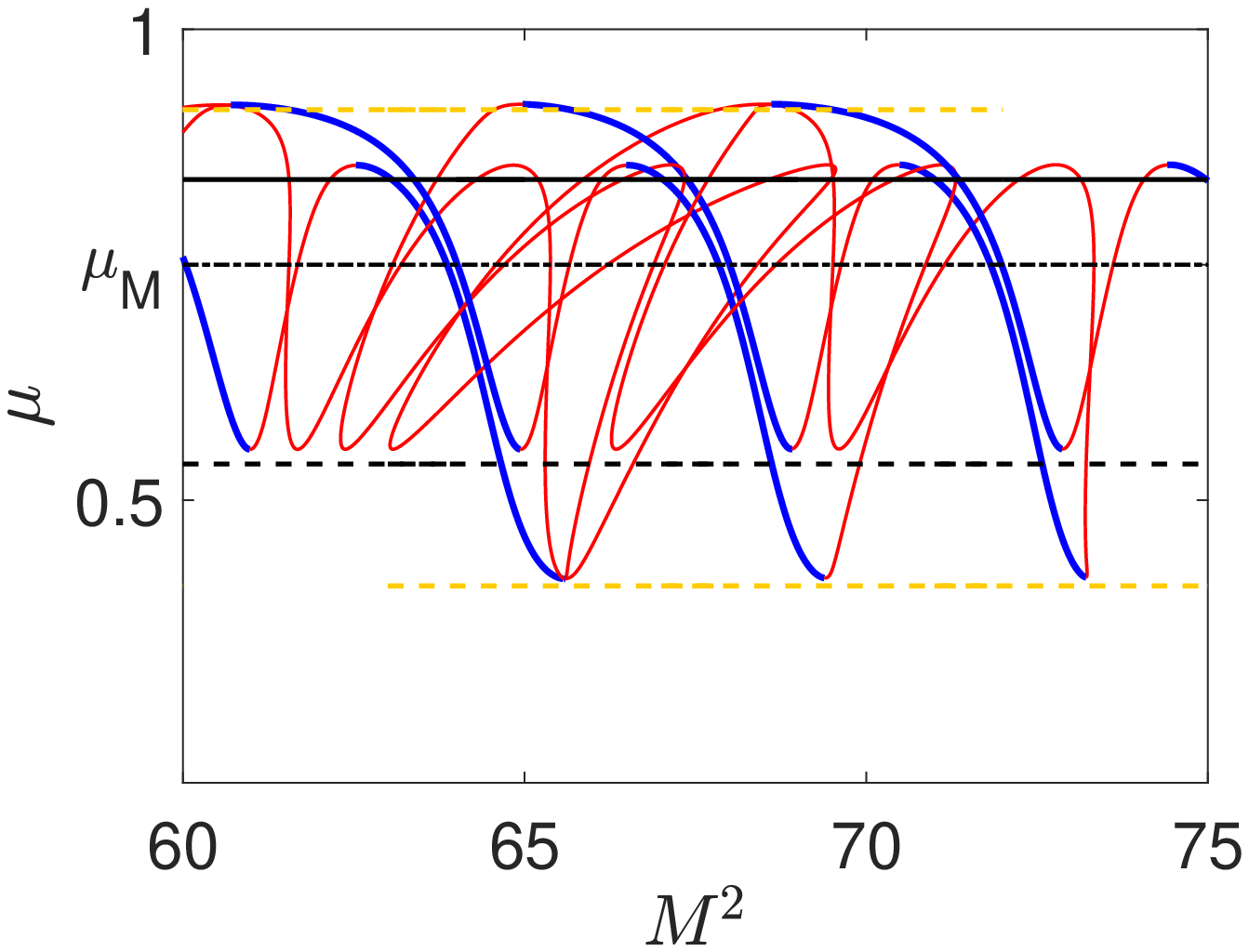}\label{subfig:snake_site_c_0_10_zoom}}
		\caption{Panels (a) and (b) show bifurcation diagrams of site-centred solutions for $c=0.05$ and $0.1$. 
			{Panels (c) and (d) zoom in on {`switchbacks'} to show the details of a complicated structure around $60<M^2<75$ for both values of $c$.}
			{The dashed} horizontal lines about the turning points are the approximation of `upper' and `lower' saddle-node bifurcations. 
			The green, black, {orange}, and purple line colors correspond to saddle-node bifurcations from our active-cell approximations of type 1, 2, 4, and 6, see section \ref{sec:saddle}. Solution profiles at the turning points labelled as (a)--(n) in the top panel are shown in Fig.\ \ref{fig:prof_site}. 
			Points (1)--(2) will be used to describe the solution stability in Fig.\ \ref{fig:eig_approx}, while points (3)--(4) will be used for time-dynamics.
		}
		\label{fig:bifur_site}
	\end{figure*}

	
	In our current work, site-centred state is a solution profile with odd number excited sites {(when $A_{m,n}$ is non-zero)} as shown in Fig.\ \ref{subfig:site_ex}.
	On the other hand, bond-centred state is a solution where the excited sites bond with other sites and form the simplest polygon. 
	Examples of bond-centred solutions are shown in Fig.\ \ref{subfig:bond_ex}.
	Herein, we use $20\times20$ lattice domain and periodic boundary conditions for the computational domain.	
	
	\begin{figure*}[htbp!]
		\centering
		\subfloat[]{\includegraphics[scale=0.16]{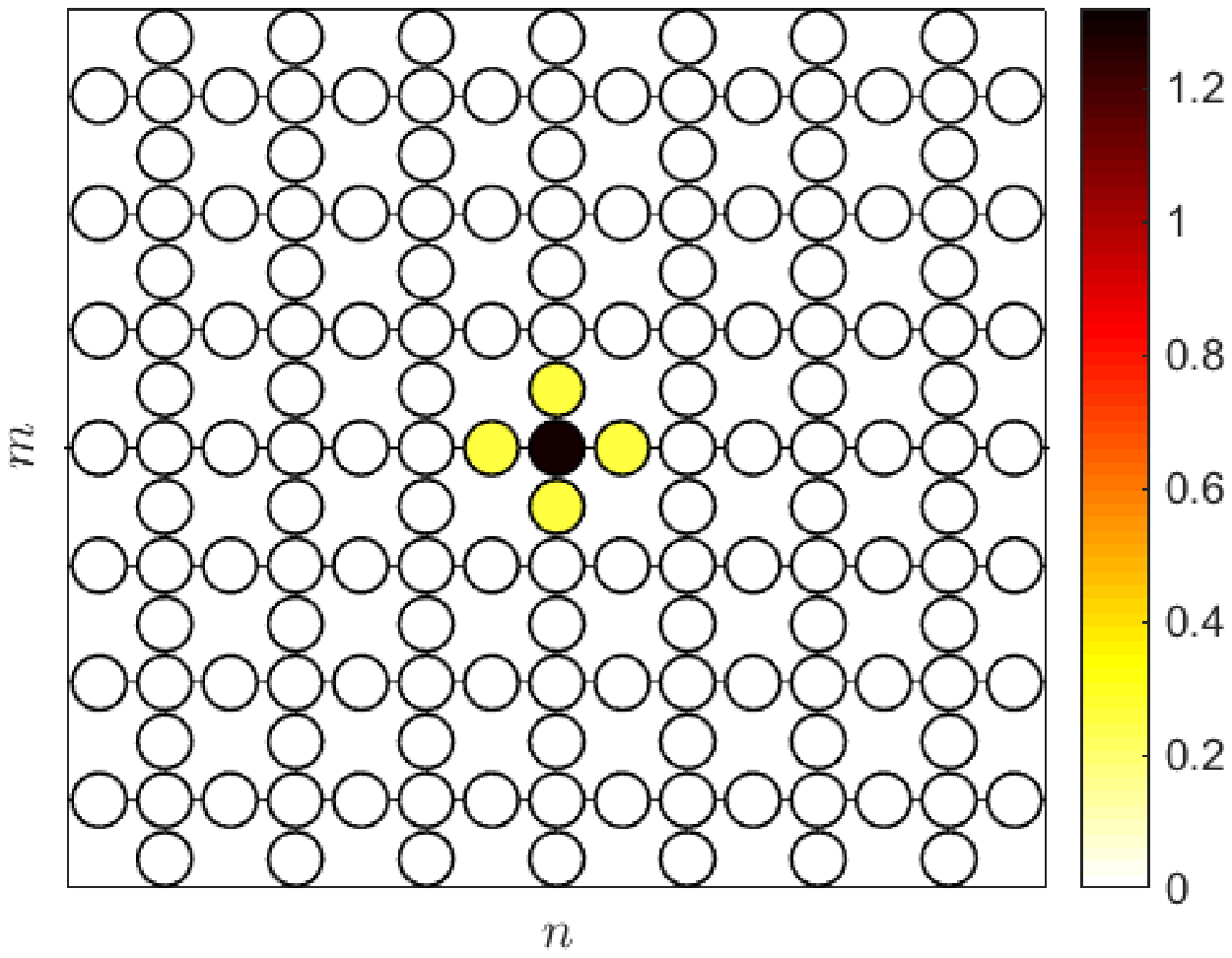}\label{subfig:prof_site_c_0_05_a}}
		\subfloat[]{\includegraphics[scale=0.16]{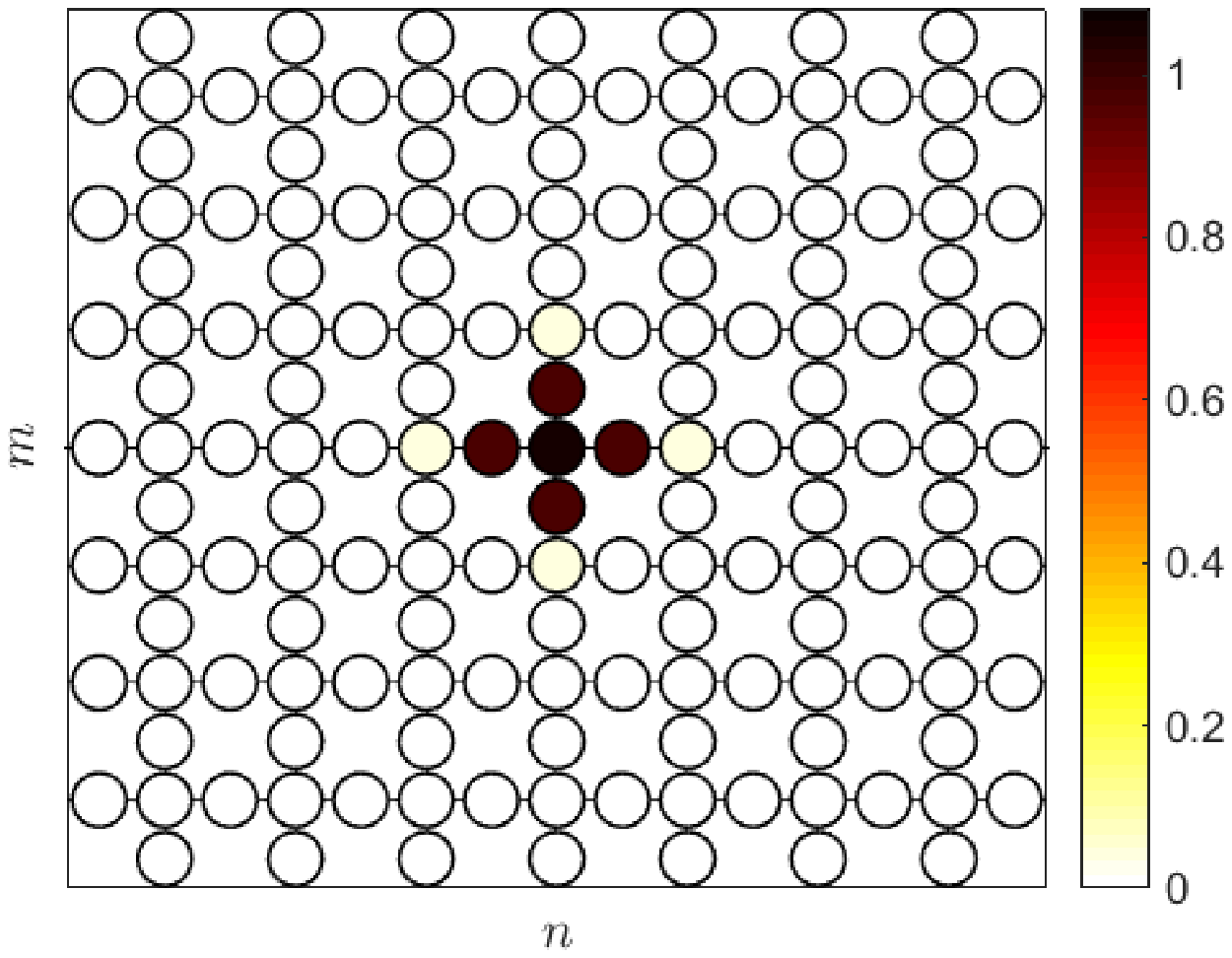}\label{subfig:prof_site_c_0_05_b}}
		\subfloat[]{\includegraphics[scale=0.16]{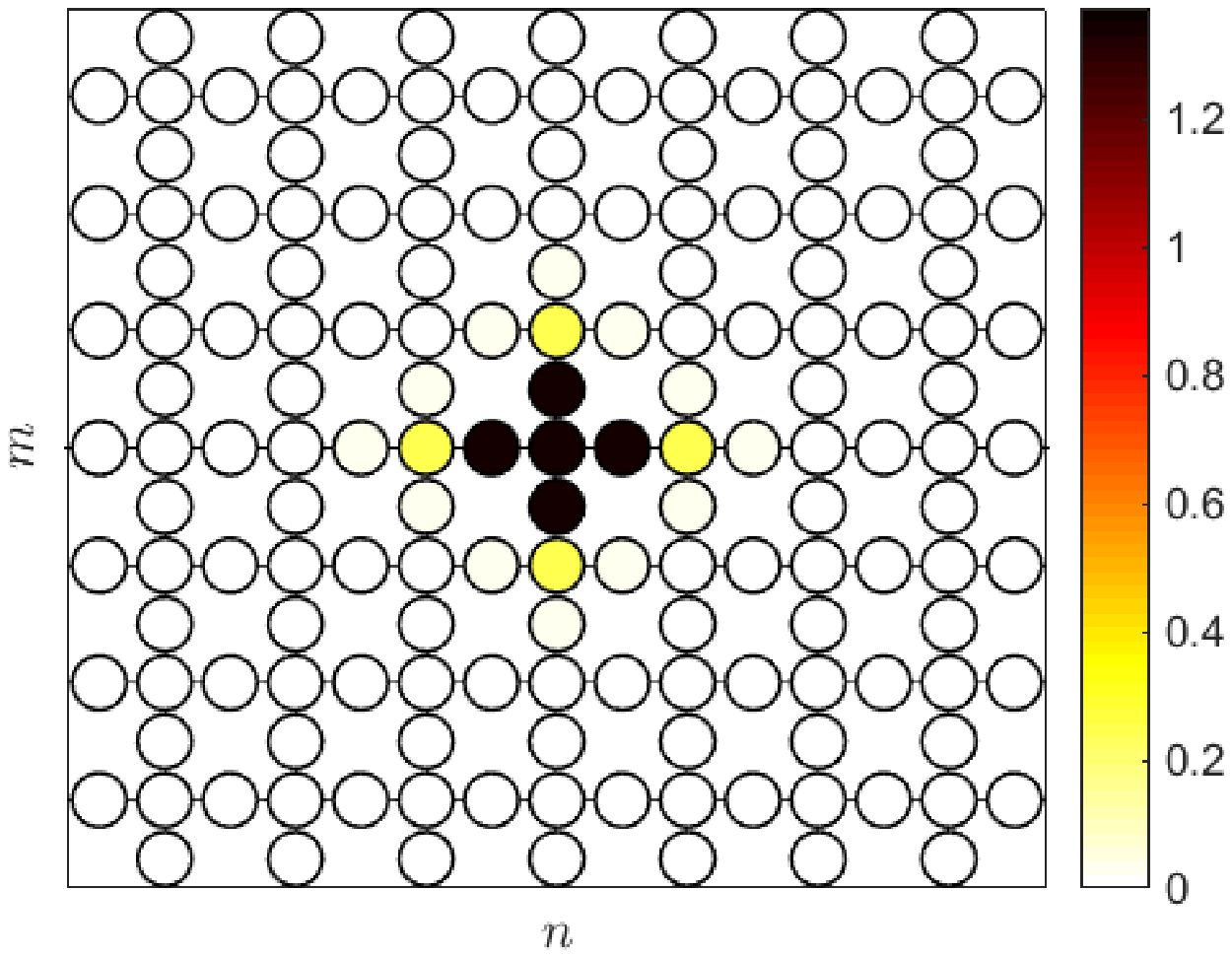}\label{subfig:prof_site_c_0_05_c}}
		\subfloat[]{\includegraphics[scale=0.16]{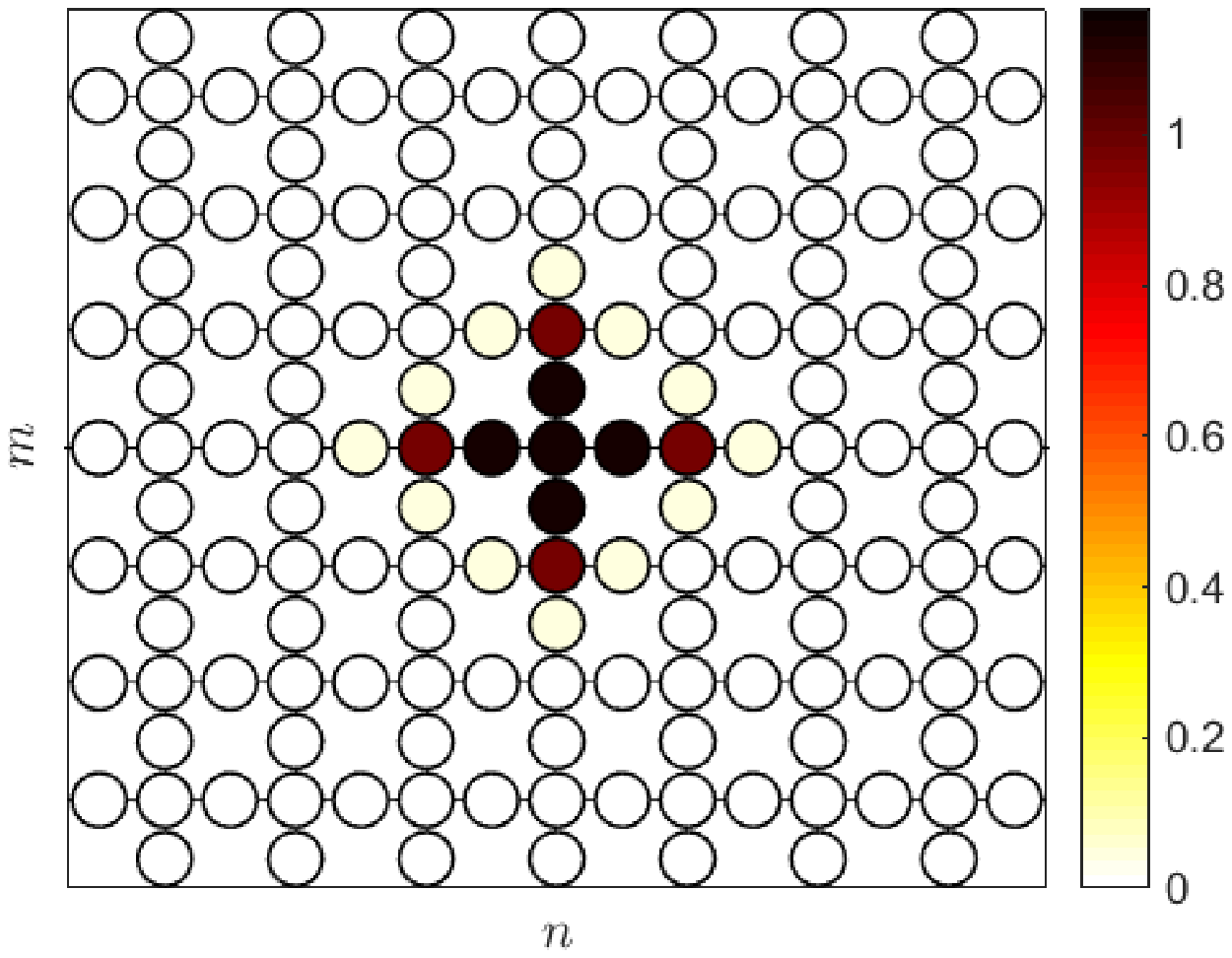}\label{subfig:prof_site_c_0_05_d}}
		\subfloat[]{\includegraphics[scale=0.16]{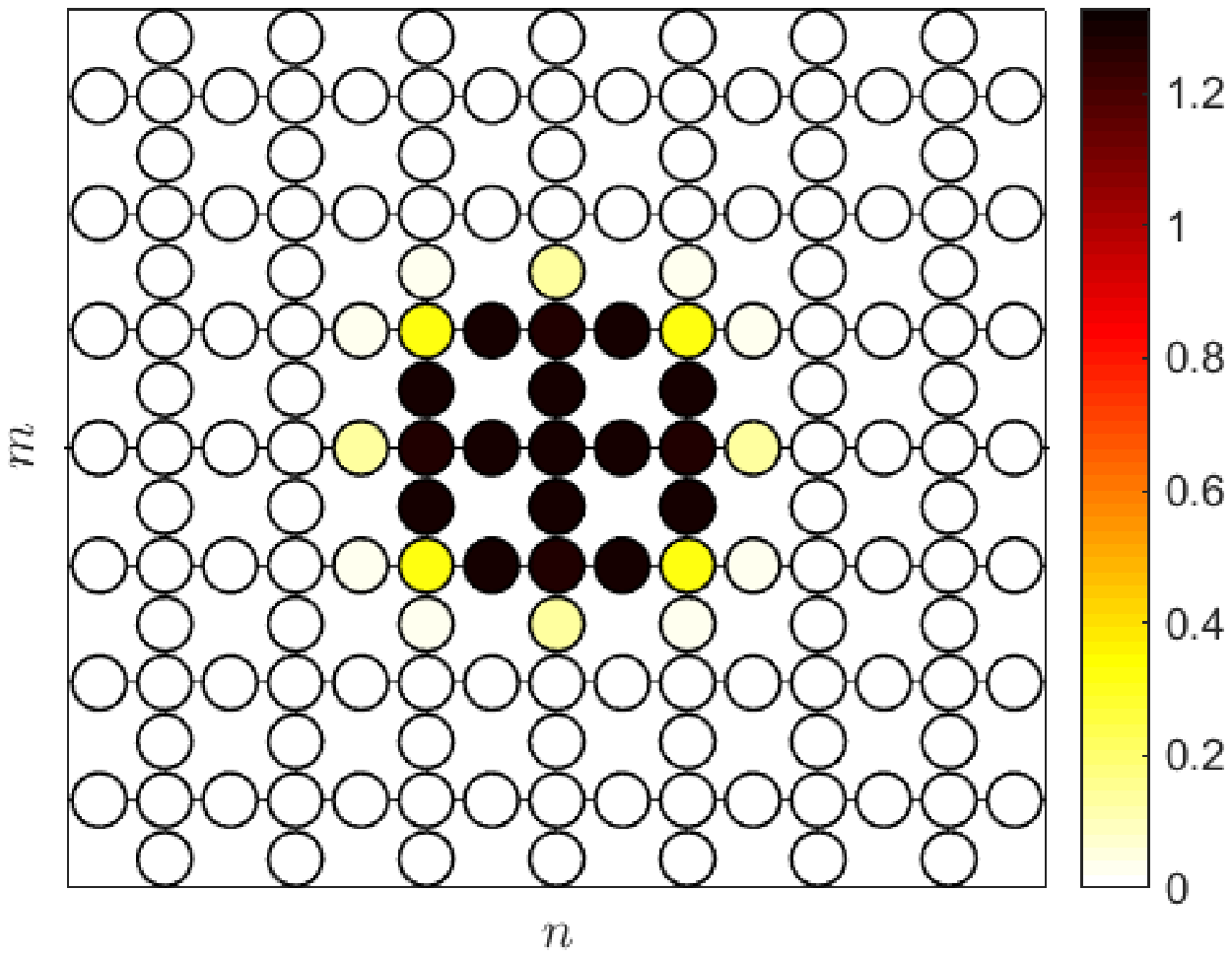}\label{subfig:prof_site_c_0_05_e}}\\
		\subfloat[]{\includegraphics[scale=0.16]{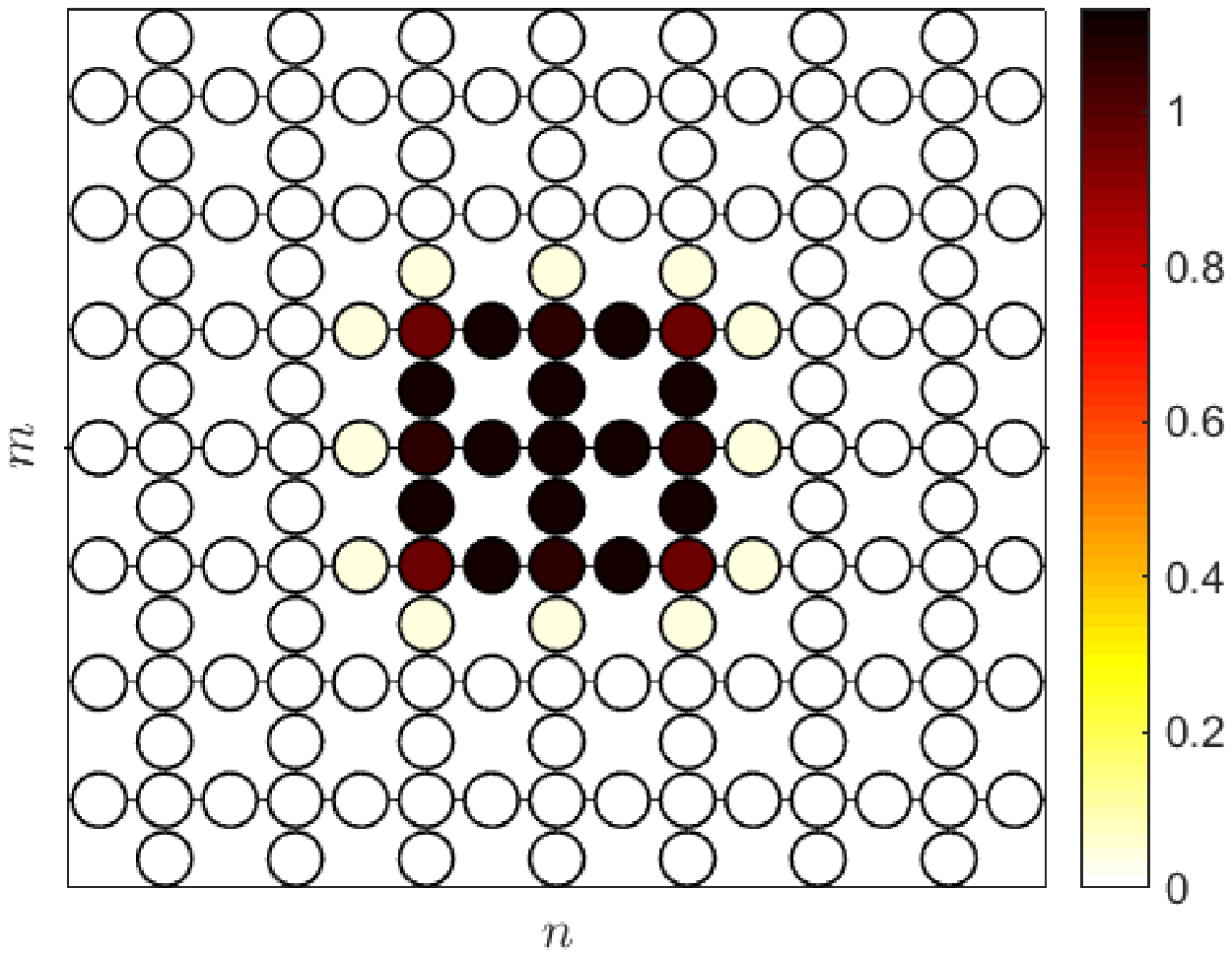}\label{subfig:prof_site_c_0_05_f}}
		\subfloat[]{\includegraphics[scale=0.16]{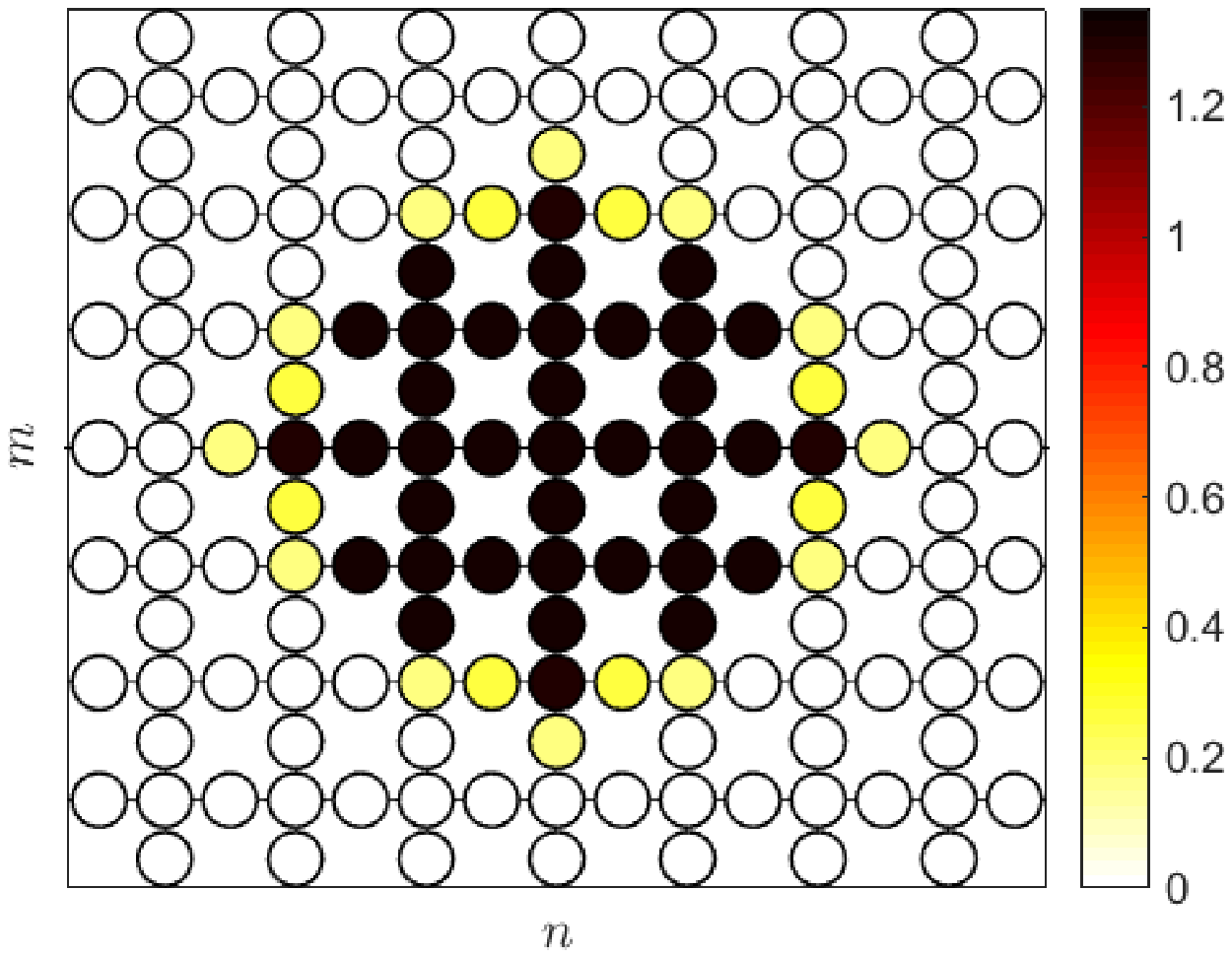}\label{subfig:prof_site_c_0_05_g}}
		\subfloat[]{\includegraphics[scale=0.16]{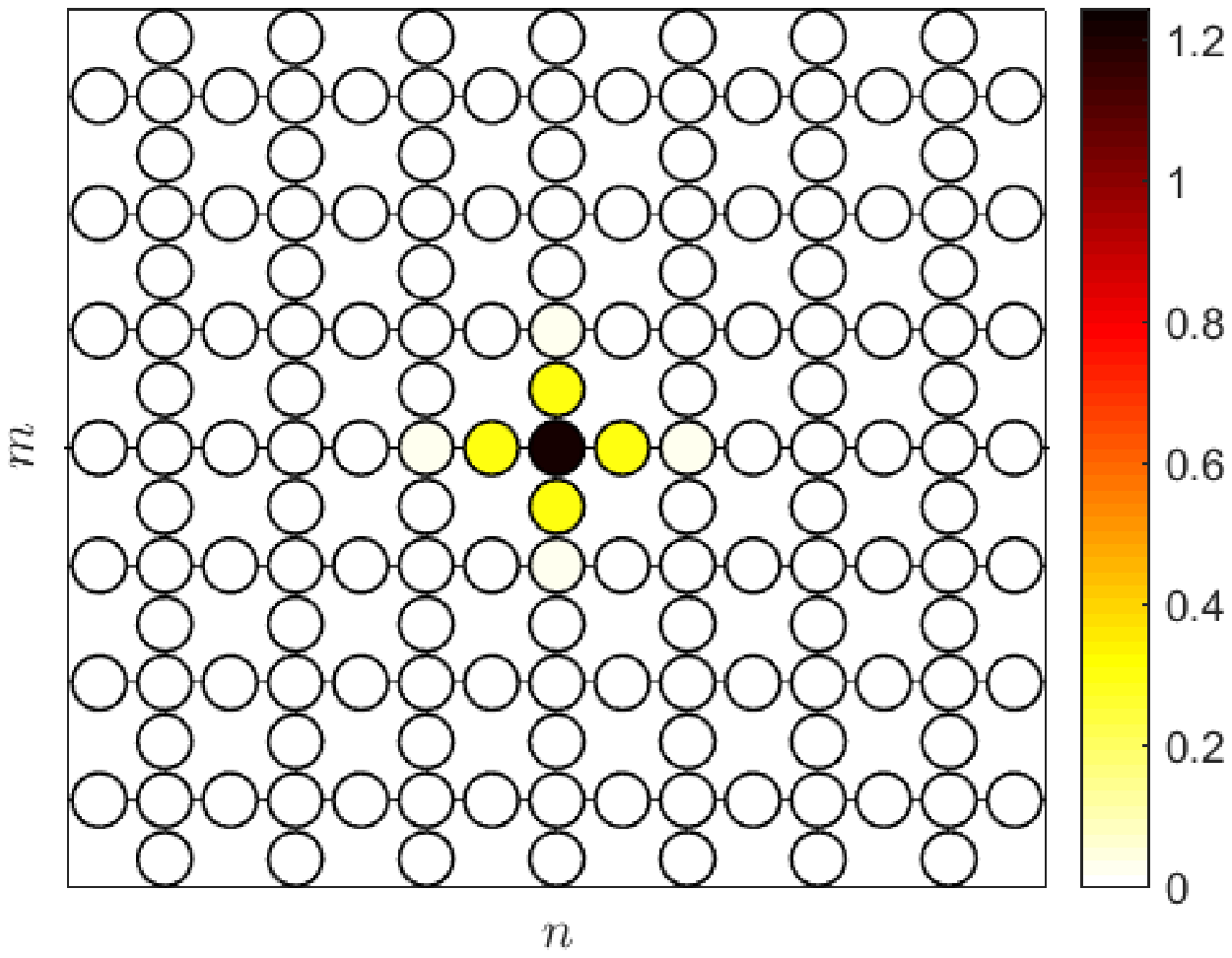}\label{subfig:prof_site_c_0_05_h}}
		\subfloat[]{\includegraphics[scale=0.16]{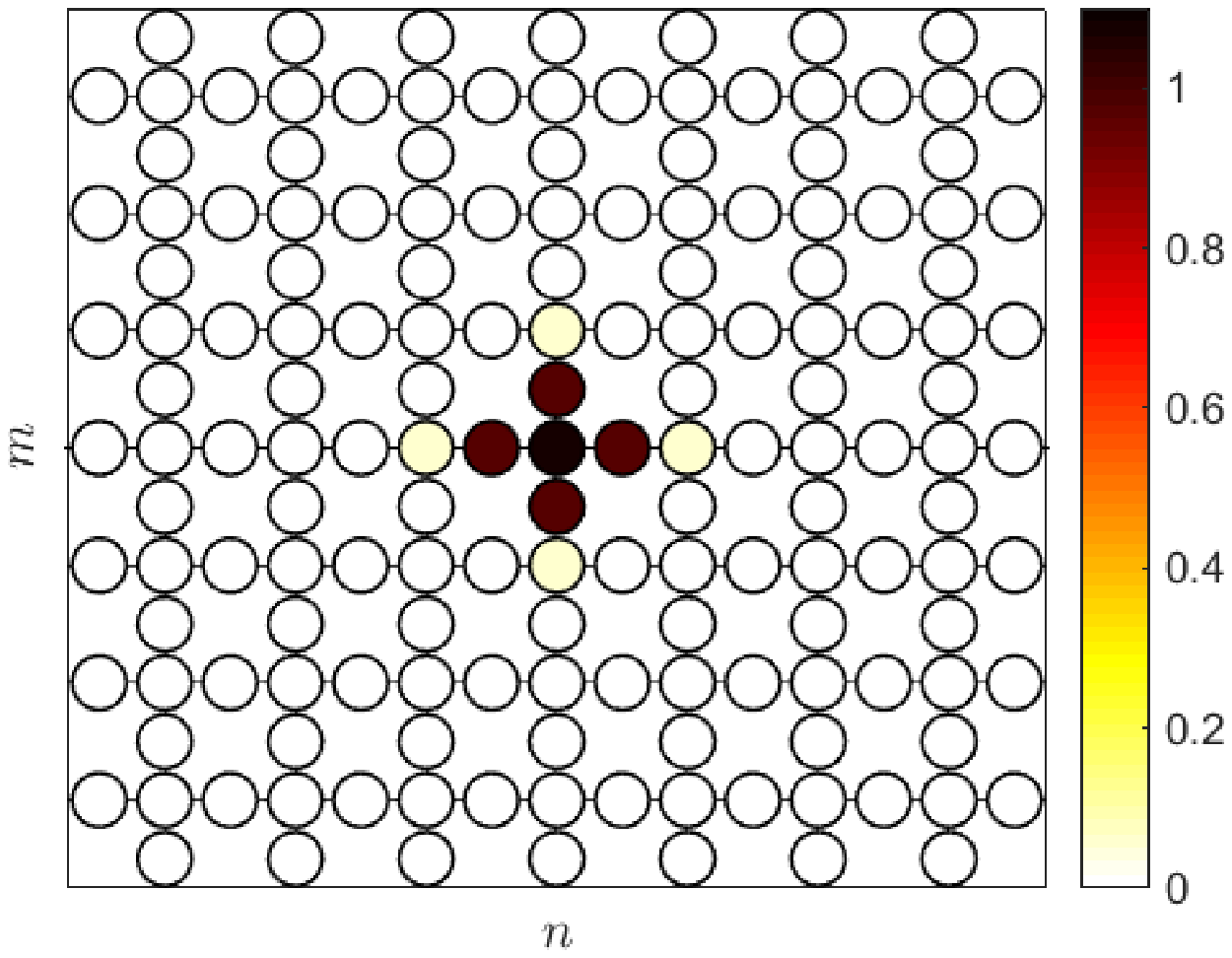}\label{subfig:prof_site_c_0_05_i}}
		\subfloat[]{\includegraphics[scale=0.16]{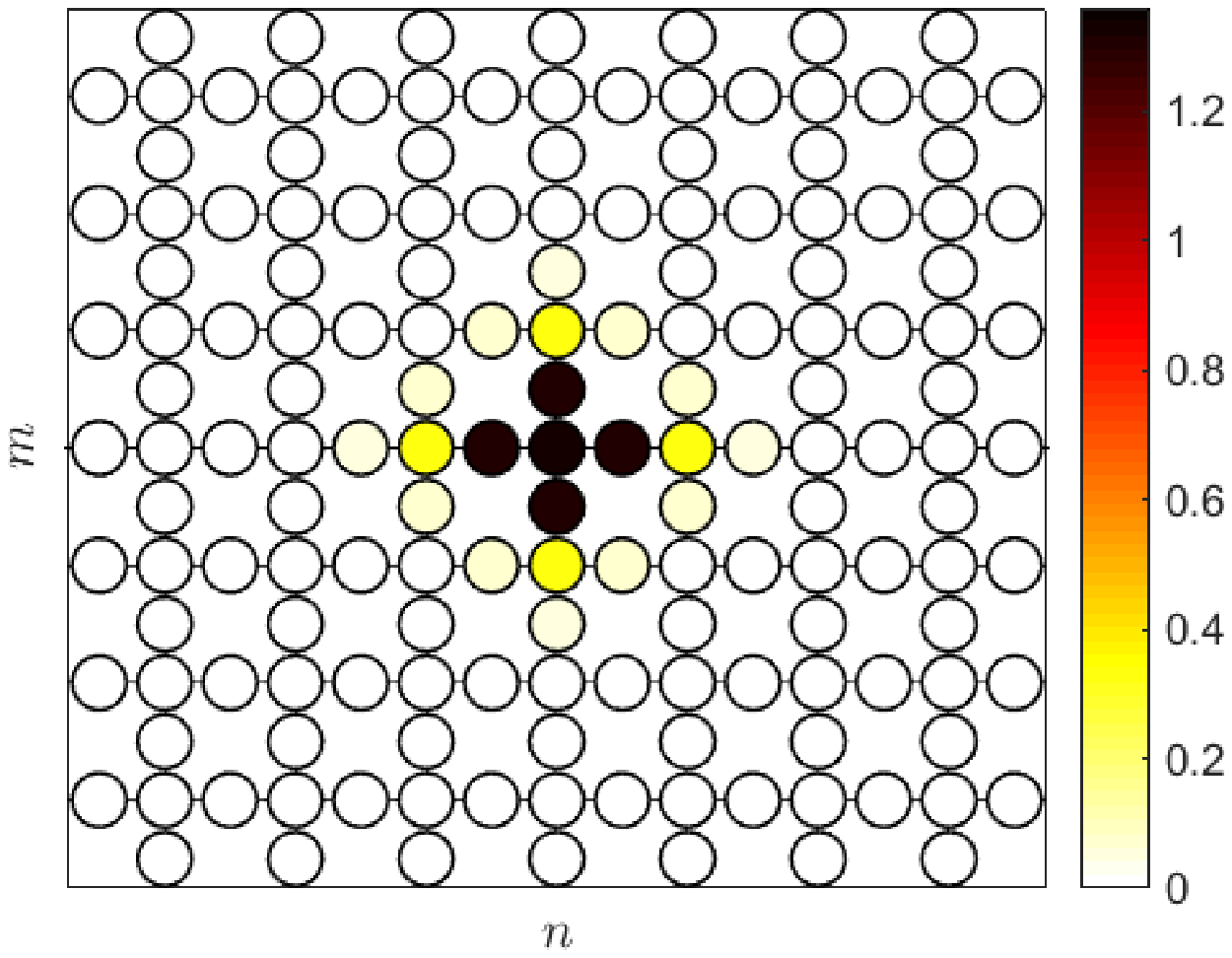}\label{subfig:prof_site_c_0_05_j}}\\
		\subfloat[]{\includegraphics[scale=0.16]{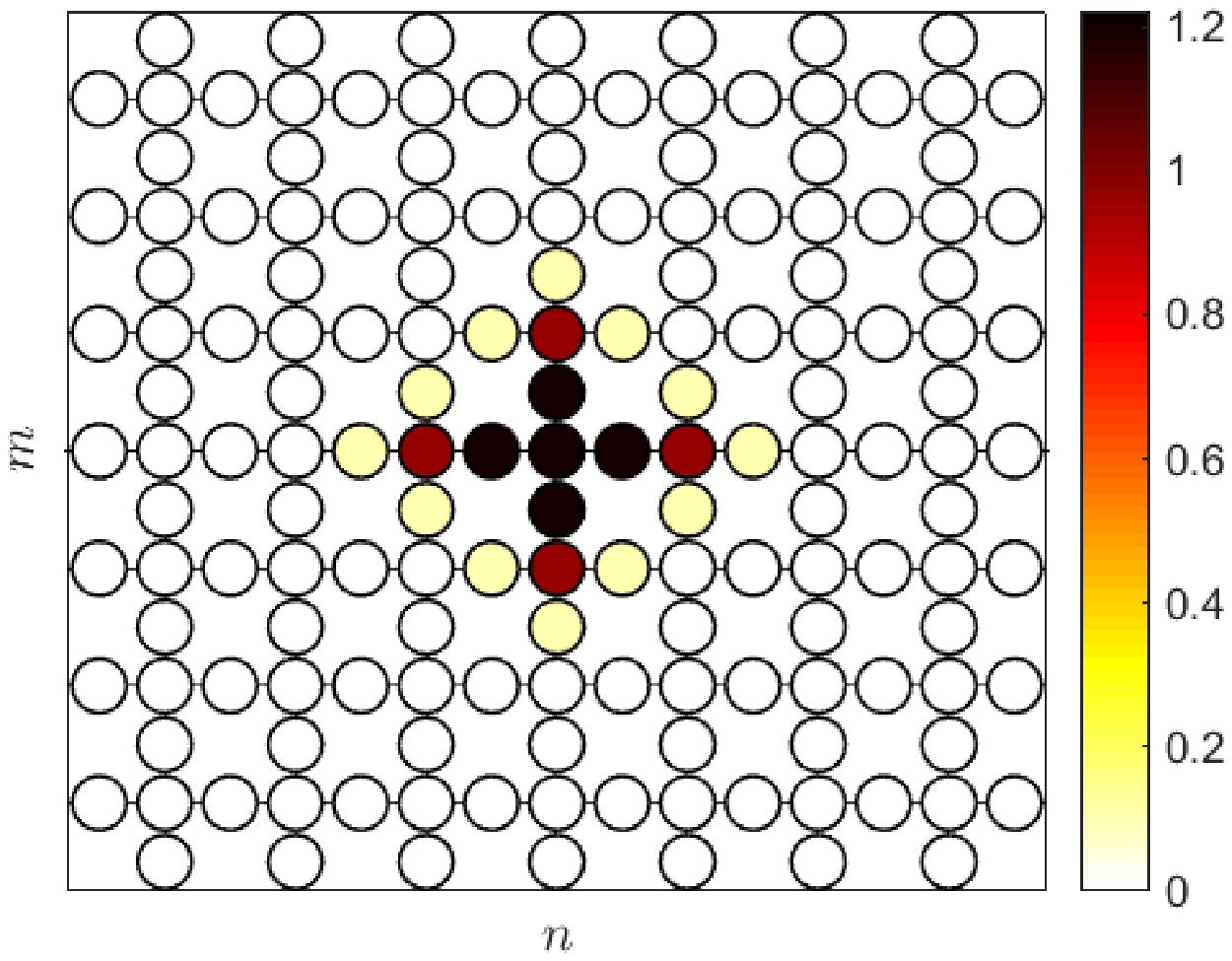}\label{subfig:prof_site_c_0_05_k}}
		\subfloat[]{\includegraphics[scale=0.16]{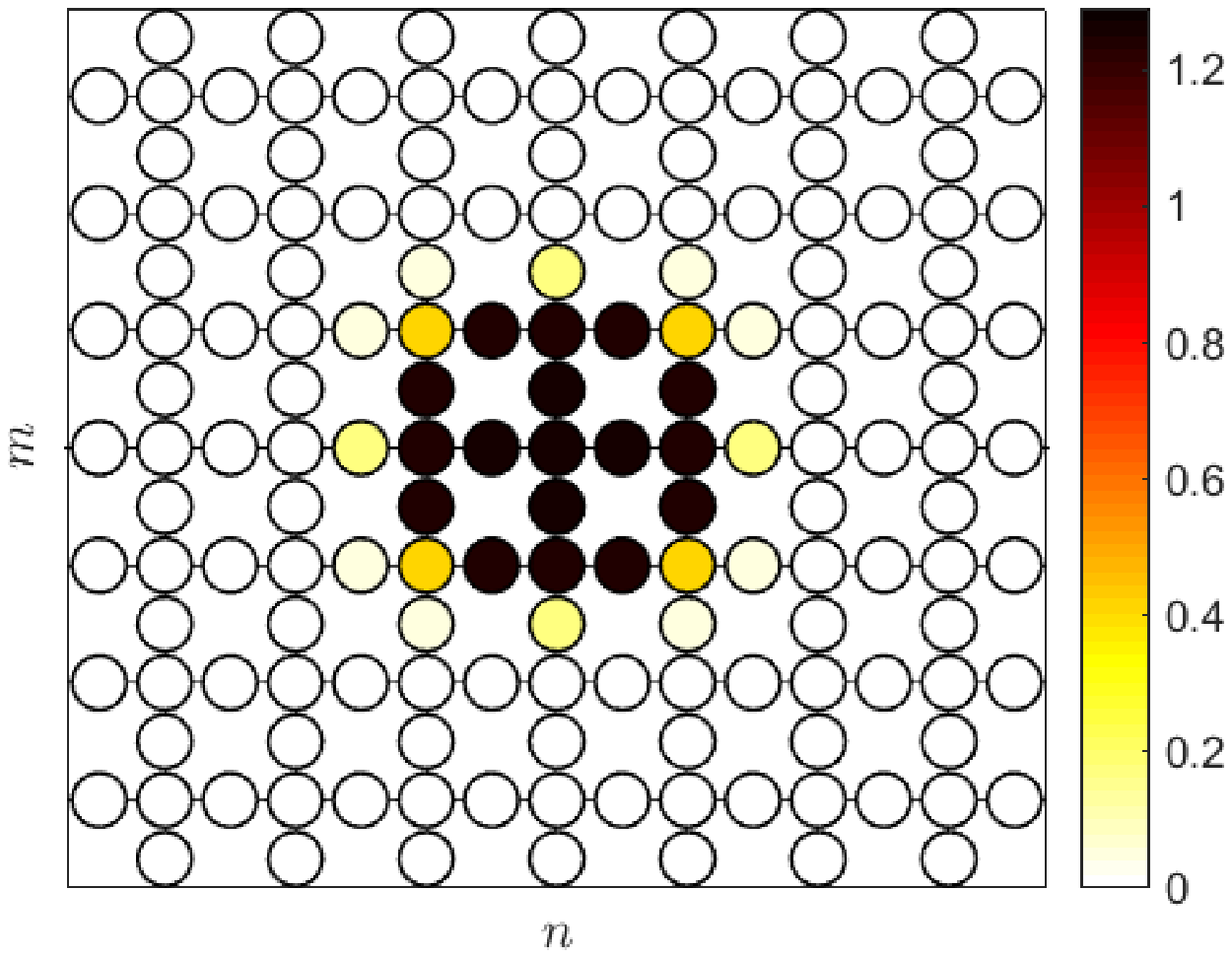}\label{subfig:prof_site_c_0_05_l}}
		\subfloat[]{\includegraphics[scale=0.16]{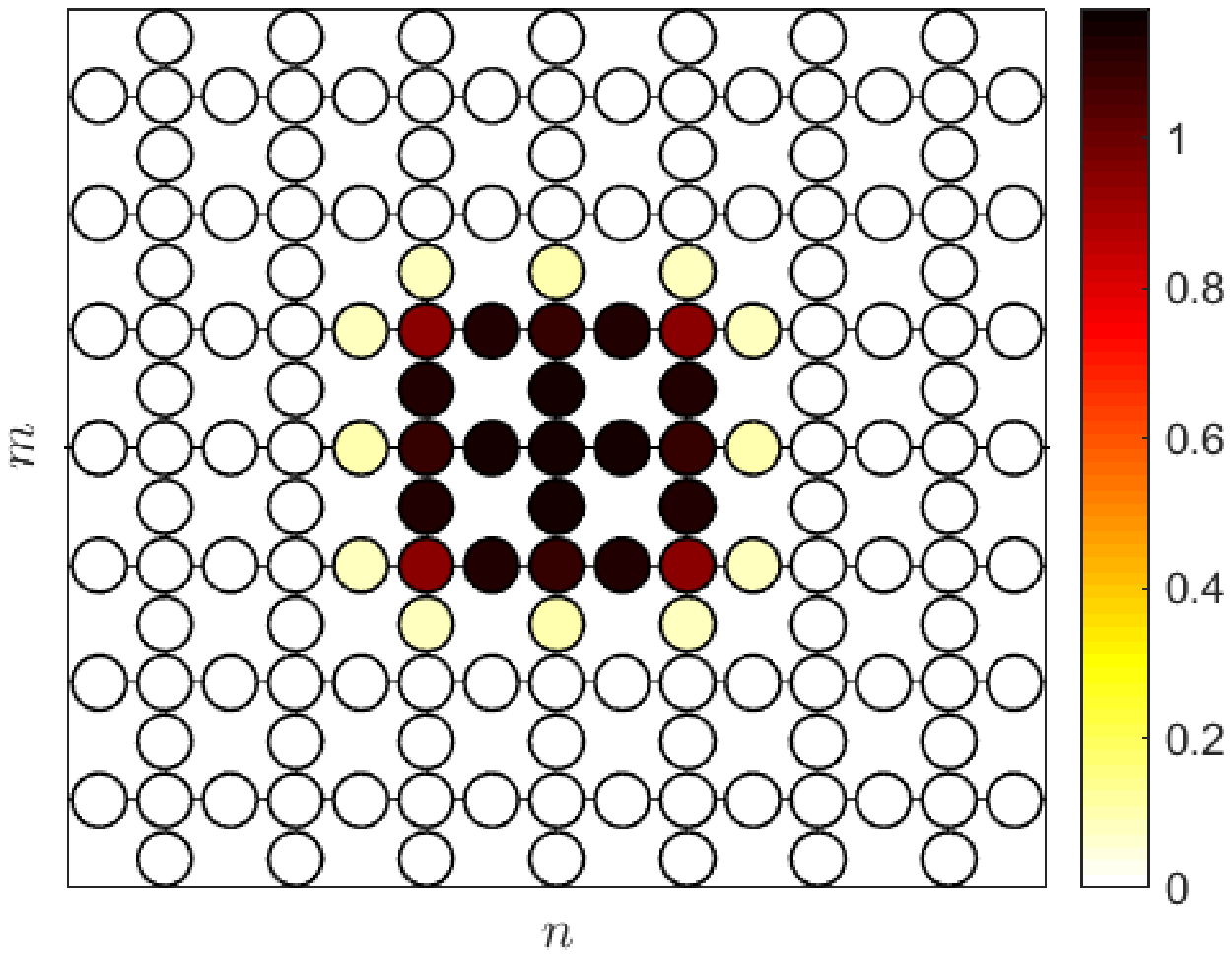}\label{subfig:prof_site_c_0_05_m}}
		\subfloat[]{\includegraphics[scale=0.16]{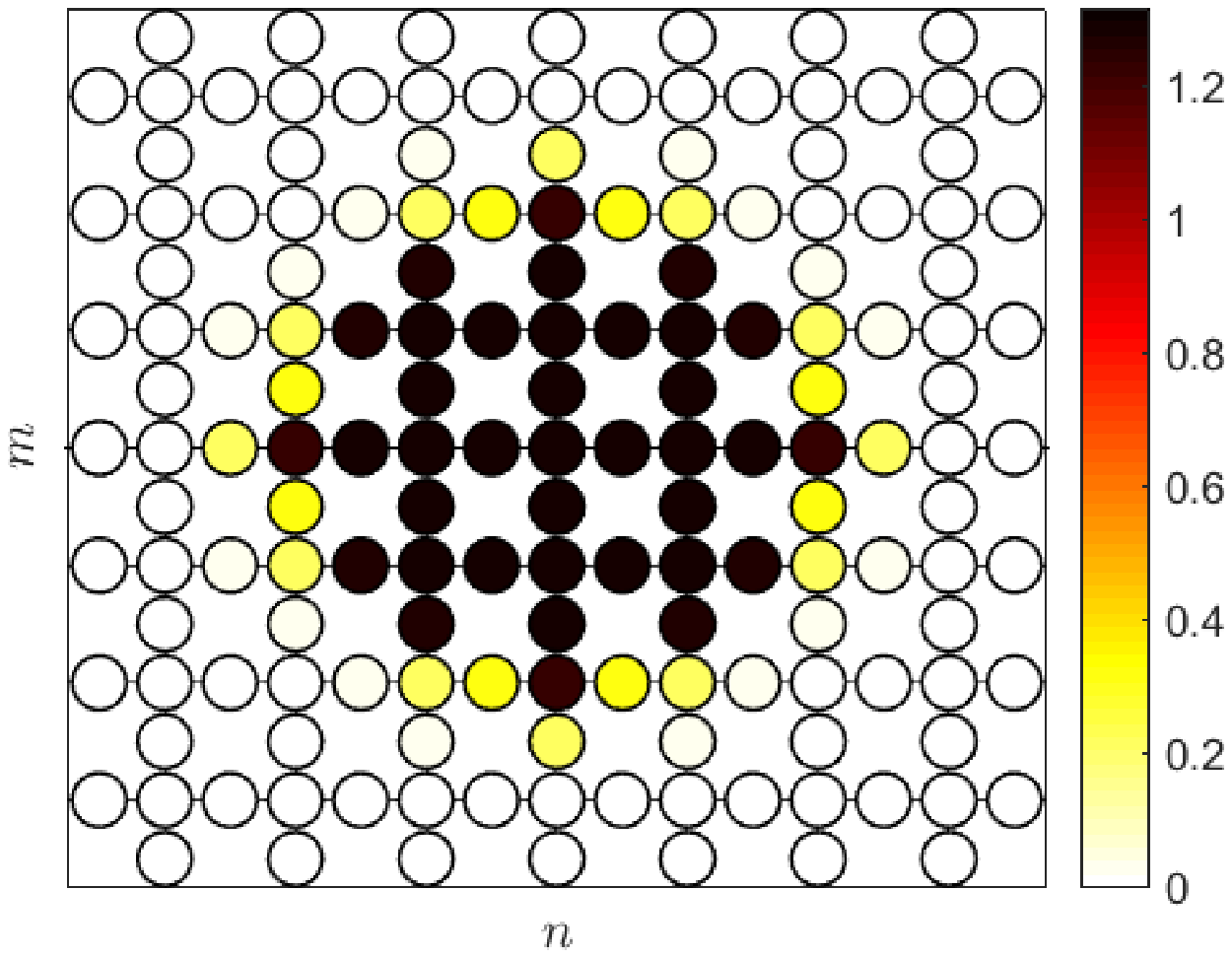}\label{subfig:prof_site_c_0_05_n}}\\
		\renewcommand{\thesubfigure}{1}
		\subfloat[]{\includegraphics[scale=0.16]{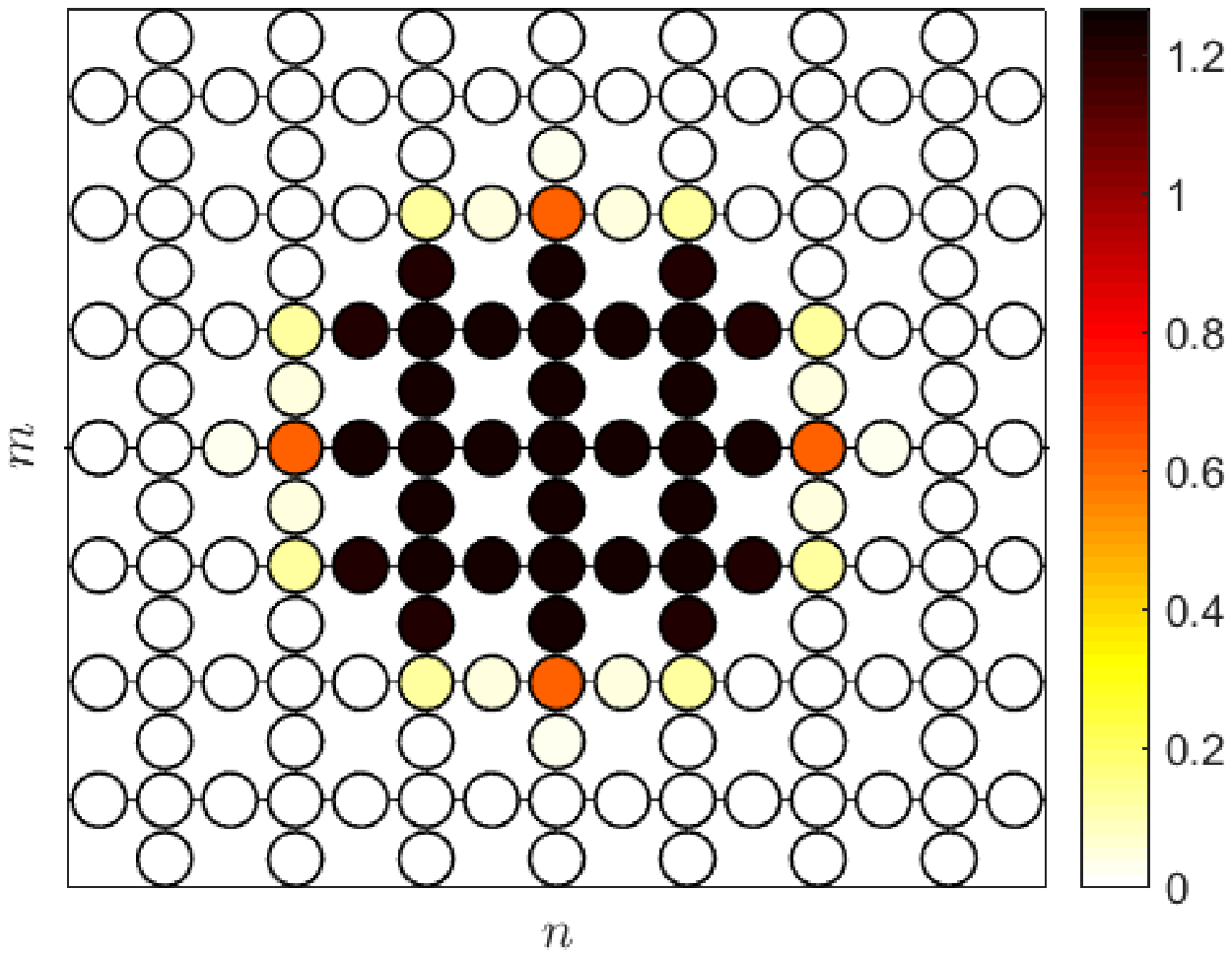}\label{subfig:prof_site_c_0_05_1}}
		\renewcommand{\thesubfigure}{2}
		\subfloat[]{\includegraphics[scale=0.16]{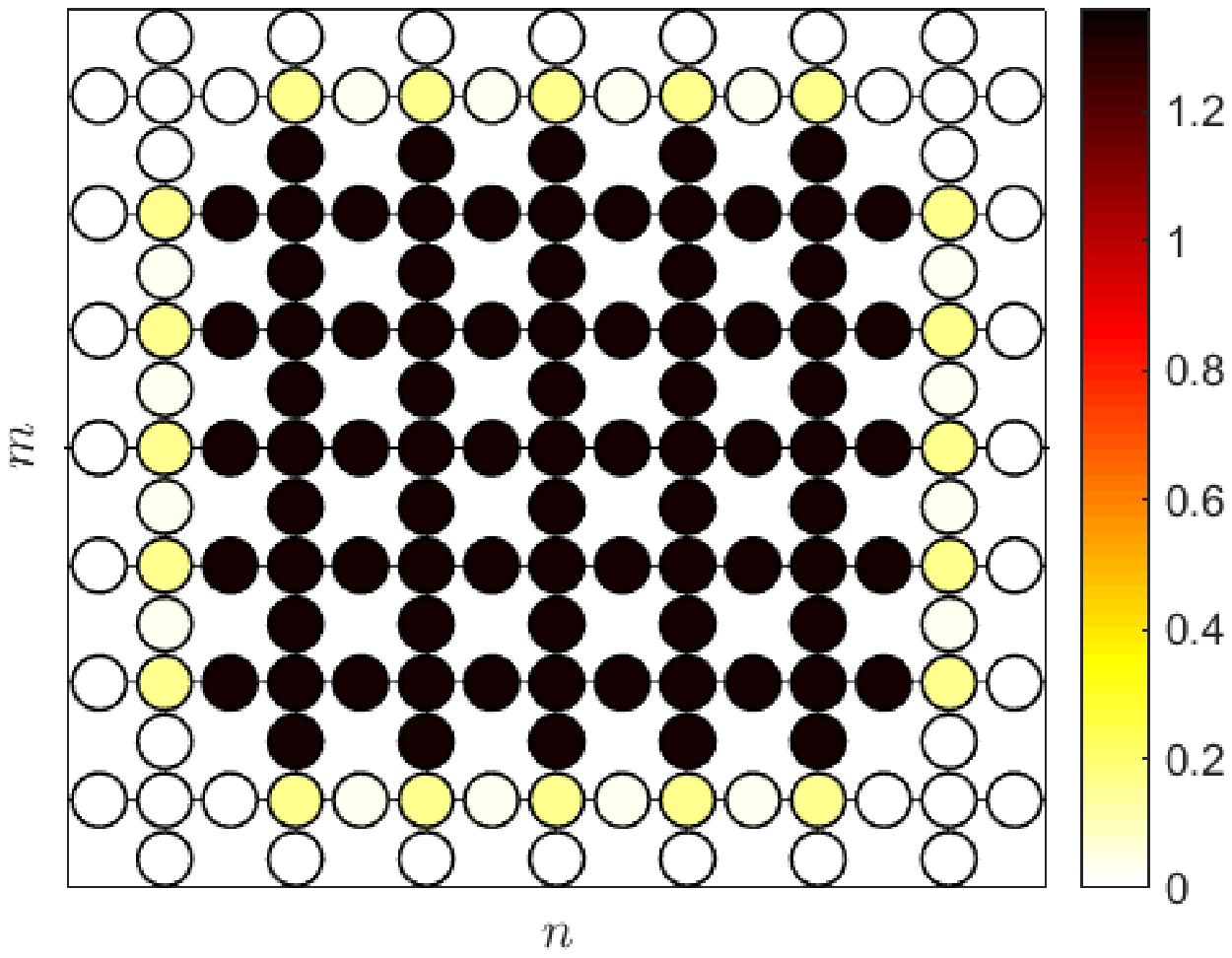}\label{subfig:prof_site_c_0_05_2}}
		\renewcommand{\thesubfigure}{3}
		\subfloat[]{\includegraphics[scale=0.16]{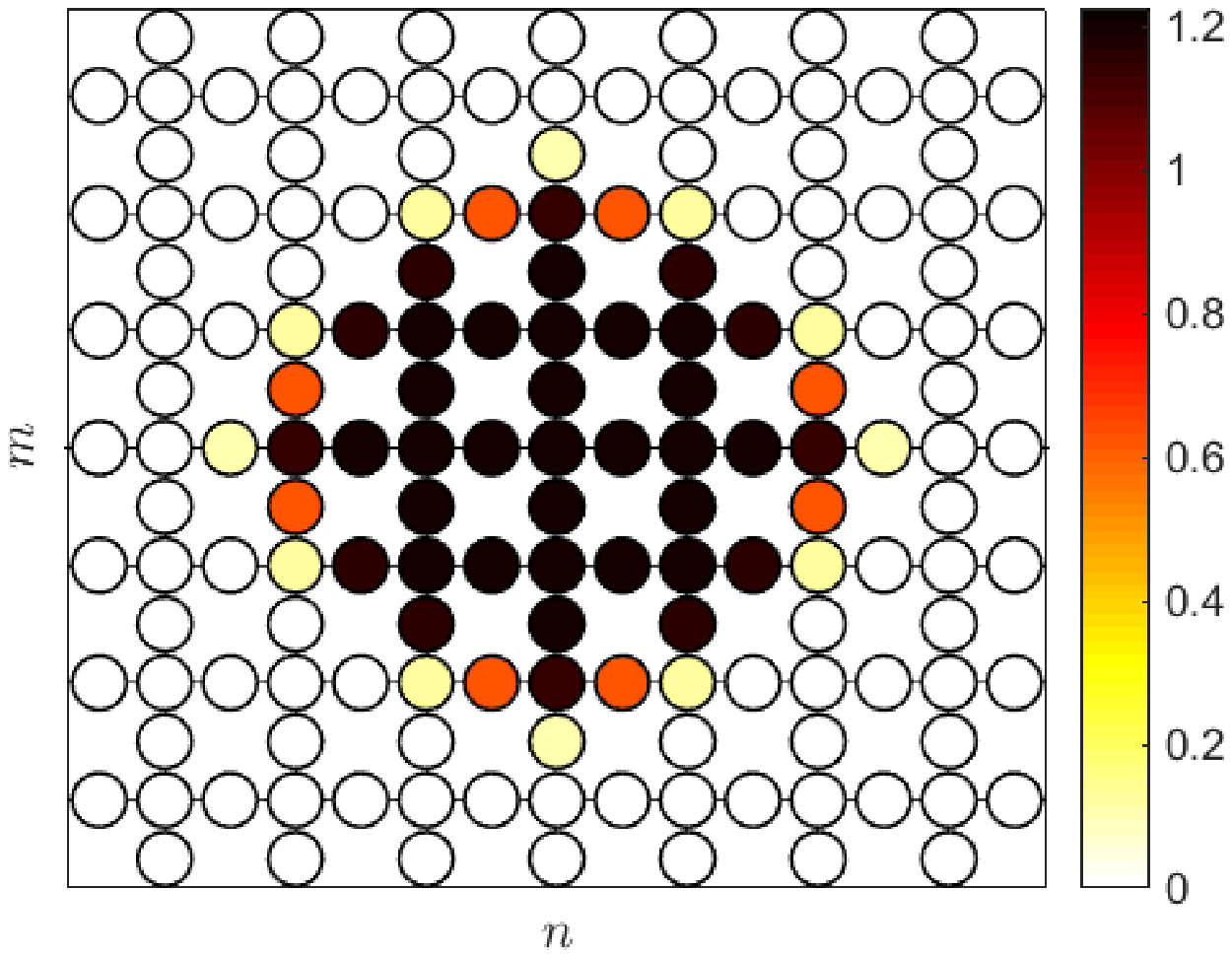}\label{subfig:prof_site_c_0_05_3}}
		\renewcommand{\thesubfigure}{4}
		\subfloat[]{\includegraphics[scale=0.16]{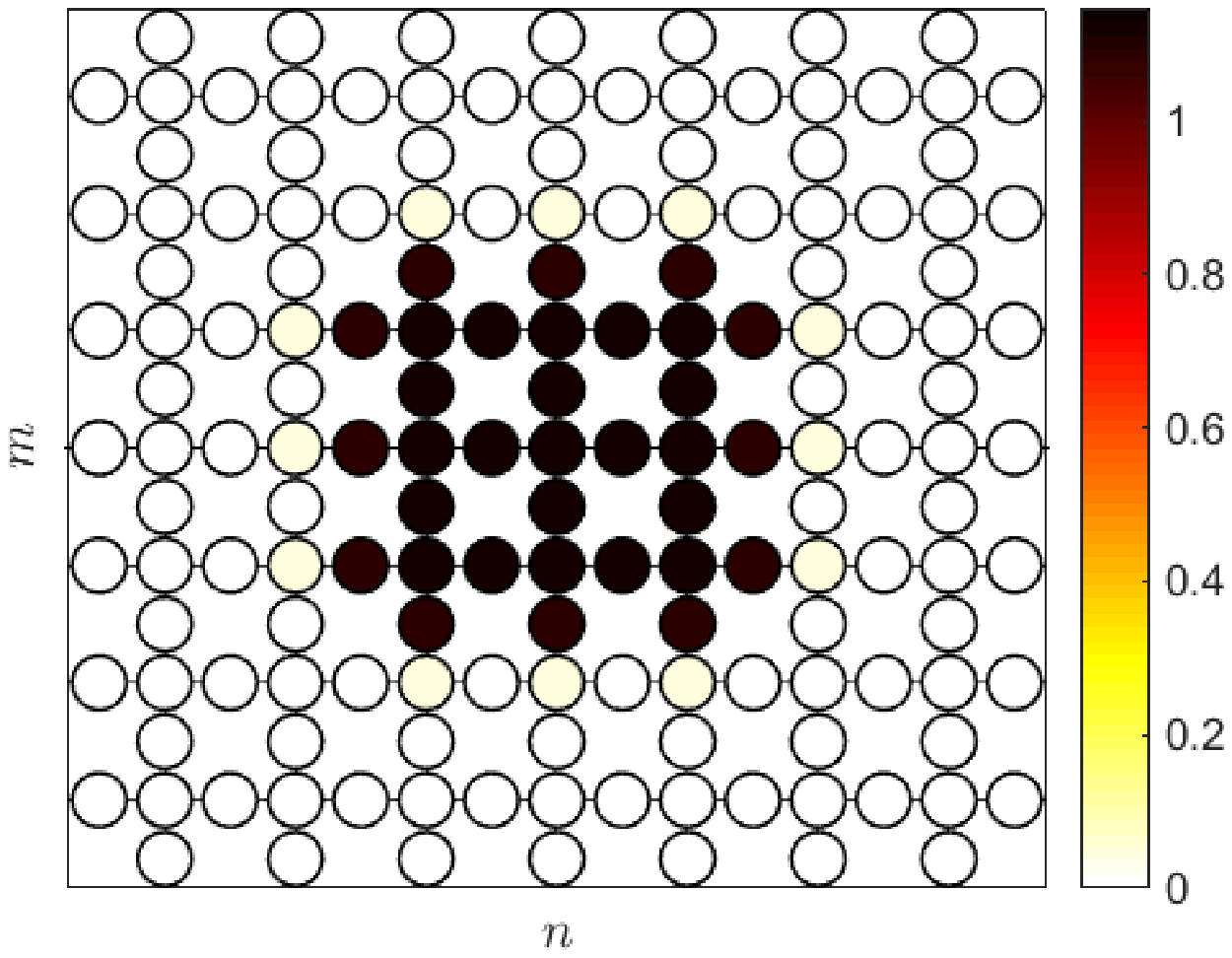}\label{subfig:prof_site_c_0_05_4}}\\
		\caption{Top-view of localized solution profiles for site-centred solutions that correspond to the indicated points in Fig.\ \ref{fig:bifur_site}.}
		\label{fig:prof_site}
	\end{figure*}
	
	\begin{figure*}[htbp!]
	\centering
	\subfloat[$c=0.05$]{\includegraphics[scale=0.44]{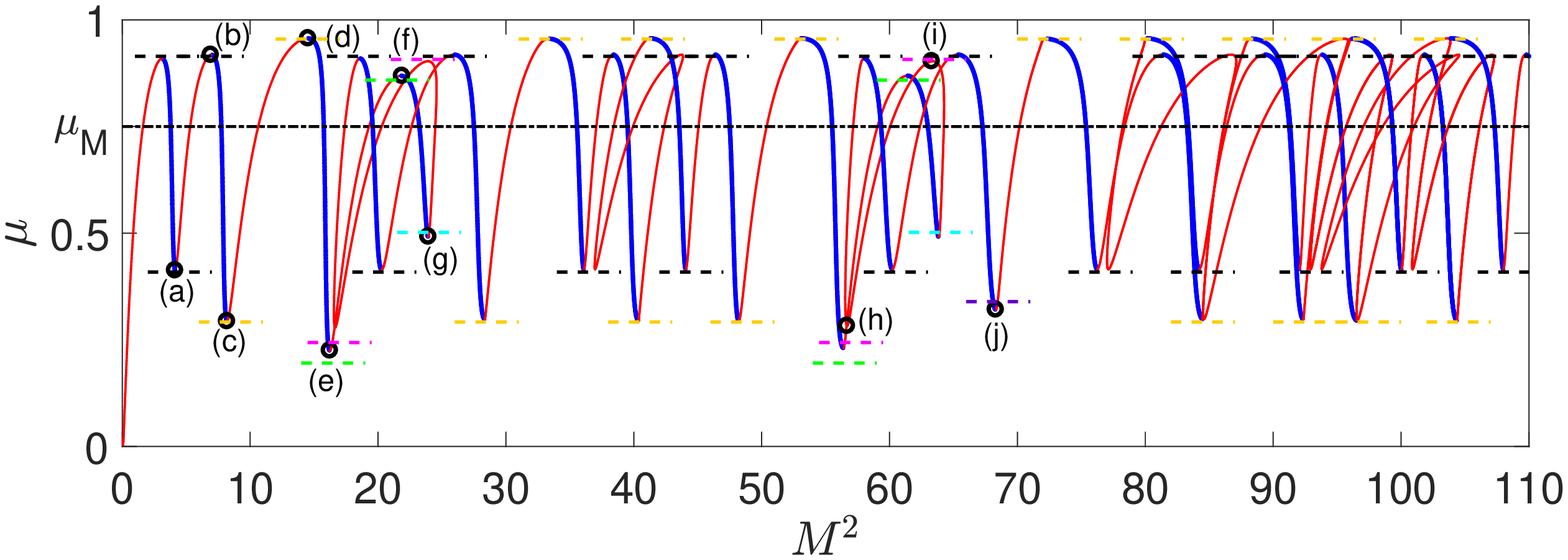}\label{subfig:snake_bond_c_0_05}}\\
	\subfloat[$c=0.1$]{\includegraphics[scale=0.44]{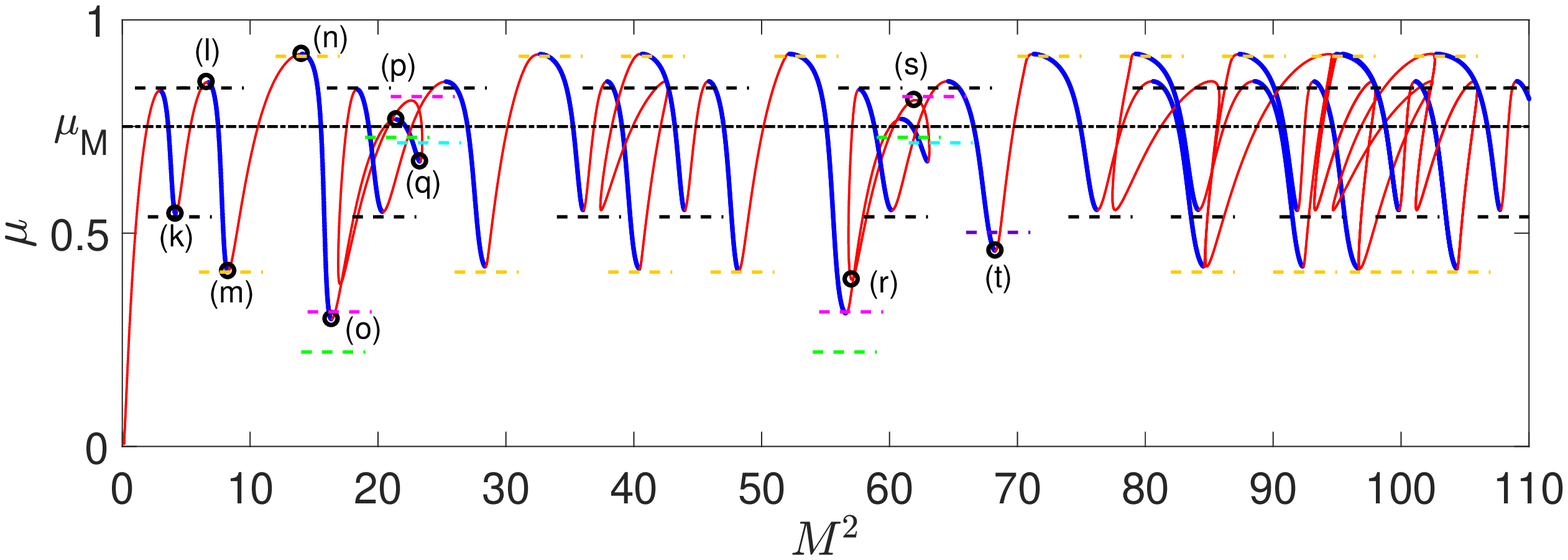}\label{subfig:snake_bond_c_0_10}}
	\caption{
		{The same as figure \ref{fig:bifur_site}, but for bond-centred solutions.  Panels (a) and (b) show bifurcation diagrams of bond-centred solutions
			for c = 0.05 and 0.1. 
			The green, black, magenta, {orange}, cyan, and purple line colors correspond to saddle-node bifurcations from our active-cell approximations of type
			1-6, see Section \ref{sec:saddle}. 
			Solution profiles at the turning points labelled as	(a)-(t) in the top panel are shown in Fig. \ref{fig:prof_bond}. 
		}
	}
	\label{fig:bifur_bond}
	\end{figure*}
	
	\begin{figure*}[h!]
		\centering
		\subfloat[]{\includegraphics[scale=0.16]{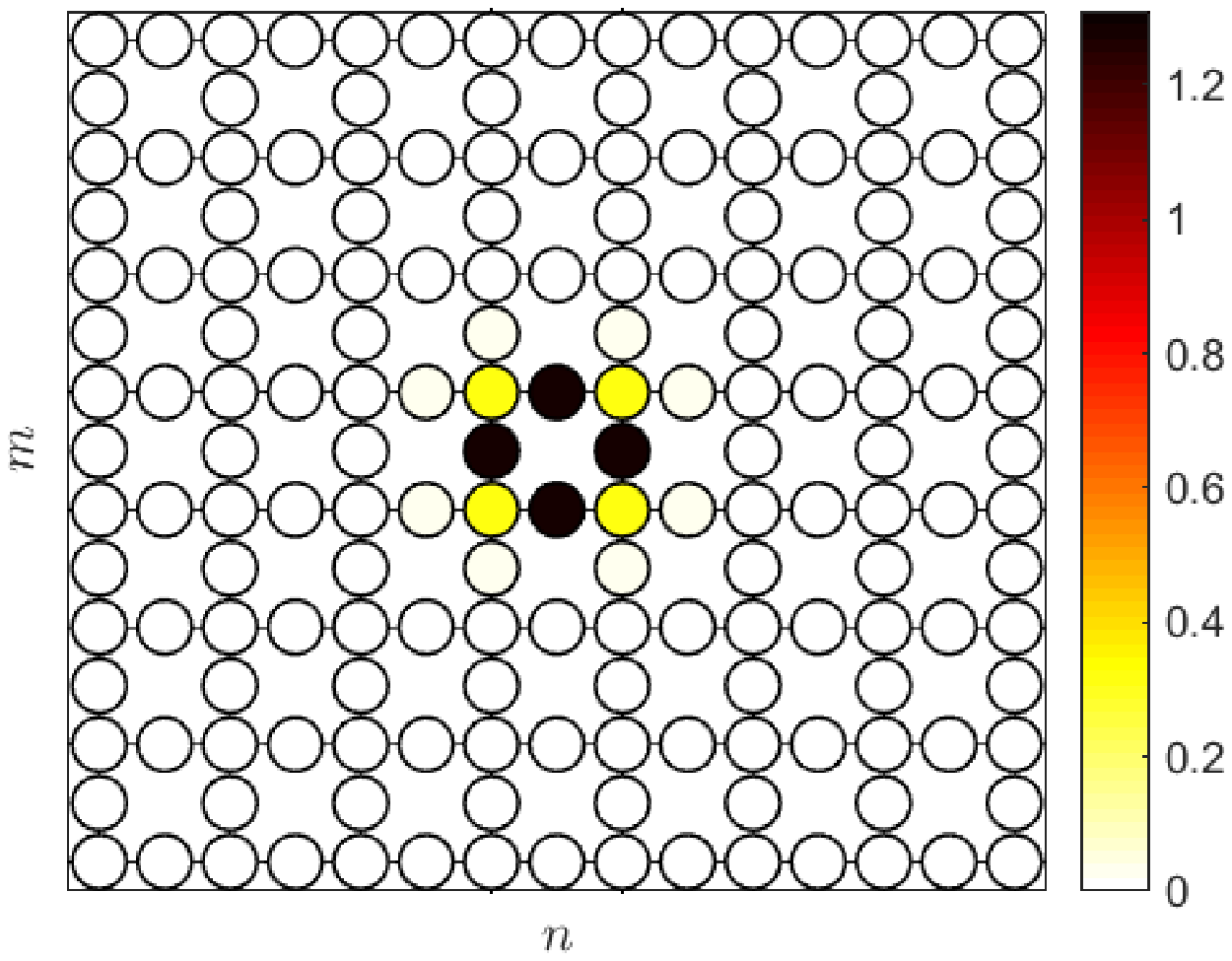}\label{subfig:prof_bond_c_0_05_a}}
		\subfloat[]{\includegraphics[scale=0.16]{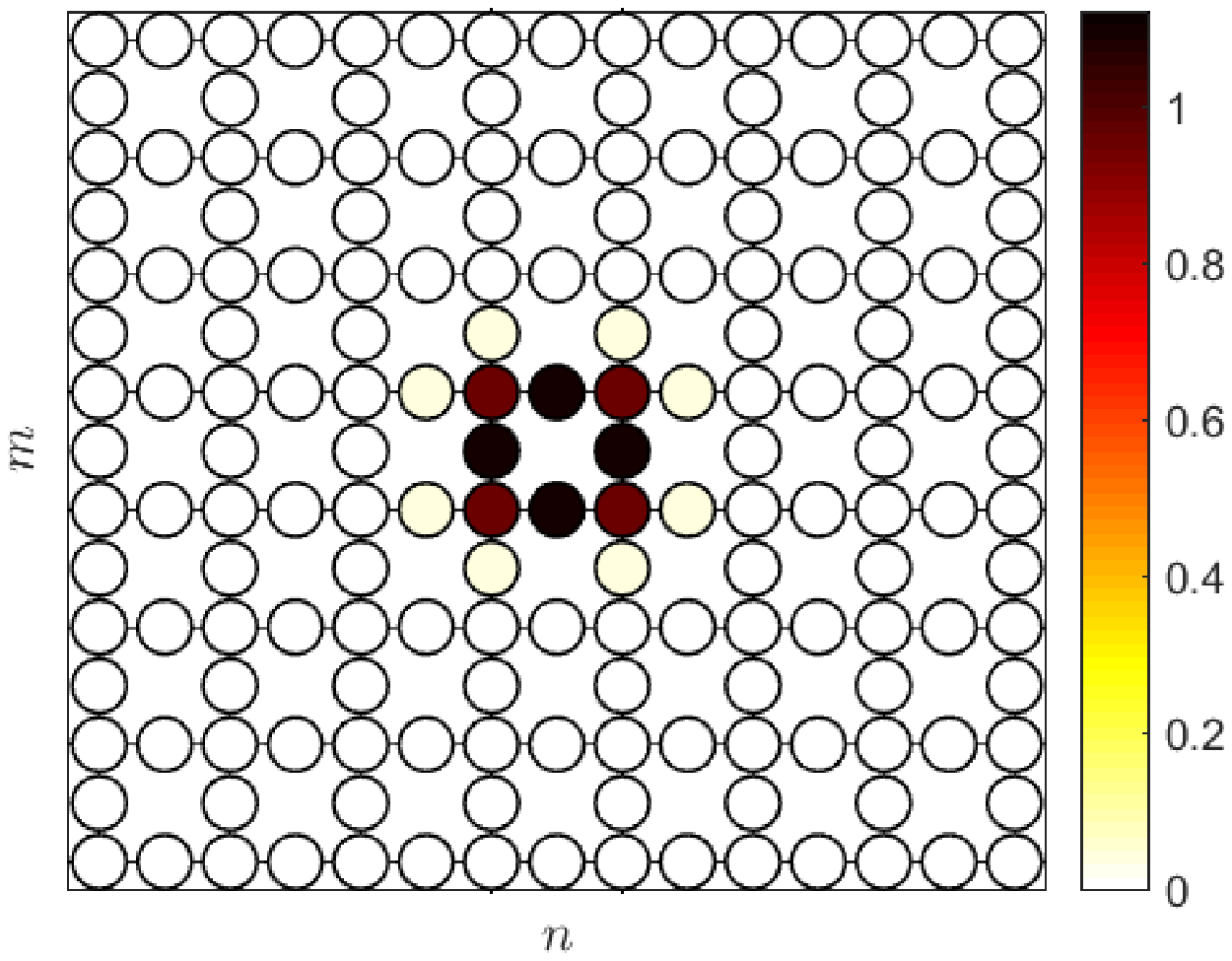}\label{subfig:prof_bond_c_0_05_b}}
		\subfloat[]{\includegraphics[scale=0.16]{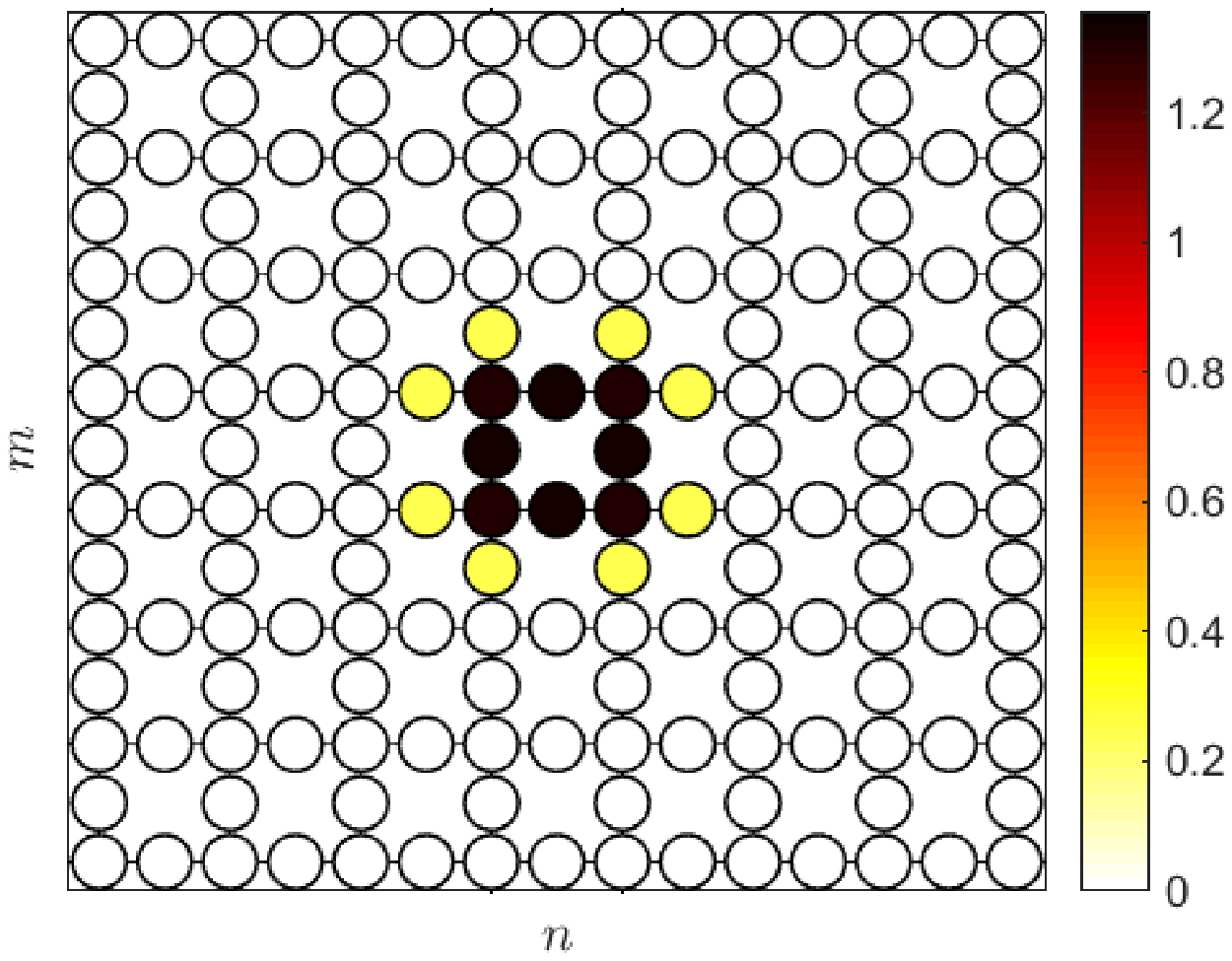}\label{subfig:prof_bond_c_0_05_c}}
		\subfloat[]{\includegraphics[scale=0.16]{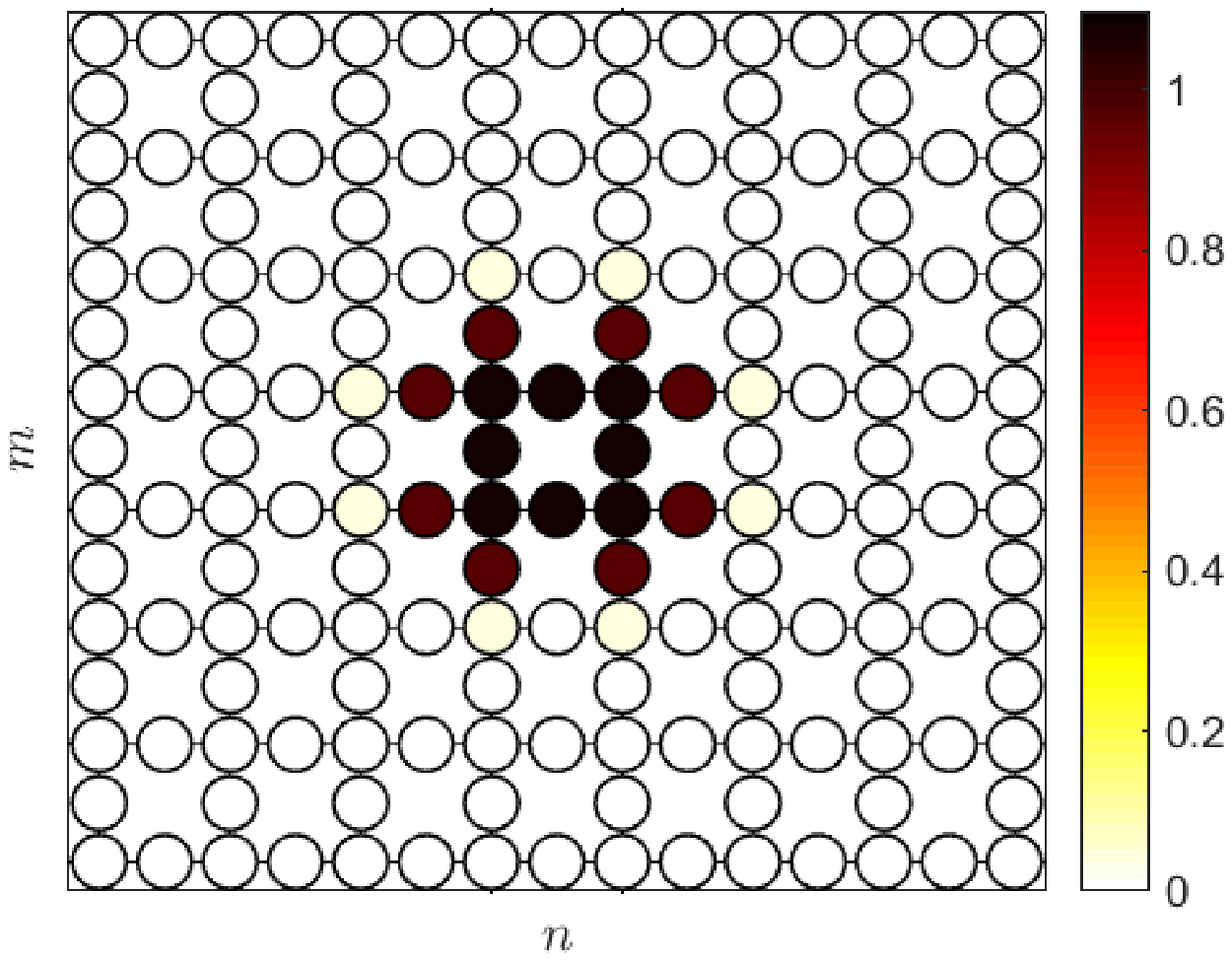}\label{subfig:prof_bond_c_0_05_d}}
		\subfloat[]{\includegraphics[scale=0.16]{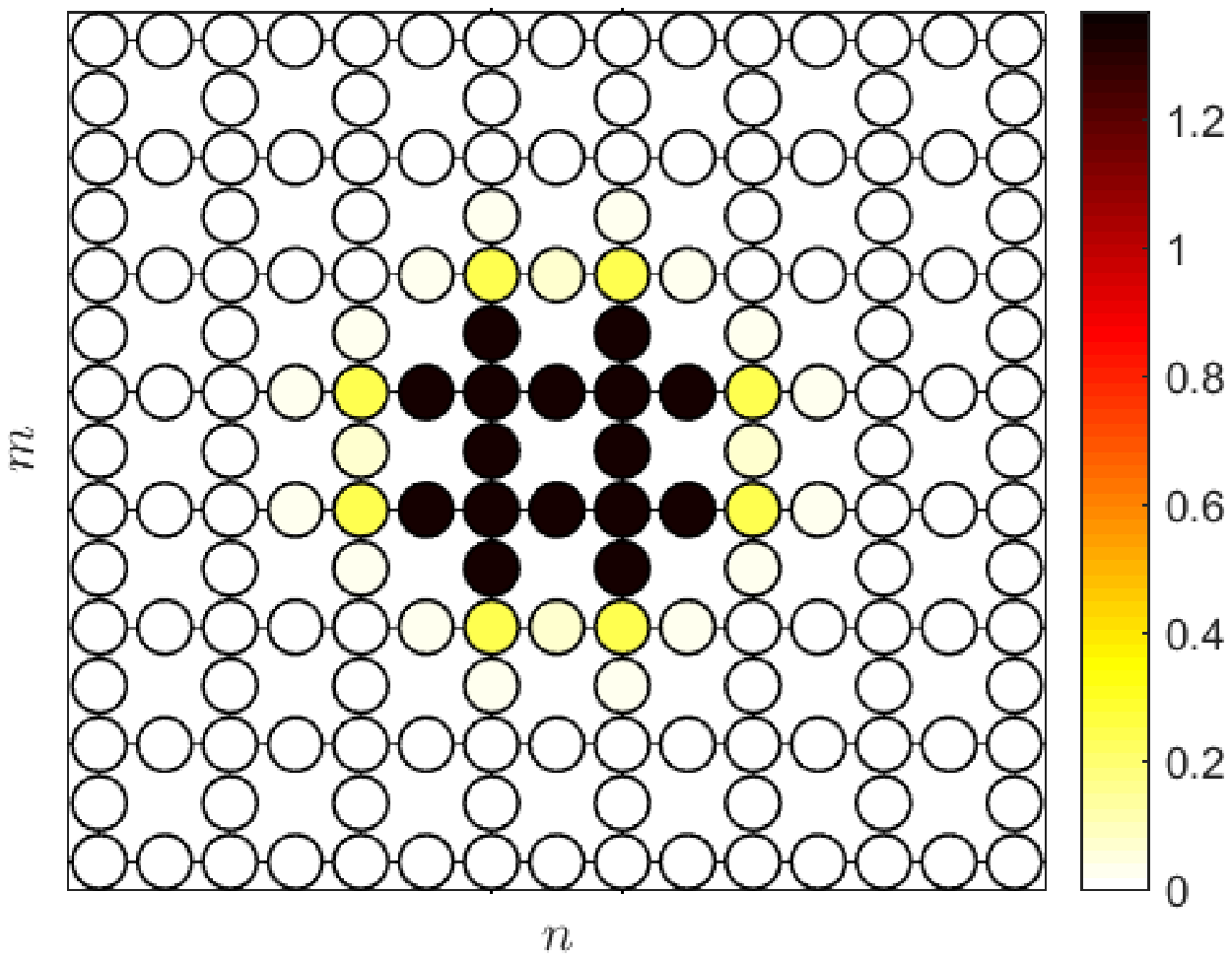}\label{subfig:prof_bond_c_0_05_e}}\\
		\subfloat[]{\includegraphics[scale=0.16]{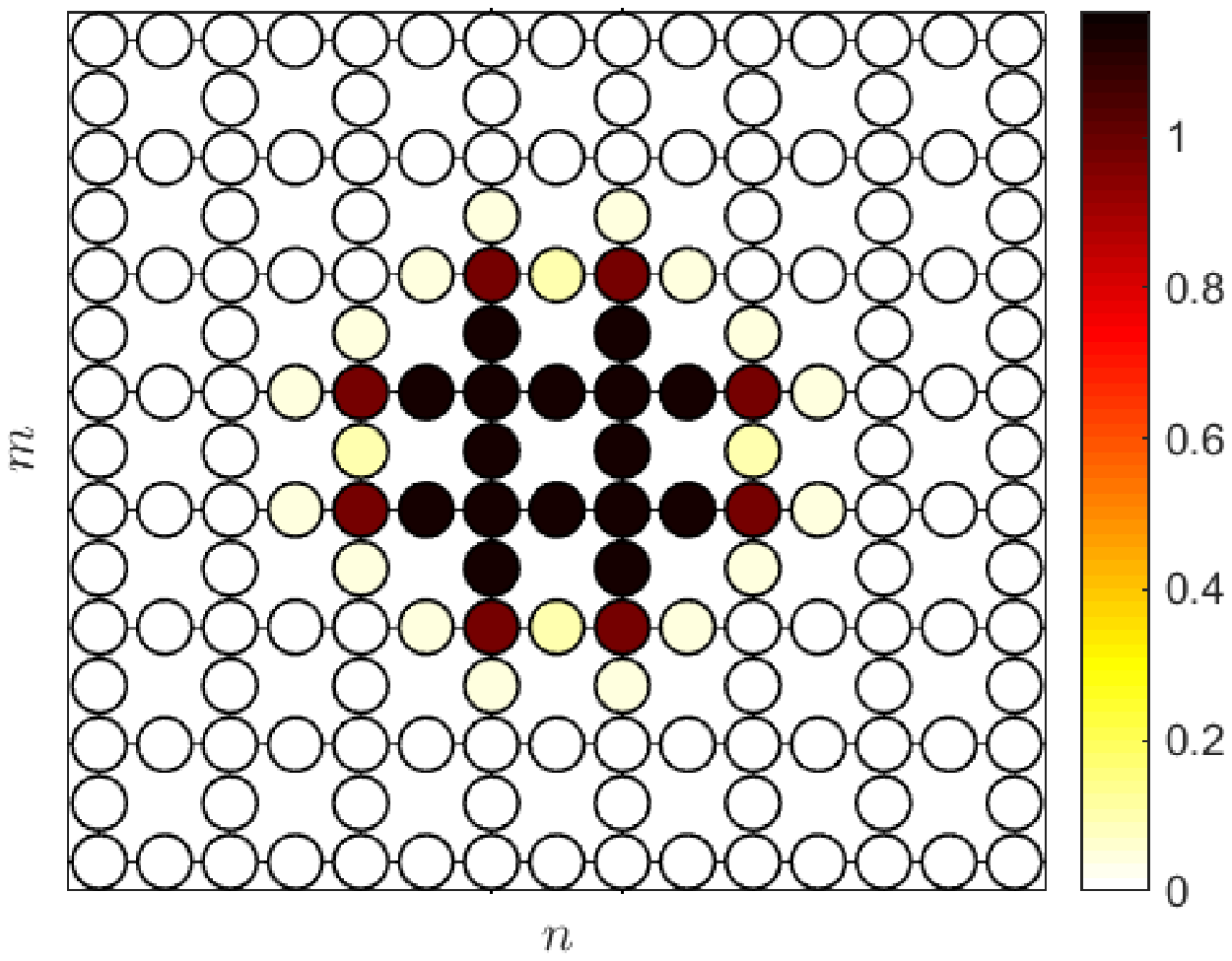}\label{subfig:prof_bond_c_0_05_f}}
		\subfloat[]{\includegraphics[scale=0.16]{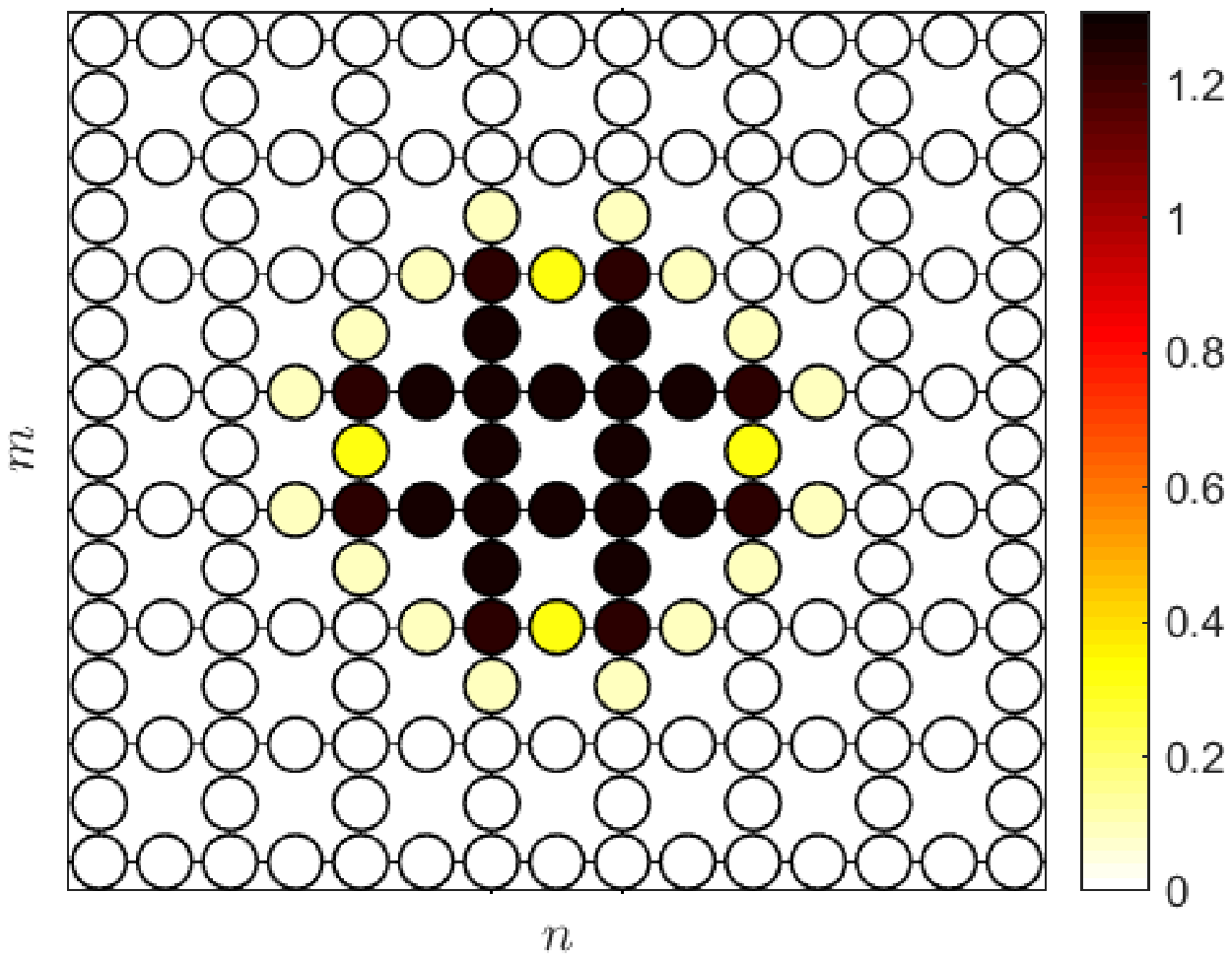}\label{subfig:prof_bond_c_0_05_g}}
		\subfloat[]{\includegraphics[scale=0.16]{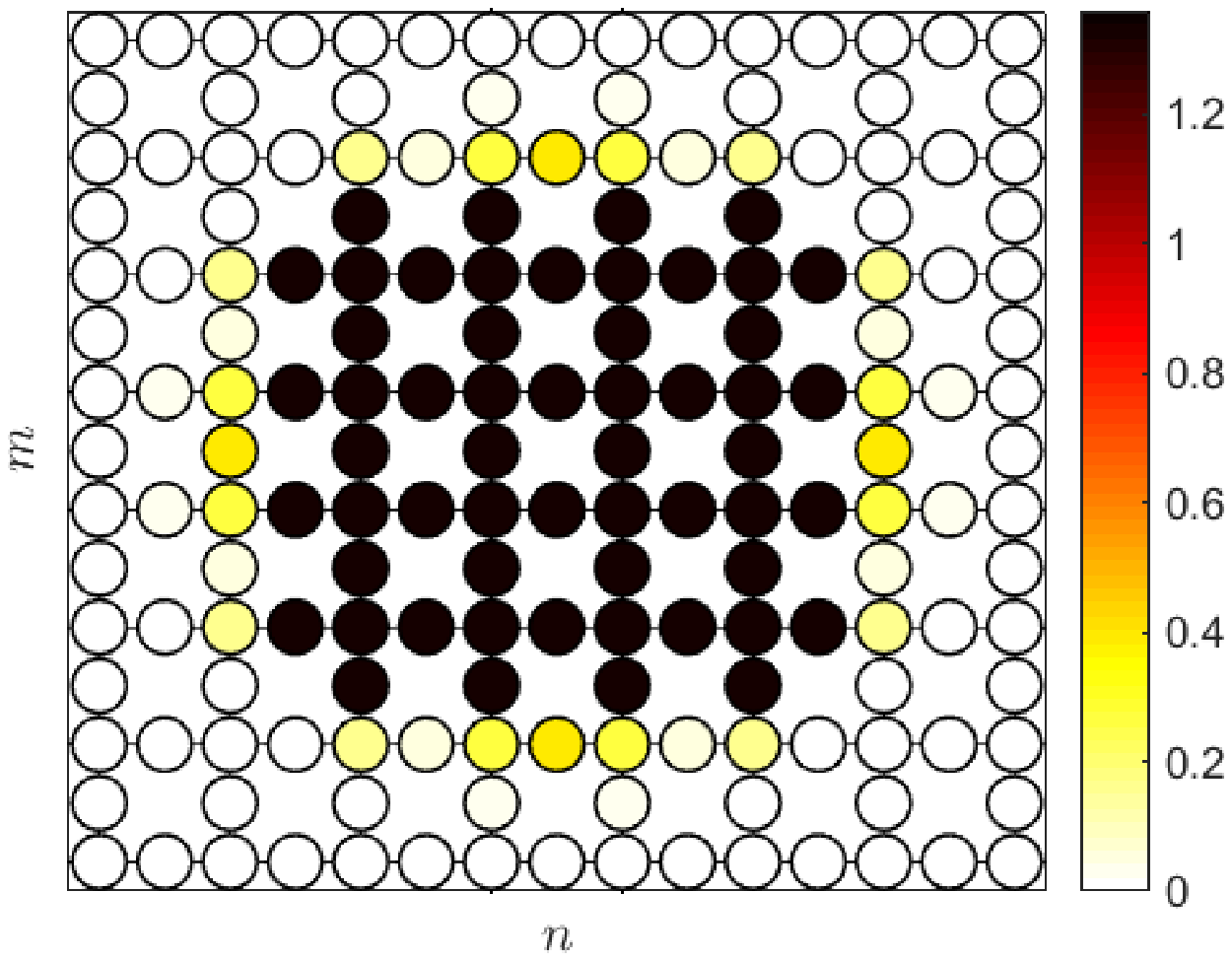}\label{subfig:prof_bond_c_0_05_h}}
		\subfloat[]{\includegraphics[scale=0.16]{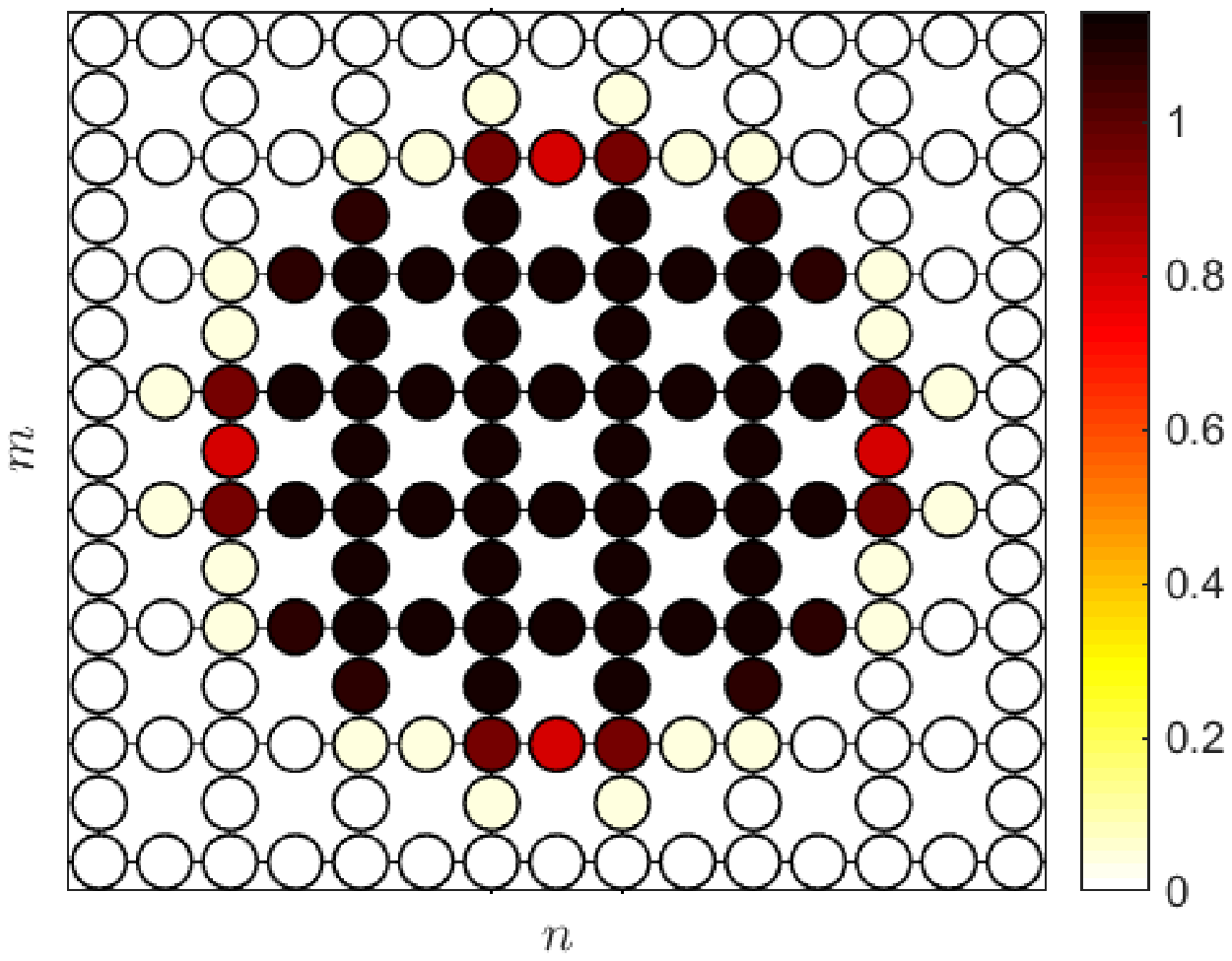}\label{subfig:prof_bond_c_0_05_i}}
		\subfloat[]{\includegraphics[scale=0.16]{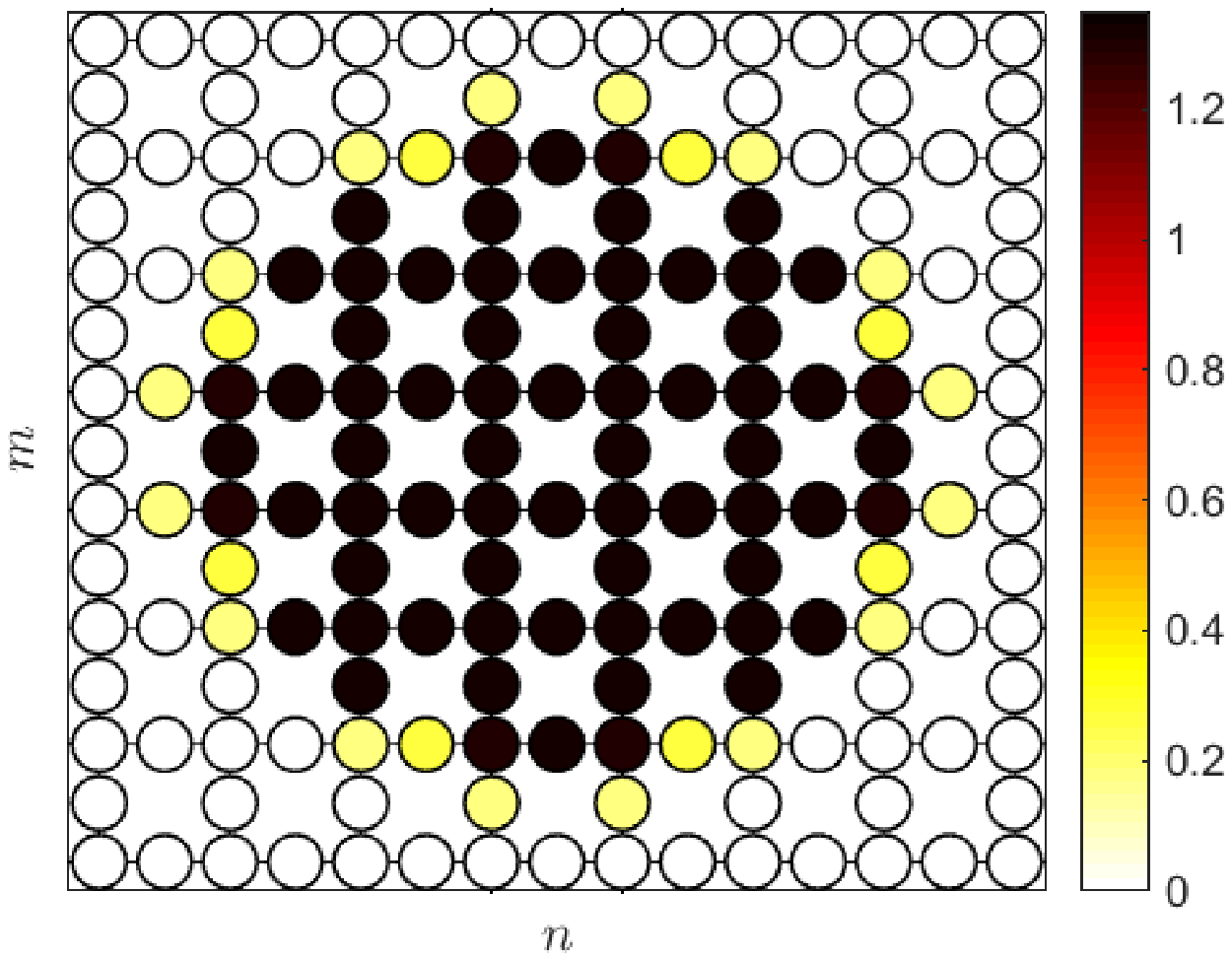}\label{subfig:prof_bond_c_0_05_j}}\\
		\subfloat[]{\includegraphics[scale=0.16]{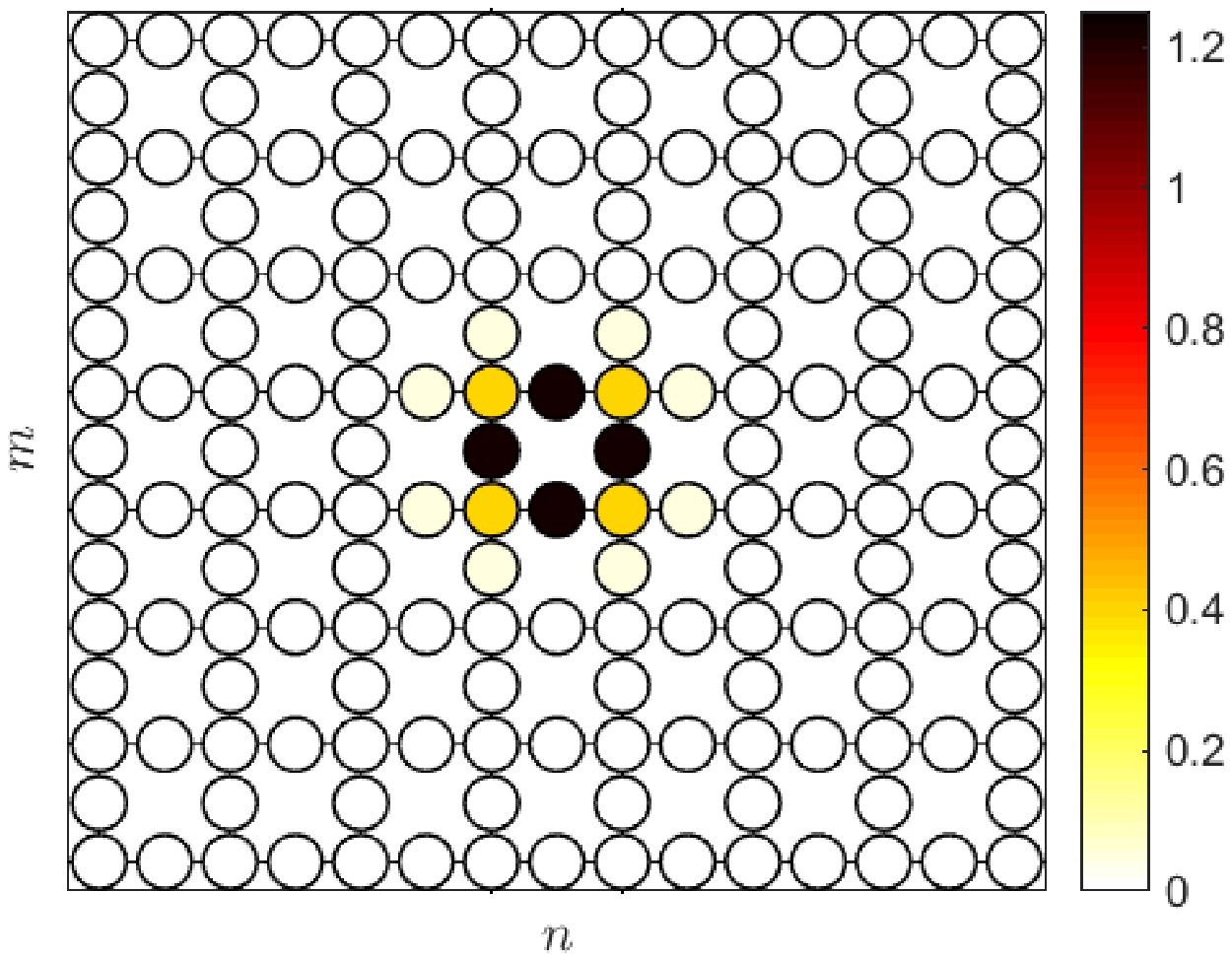}\label{subfig:prof_bond_c_0_10_k}}
		\subfloat[]{\includegraphics[scale=0.16]{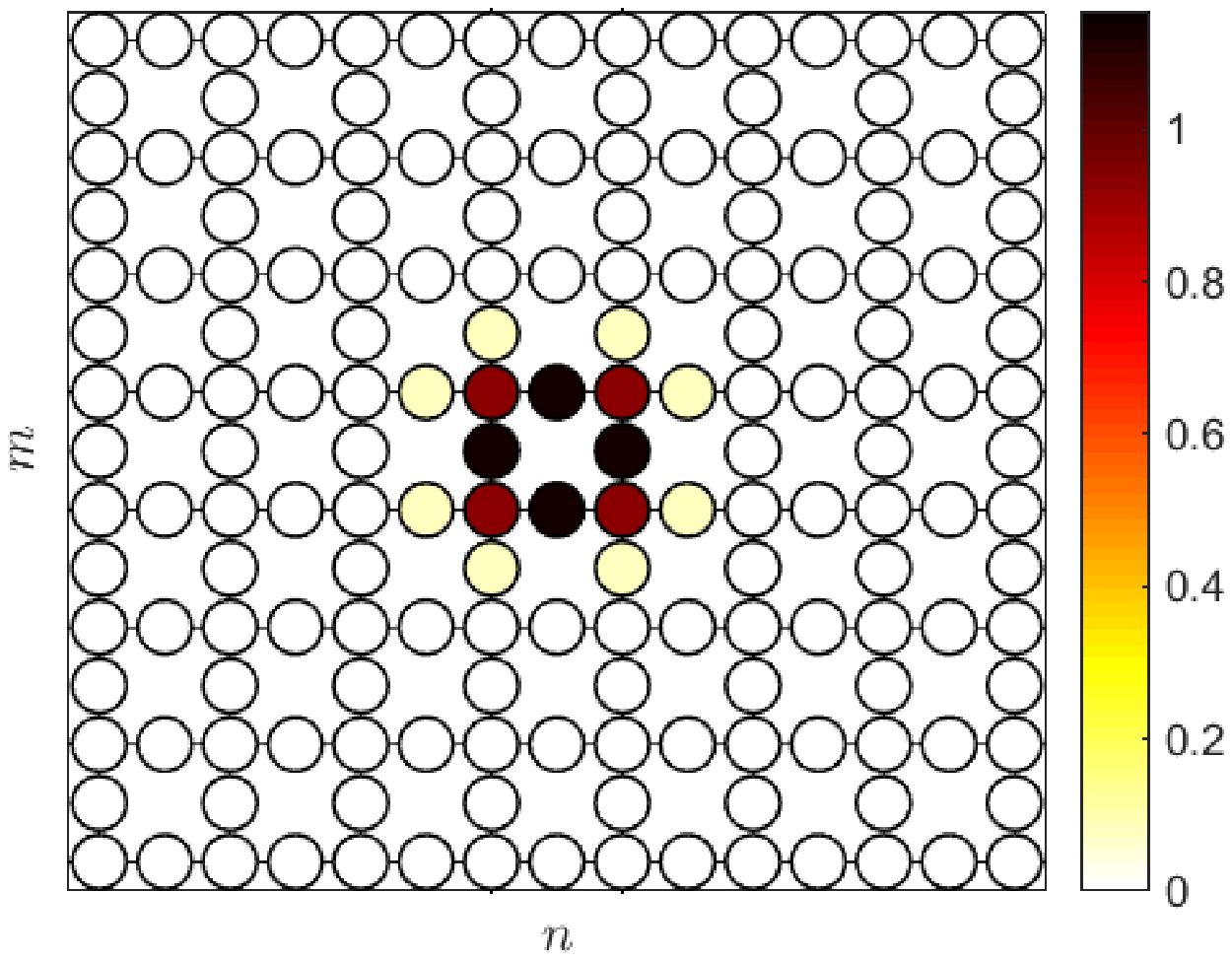}\label{subfig:prof_bond_c_0_10_l}}
		\subfloat[]{\includegraphics[scale=0.16]{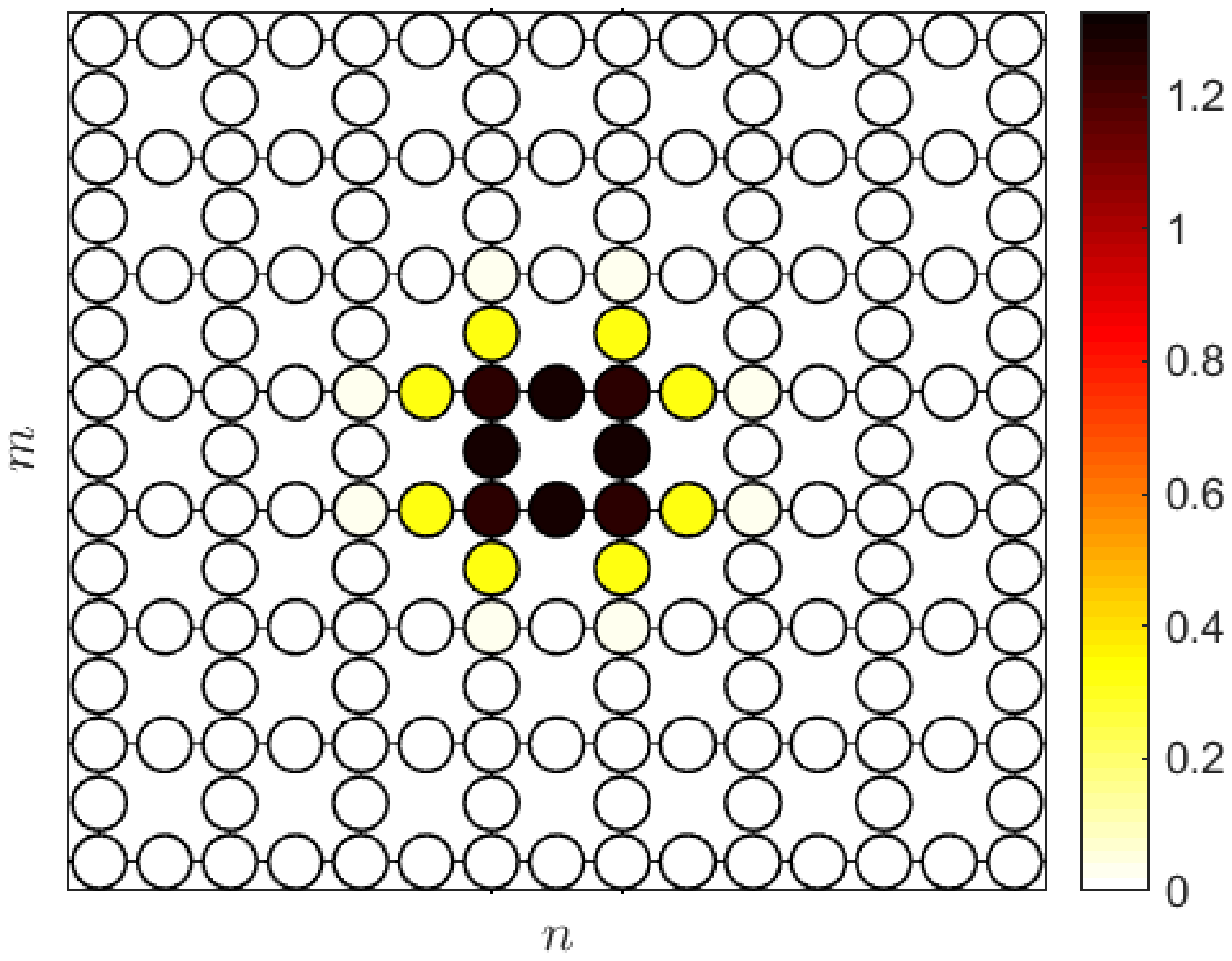}\label{subfig:prof_bond_c_0_10_m}}
		\subfloat[]{\includegraphics[scale=0.16]{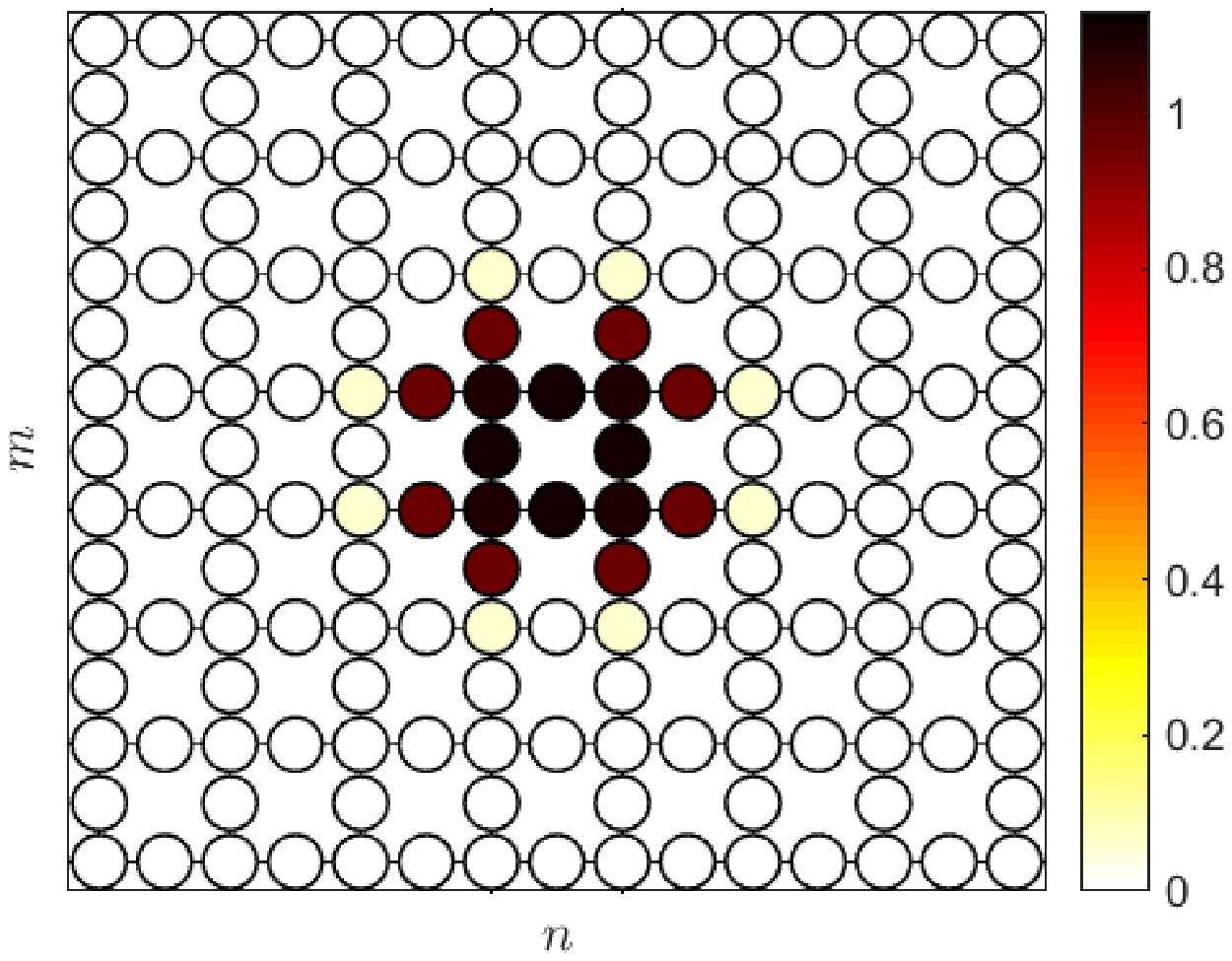}\label{subfig:prof_bond_c_0_10_n}}
		\subfloat[]{\includegraphics[scale=0.16]{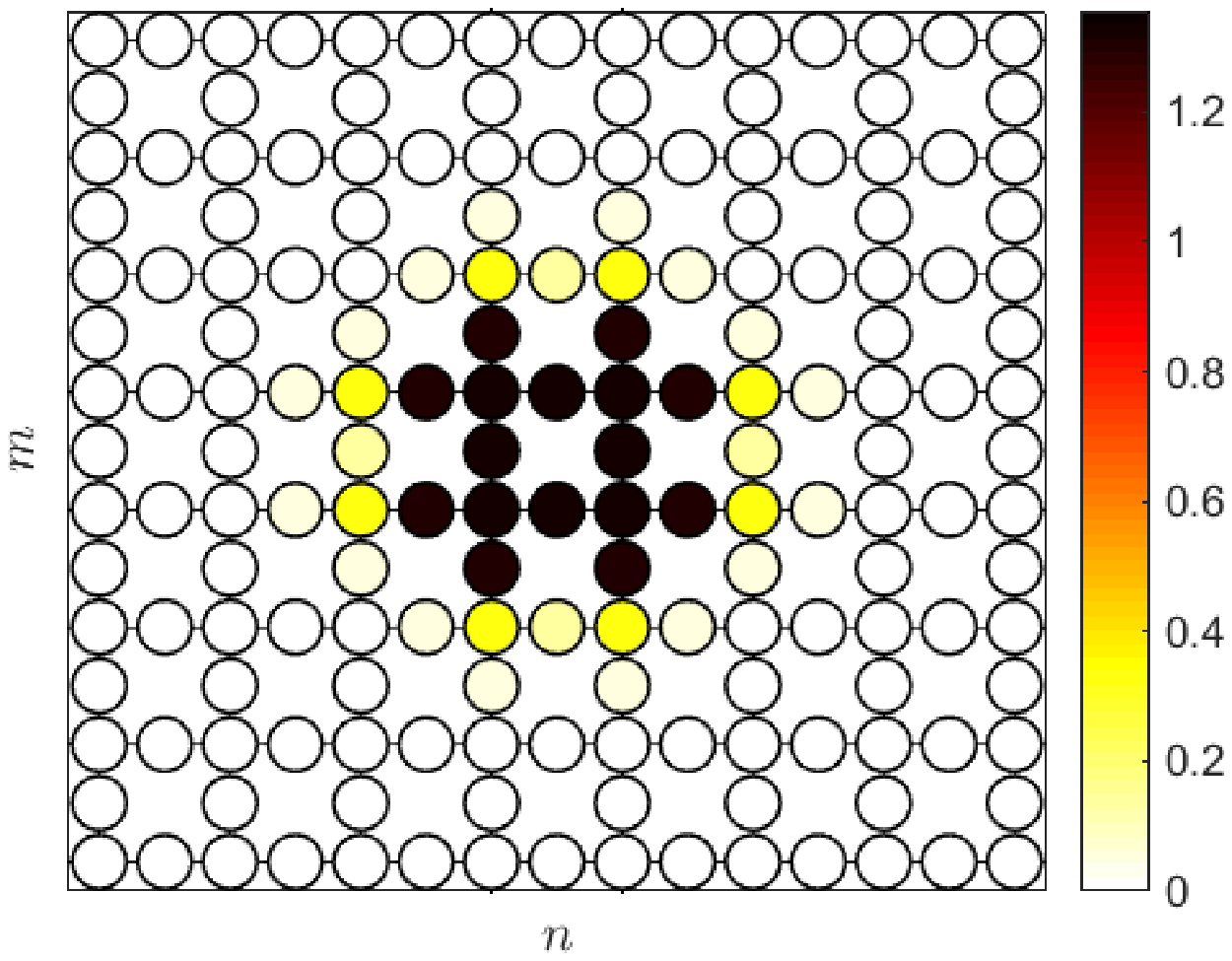}\label{subfig:prof_bond_c_0_10_o}}\\
		\subfloat[]{\includegraphics[scale=0.16]{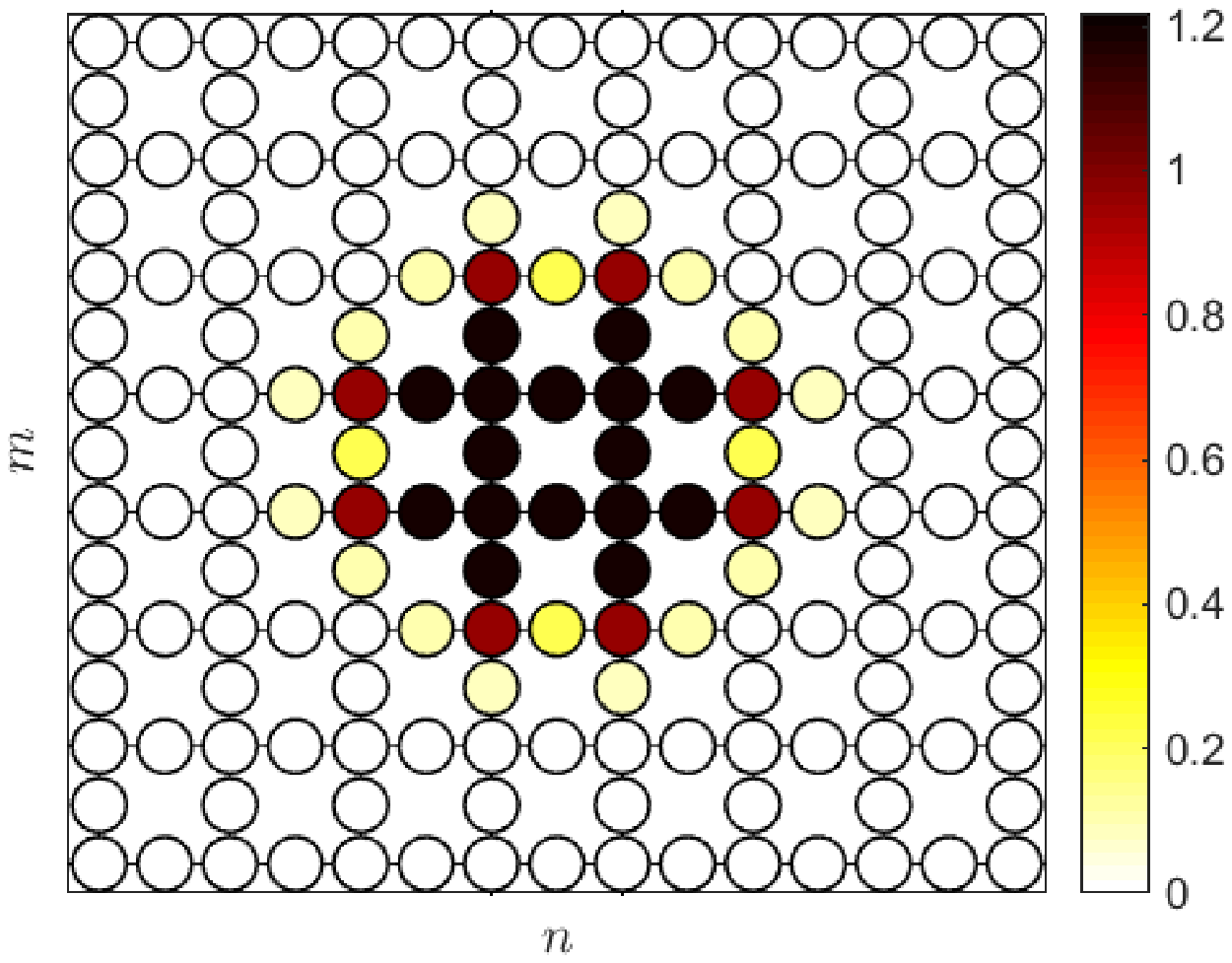}\label{subfig:prof_bond_c_0_10_p}}
		\subfloat[]{\includegraphics[scale=0.16]{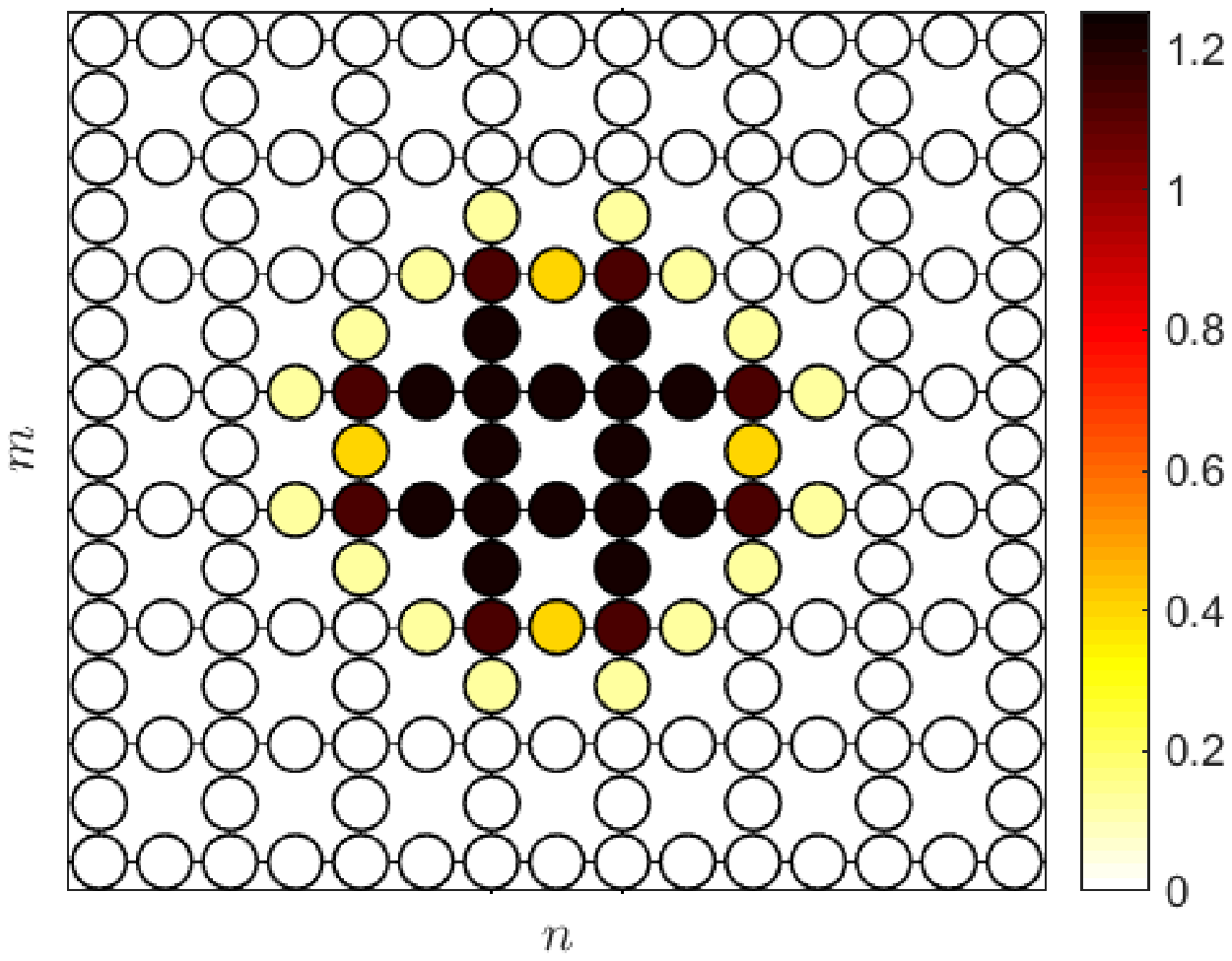}\label{subfig:prof_bond_c_0_10_q}}
		\subfloat[]{\includegraphics[scale=0.16]{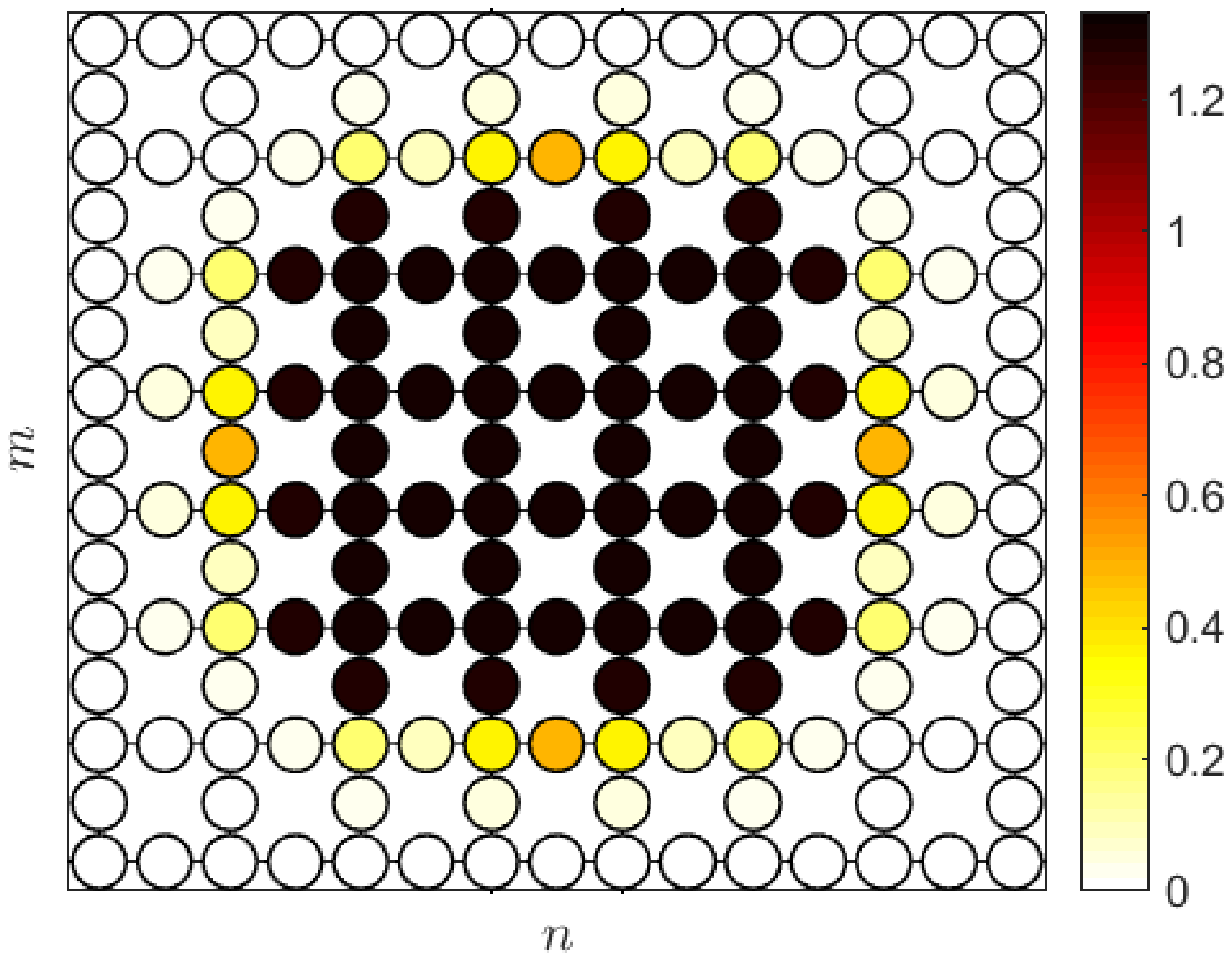}\label{subfig:prof_bond_c_0_10_r}}
		\subfloat[]{\includegraphics[scale=0.16]{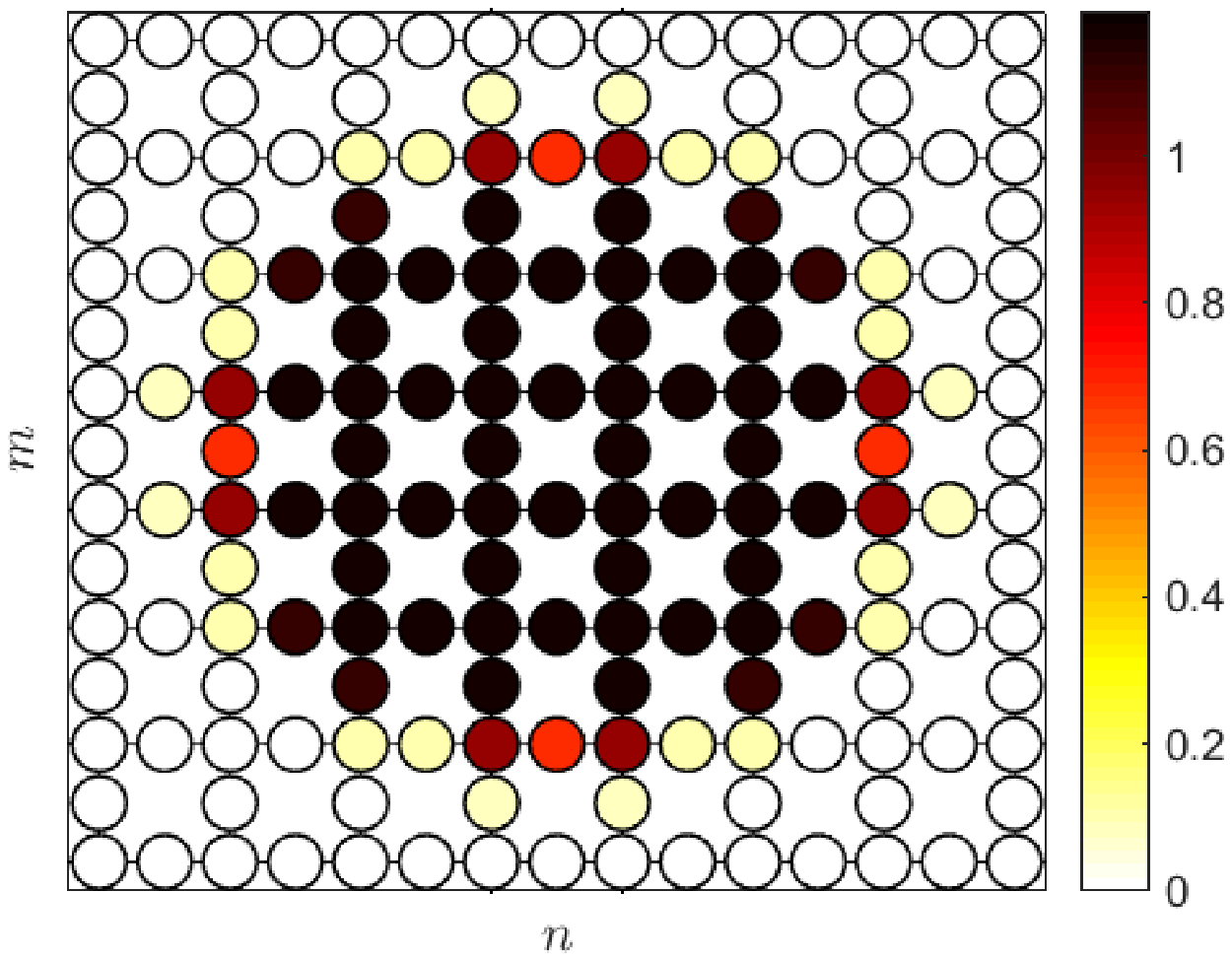}\label{subfig:prof_bond_c_0_10_s}}
		\subfloat[]{\includegraphics[scale=0.16]{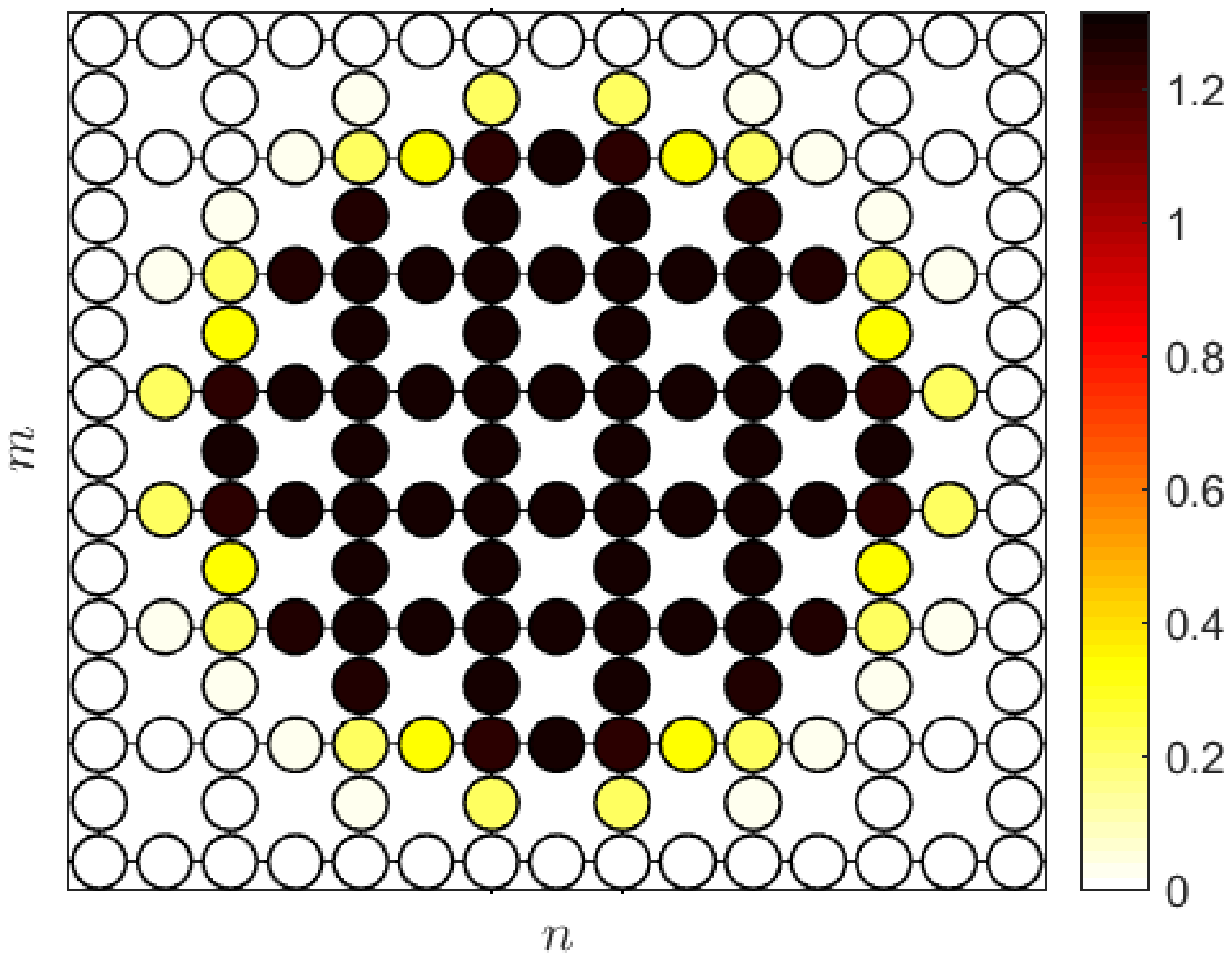}\label{subfig:prof_bond_c_0_10_t}}\\
		\caption{Top-view of localized solution profiles for bond-centred solutions that correspond to the points indicated in Fig.\ \ref{fig:bifur_bond}.}
		\label{fig:prof_bond}
	\end{figure*}
	
	By using site-centred and bond-centred solutions in Fig.\ \ref{fig:site_bond_example} as initial guess and performing numerical continuation for varying $\mu$,
	we solve the time--independent solution of Eq.\ \eqref{eq:ac_ori}, i.e.,\ Eq.\ \eqref{eq:ac_ti} and use a Newton-Raphson method combined with a numerical continuation, i.e.,\ pseudo-arclength method. 	
	We obtain bifurcation diagrams of the localized solutions that show a snaking structure, see Figs.\ \ref{fig:bifur_site}-\ref{fig:prof_bond}, where we use a scaled version of the $\mathbb{L}_2$ norm or ``mass'' norm \cite{Taylor2010} for the horizontal axis, i.e.,\
	\begin{equation}
		M=\left(\sum_{n,m}\frac{u_{n,m}^2}{\left(1+\sqrt{1-\mu}\right)}\right)^\frac{1}{2}.
	\end{equation}
	
	The snaking structure in the bifurcation diagrams exists within a certain interval called pinning region \cite{Pomeau1986}.
	In the 1D case, the pinning region in the limit $M\rightarrow\infty$ is bounded by two saddle-node bifurcations \cite{Taylor2010,Chong2009,Matthews2011,Susanto2011}.
	In our case, the bifurcations occur at several values of bifurcation parameter due to the presence of different types of localized solutions at the turning points, as we will show below. One can define that in this 2D case, the pinning region is formed by the largest distance between the upper and lower saddle-node bifurcations.
	
	Figures \ref{fig:bifur_site} and \ref{fig:bifur_bond} show the bifurcation diagram of site and bond-centred solutions at $c=0.05$ and $0.1$, respectively. The saddle-node bifurcations indeed occur at different critical parameter $\mu$. Moreover, the distance between the `upper' and `lower' saddle-node bifurcations are getting smaller when the coupling strength $c$ increases. 	
	In the continuum limit $c\rightarrow\infty$, the site-centred and bond-centred solutions merge as the snaking disappears in the Maxwell point, which also occurs in the 1D case. 

	We can see that the snaking has a general behaviour where there is an interchange of stability between localized solutions at the turning points. Nevertheless, it is also possible to obtain	a condition where the stability interchange does not occur. In this case, we obtain a {`switchback'} phenomenon around $M^2\approx 20$ and $60-75$ in both Figs.\ \ref{fig:bifur_site} and \ref{fig:bifur_bond}. {From our observation, this happens because the corresponding solutions have many ``fronts'' (connecting state between the ``zero'' and ``upper'' states) which result in a complicated interaction between neighbouring sites as we vary a parameter (for example, $\mu$). We show these in details in Figs.\ \ref{subfig:snake_site_c_0_05_zoom} and \ref{subfig:snake_site_c_0_10_zoom}. 
	It is also possible that isolated bifurcation curves (i.e., isolas) may also form when $c$ is being varied \cite{Taylor2010}.
	Figures \ref{fig:prof_site} and \ref{fig:prof_bond} show top-views (2D projection) of the solution profiles at several turning points in the bifurcation diagrams in Fig.\ \ref{fig:bifur_site}.} We can see that, as the norm $M$ increases, the non-zero plateau becomes wider.

	{
	On the flat band of the lattice mentioned previously, cf.\ Fig.\ \ref{fig:disrel}, we comment that it does not have any effect on homoclinic snaking simply because the exotic band is located in the negative value of $\mu$, while the snaking exists in the positive value of the parameter.    }
	
	\section{Saddle-node bifurcation analysis}\label{sec:saddle}
	
	In this section, we will derive an analytical approximation to the numerical results reported above.

	\begin{figure}
		\centering
		\includegraphics[scale=0.4]{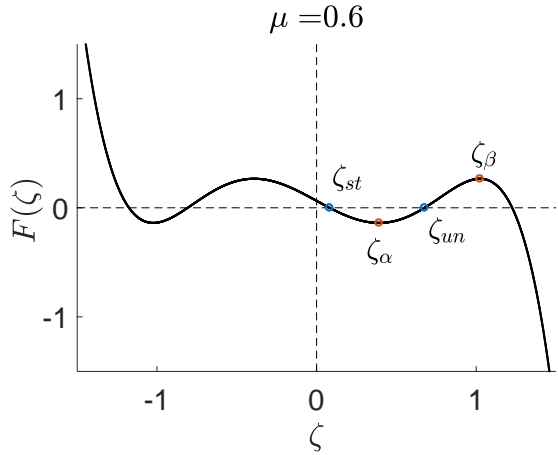}
		\caption{Active-cell function type 1 at $c=0.05$,
			$\zeta_\alpha$ and $\zeta_\beta$ represent as `upper' and `lower' saddle-node bifurcations.
			$\zeta_{st}$ and $\zeta_{un}$ represent the stable and unstable cell solution.}
		\label{fig:one_active_lieb}
	\end{figure}
	\begin{figure*}[t!]
		\begin{flushleft}
			\subfloat[$Z(\zeta)=c\left(U_1-4\zeta\right)$]{\includegraphics[scale=0.21]{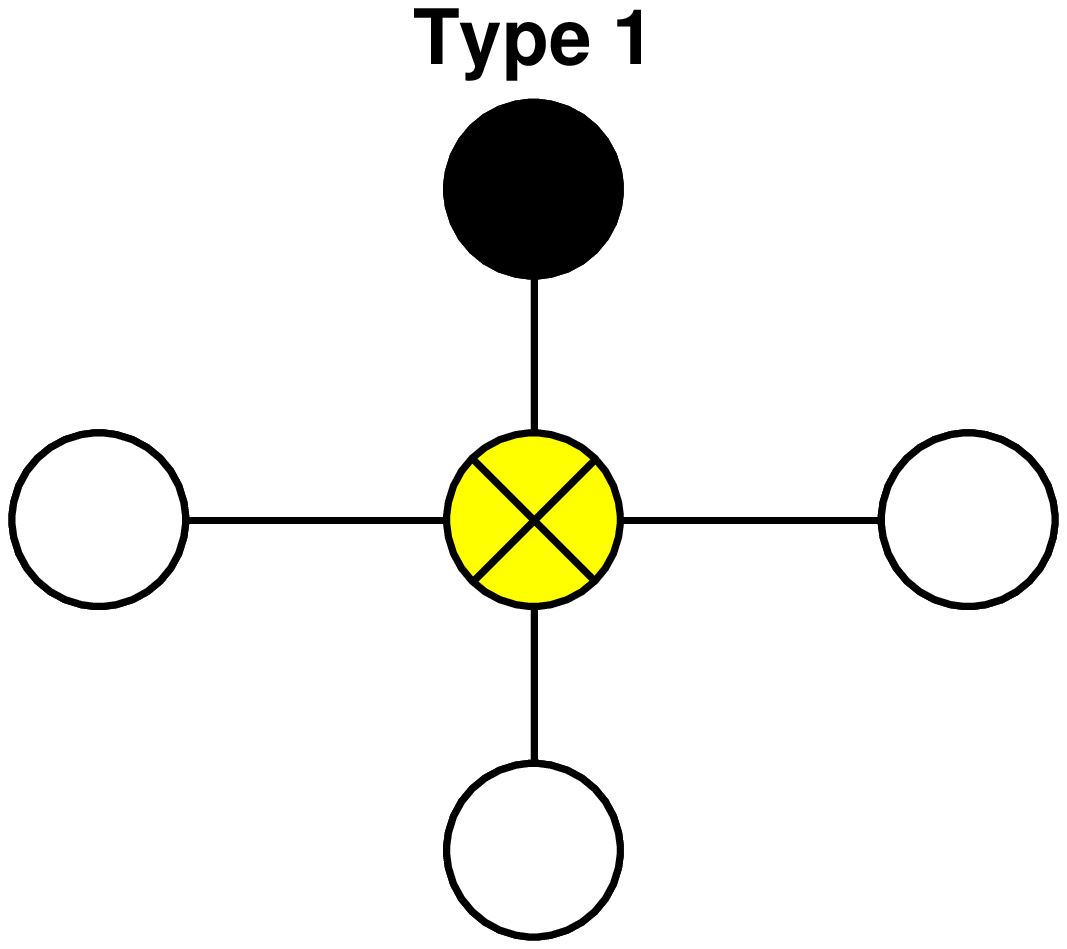}
				\label{subfig:Lieb_type_1}}
			\subfloat[$Z(\zeta)=c\left(2U_1-4\zeta\right)$]{\includegraphics[scale=0.21]{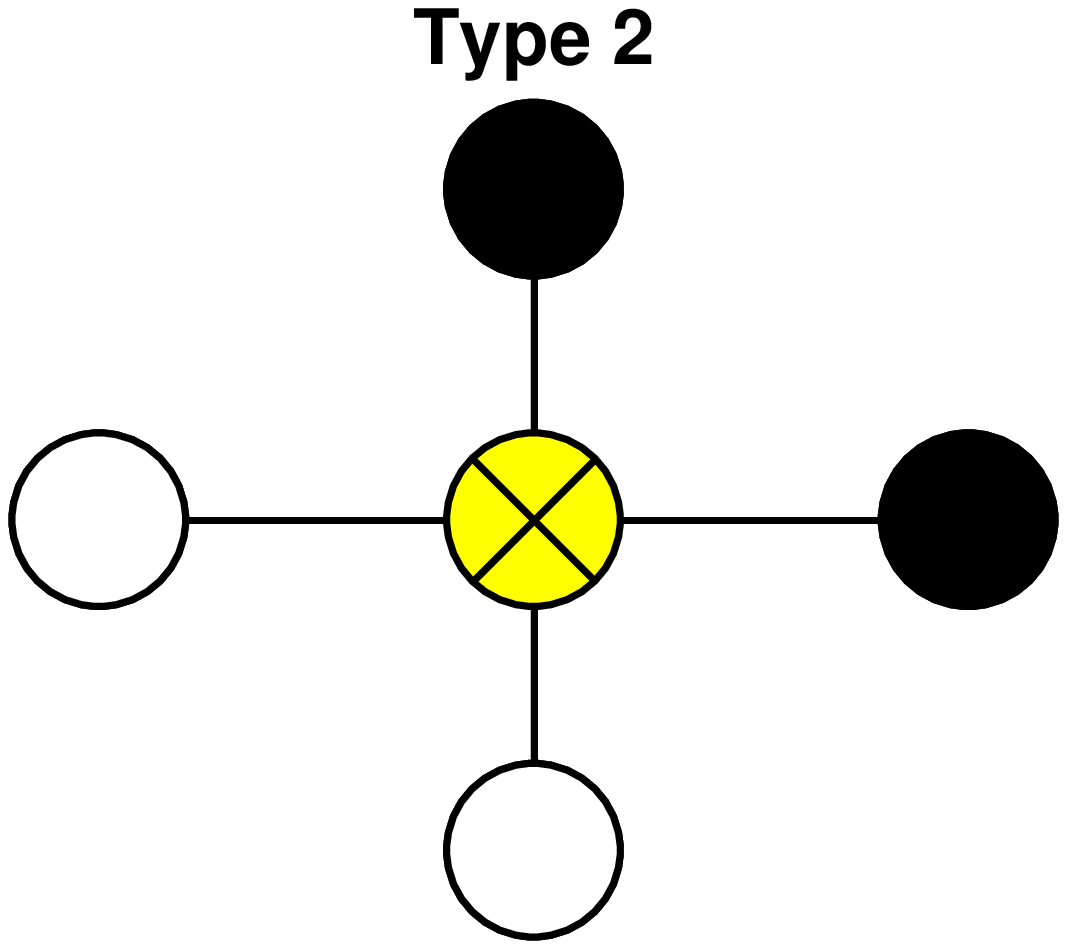}
				\label{subfig:Lieb_type_2}}
			\subfloat[$Z(\zeta)=c\left(U_1-3\zeta\right)$]{\includegraphics[scale=0.21]{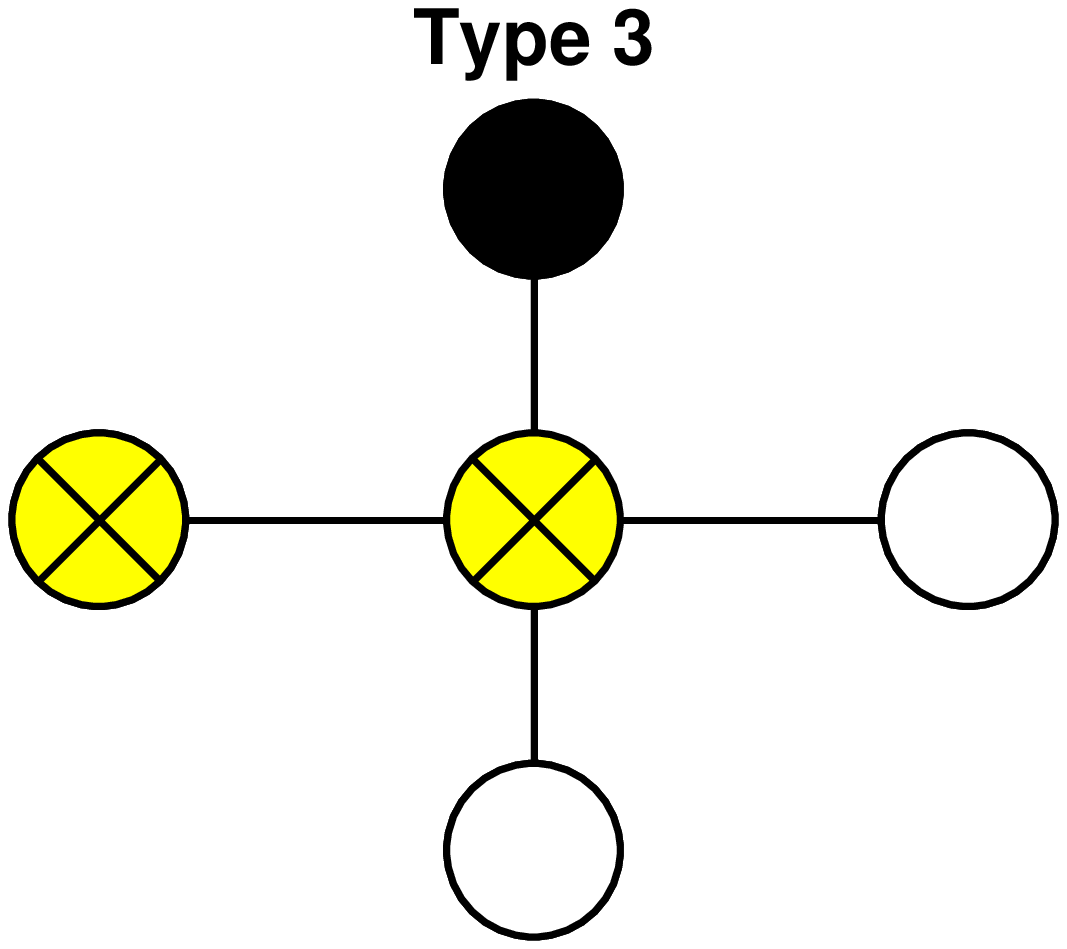}
				\label{subfig:Lieb_type_3}}
			\includegraphics[scale=0.21]{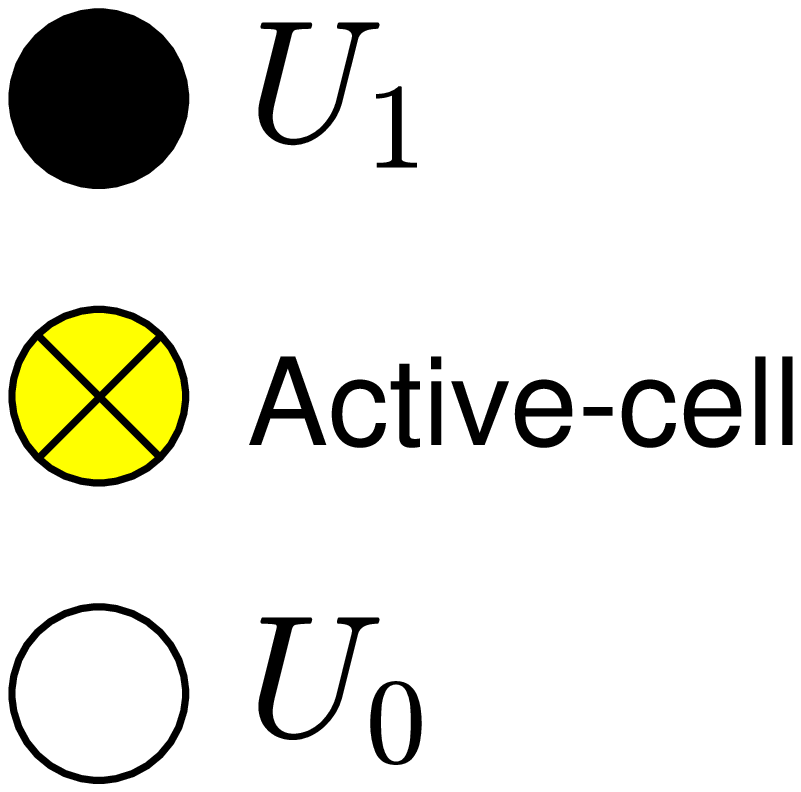}
			\subfloat[$Z(\zeta)=c\left(U_1-2\zeta\right)$]{\includegraphics[scale=0.21]{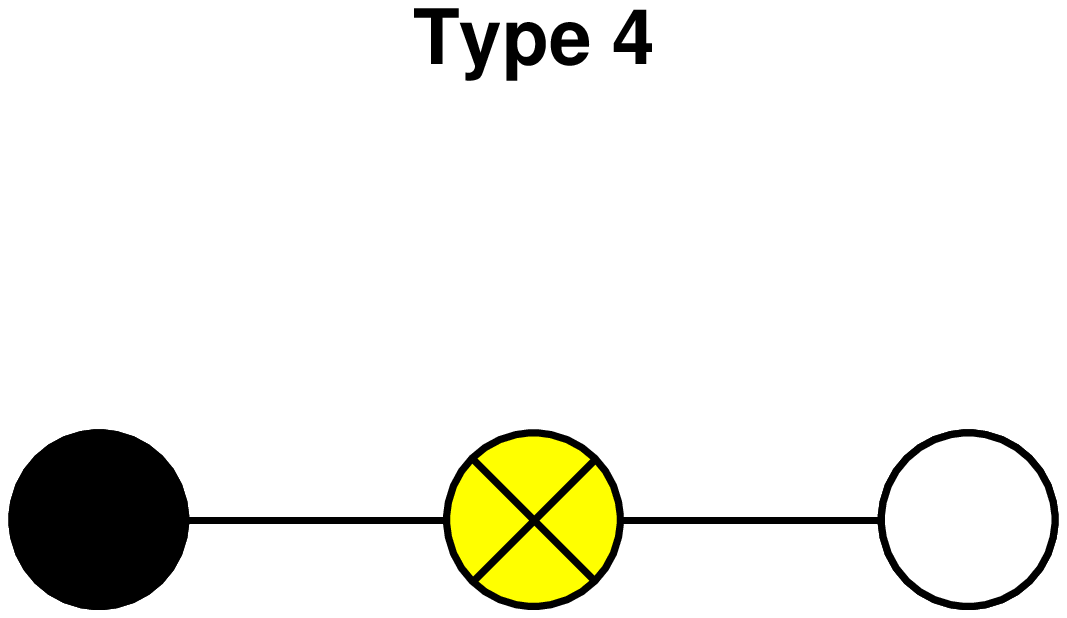}
				\label{subfig:Lieb_type_4}}
			\subfloat[$Z(\zeta)=c\left(2U_1-2\zeta\right)$]{\includegraphics[scale=0.21]{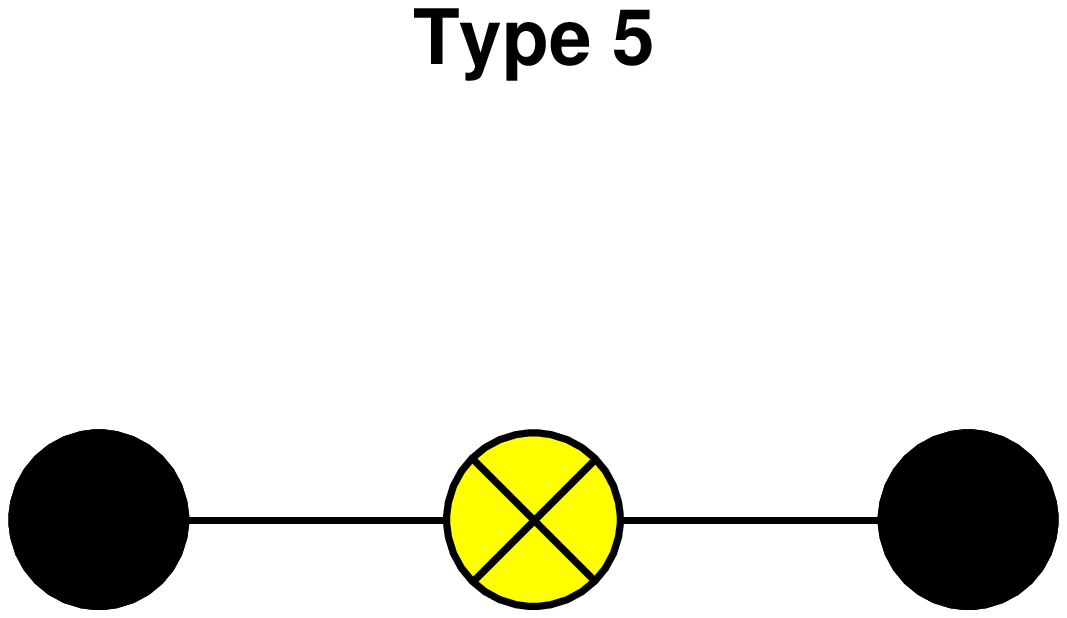}
				\label{subfig:Lieb_type_5}}
			\subfloat[$Z(\zeta)=c\left(U_1-\zeta\right)$]{\includegraphics[scale=0.21]{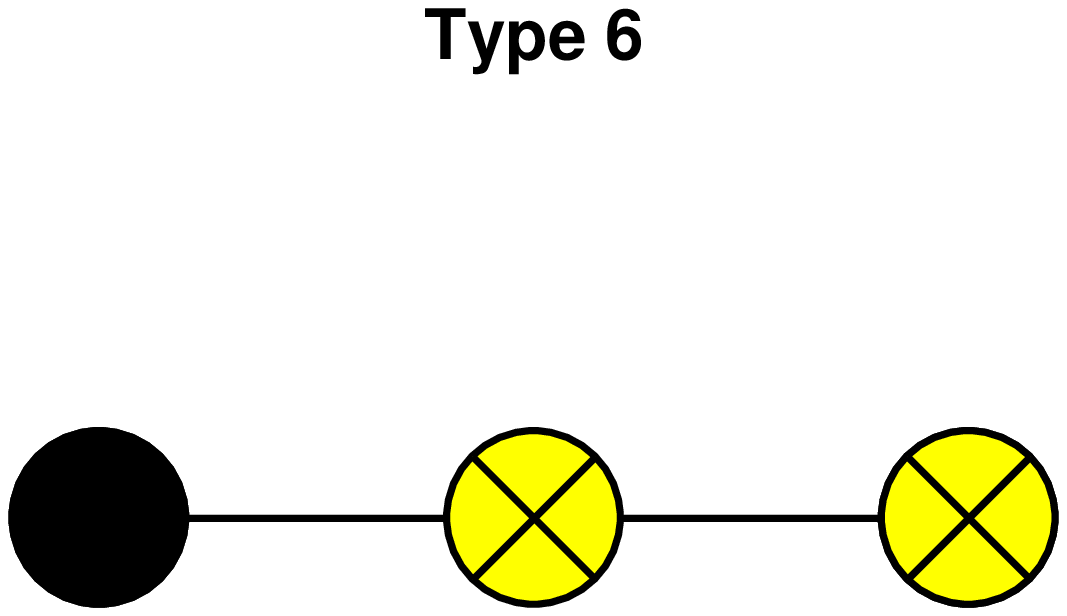}
				\label{subfig:Lieb_type_6}}
		\end{flushleft}
		\caption{Sketch of different types of active-cell approximation to the saddle-node bifurcations that occur in the system.}
		\label{fig:sol_type}
	\end{figure*}
	
	\subsection{Active-cell approximation}
	In general, when the coupling strength $c$ is weak, we assume that to the leading order the solution effectively consists of
	three states only, i.e.,\ homogenous state $U_1$, zero state $U_0$, and interface (active-cell). By using this assumption, there will be only three states involved in the dynamics for small coupling $c$. 
	Thus, we can re-write equation (3) into a simple ordinary differential equation \cite{Kusdiantara2017}
	\begin{equation}
		\frac{d \zeta}{dt}=F(\zeta)=-\mu\zeta+2\zeta^3-\zeta^5+cZ(\zeta), 
		\label{eq:active_cell}
	\end{equation}
	where
	\begin{equation}
		Z(\zeta)=aU_1-b\zeta,
	\end{equation}
	as the replacement for the Laplacian term and $\zeta$ is the active cell or interface. 
	The coefficients $a$ and $b$ are determined by the number of homogeneous state $U_1$, zero state $U_0$, and active-cell at the ``fronts''/interface. 
	
	Generally, $F(\zeta)$ can have five real roots, see Fig.\ \ref{fig:one_active_lieb}. 
	Note that only two of them are related to the snaking boundaries that correspond to the `upper' and `lower' saddle-node bifurcations. 
	We can recognise that a saddle-node bifurcation is a condition when $F(\zeta)$ at the local minimum $\zeta = \zeta_\alpha$ and local maximum $\zeta = \zeta_\beta$ disappears, which corresponds to the `upper' and `lower' saddle-node bifurcations, respectively. 
	It is quite straightforward to obtain that
	\begin{equation}
		\zeta_{\alpha,\beta}=\left(\frac{3}{5}\pm\frac{1}{5}\sqrt{9-5\left(\mu+cb\right)}\right)
	\end{equation}
	
	We classify that there are several types of saddle-node bifurcations in the snaking diagrams.
	By identifying them, we can obtain the correct active-cell approximation to the solution profiles. In particular, we have six types, which are characterised by the numbers and positions of the homogenous state $U_1$, zero state $U_0$, and active-cell in their solution profiles, see Fig.\ \ref{fig:sol_type}. The list of coefficients $a$ and $b$ for each type is shown in Table \ref{tab:coef}. 
	
	\begin{table}[h!]
		\centering
		\caption{Values of the coefficient $a$ and $b$ in the active-cell approximation \eqref{eq:active_cell} for the different types of solution at the turning point.}
		\begin{tabular}{|c|c|c|}
			\hline
			\textbf{Type} & $a$ & $b$ \\ \hline
			\textbf{1}    & 1   & 4   \\ 
			\textbf{2}    & 2   & 4   \\ 
			\textbf{3}    & 1   & 3   \\ 
			\textbf{4}    & 1   & 2   \\ 
			\textbf{5}    & 2   & 2   \\ 
			\textbf{6}   & 1   & 1   \\ \hline
		\end{tabular}
		\label{tab:coef}
	\end{table}
	
	We also note that the active-cell approximation is a rotation invariant at their center or axes. 
	Approximations of the saddle-node bifurcations in Figs.\ \ref{fig:bifur_site} and \ref{fig:bifur_bond} have been depicted in the same figures.

	Figure \ref{subfig:snake_site_c_0_05} shows several types of saddle-node bifurcations and their approximations for the site-centred solutions at $c = 0.05$. 
	In general, there are four types of saddle-node bifurcations for the site-centred solutions as shown in Fig.\ \ref{fig:sol_type}.
	The bifurcations at points (a)-(n) belong to type 1,2,4 and 6 as indicated in the caption of Fig.\ \ref{fig:bifur_site}.
	These types of saddle-node bifurcations only appear in site-centred solutions. Type 6 mostly appear in the relatively large value of norm $M^2$.
	Note that the approximations are in good agreement with the numerics. 
	
	\begin{figure}[t!]
		\centering
		\subfloat[{Type 1}]{\includegraphics[scale=0.27]{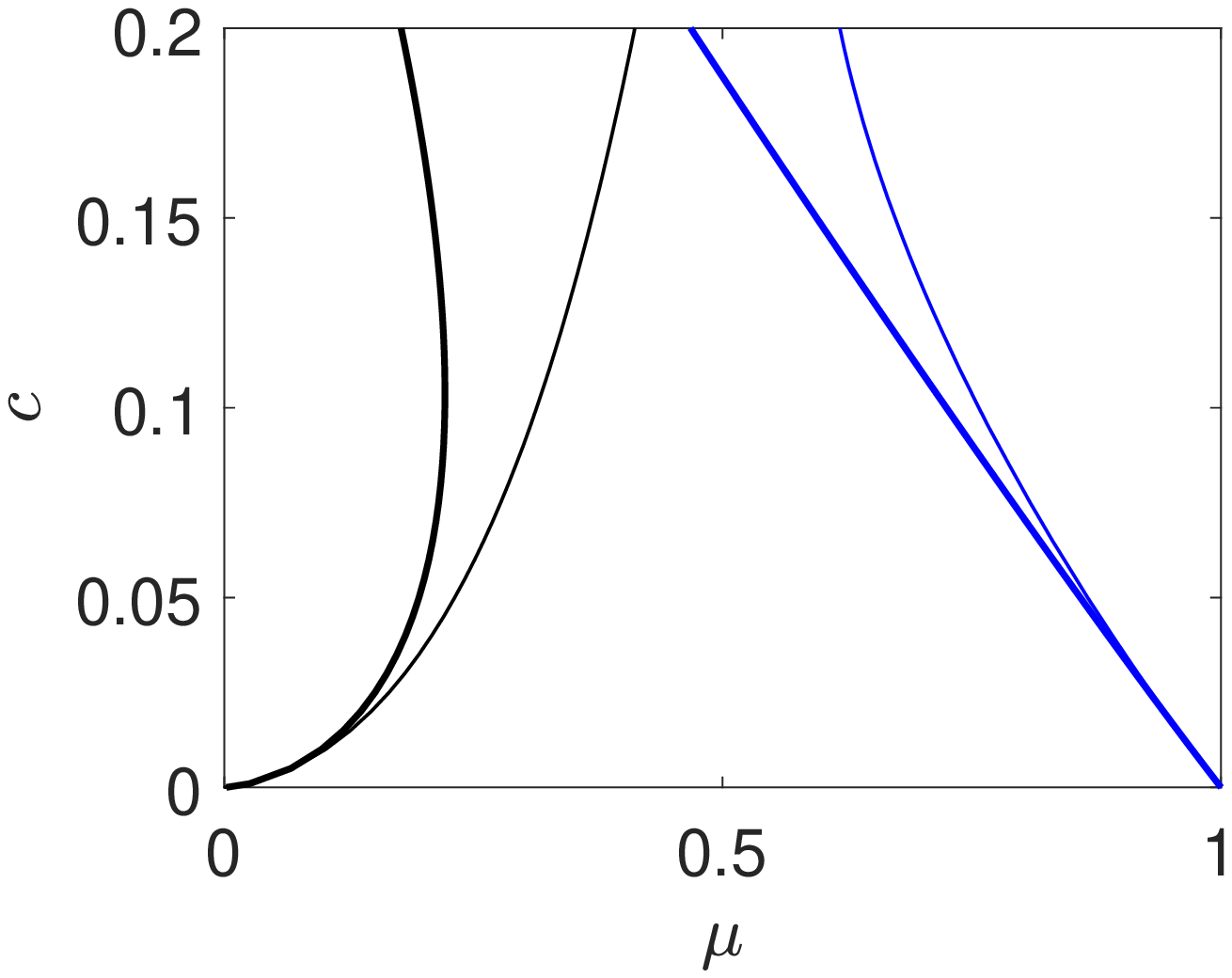}\label{subfig:pinning_type_1}}\,\,
		\subfloat[{Type 2}]{\includegraphics[scale=0.27]{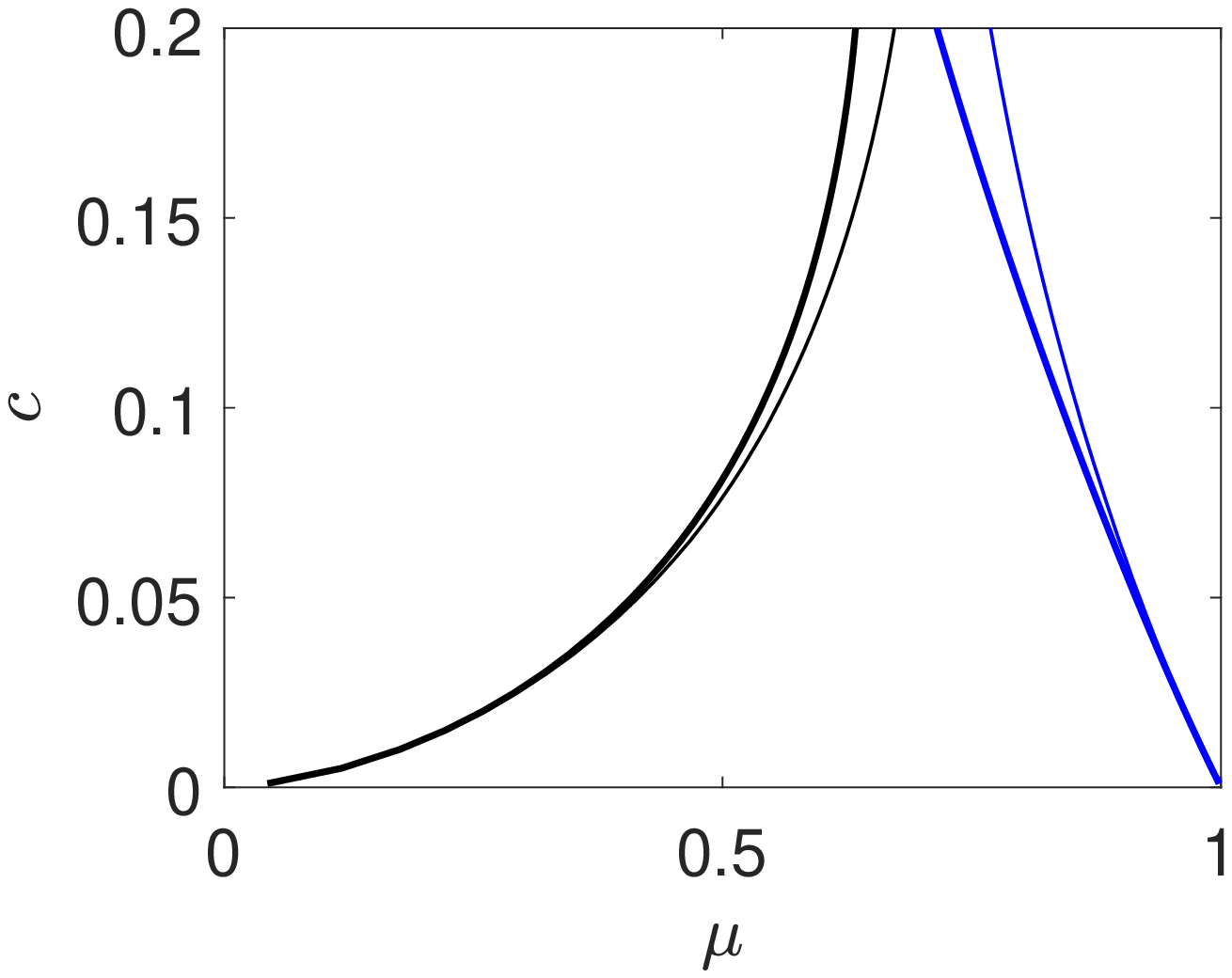}\label{subfig:pinning_type_2}}\,\,
		\subfloat[{Type 3}]{\includegraphics[scale=0.27]{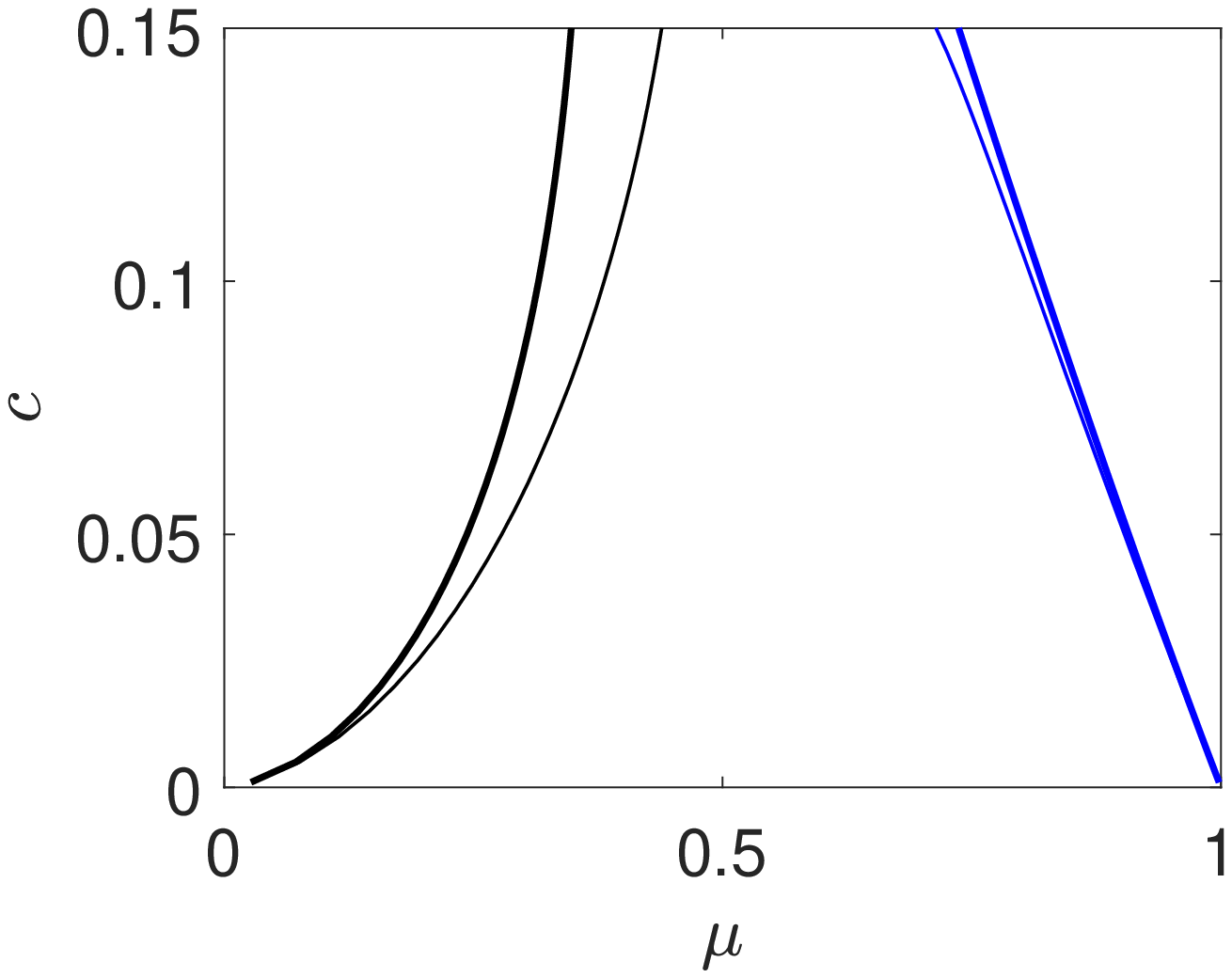}\label{subfig:pinning_type_3}}\\
		\subfloat[{Type 4}]{\includegraphics[scale=0.27]{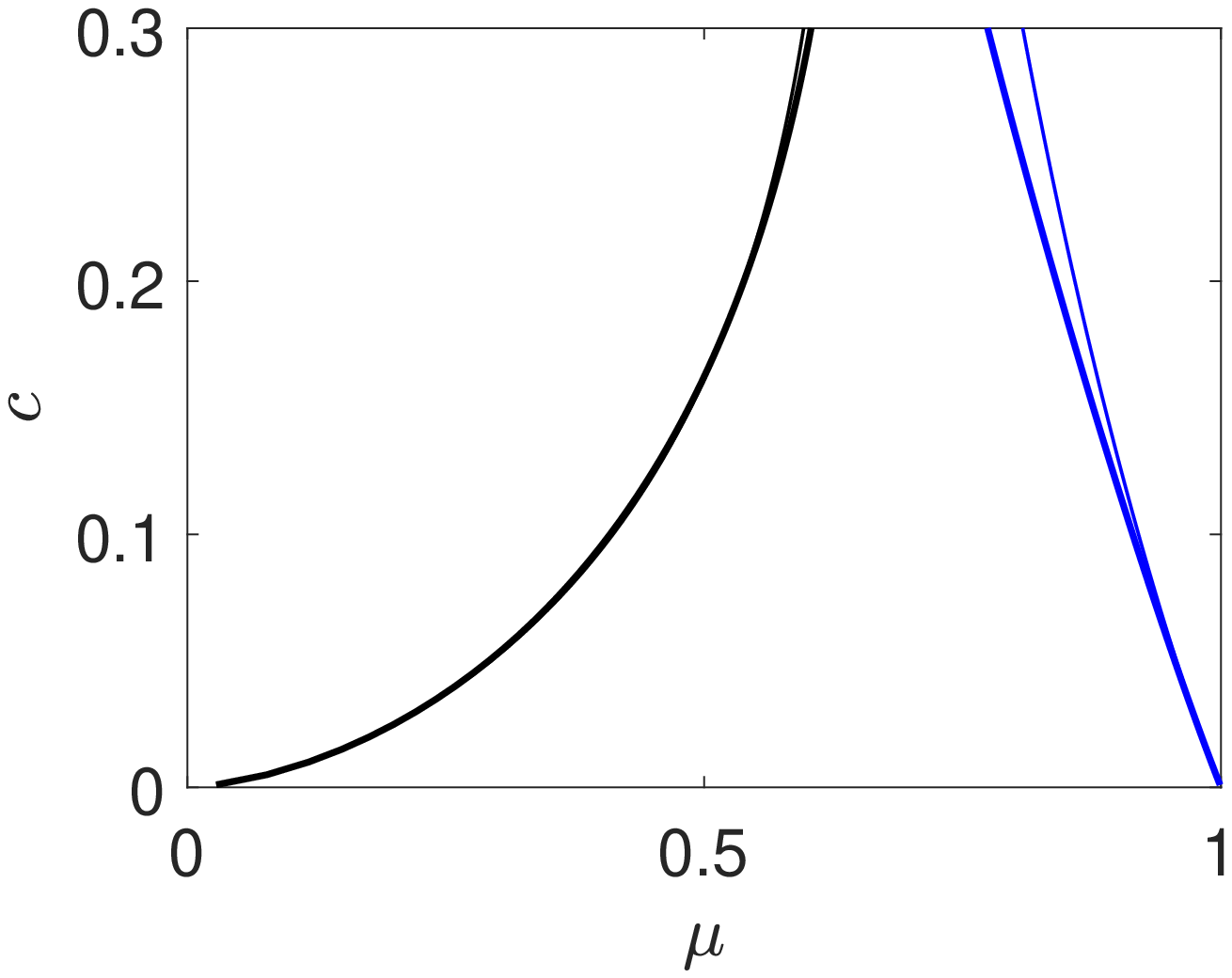}\label{subfig:pinning_type_4}}\,\,
		\subfloat[{Type 5}]{\includegraphics[scale=0.27]{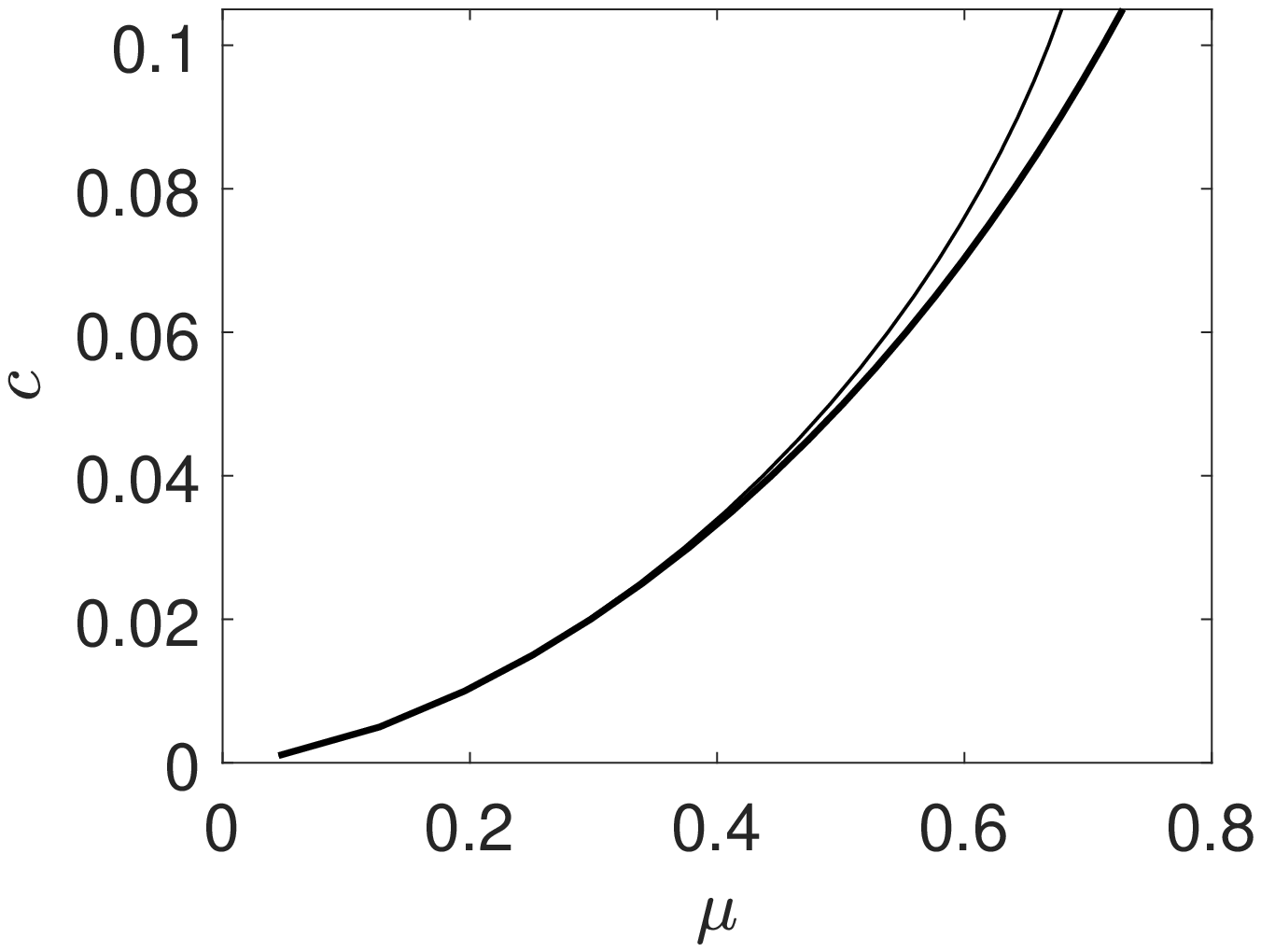}\label{subfig:pinning_type_5}}\,\,
		\subfloat[{Type 6}]{\includegraphics[scale=0.27]{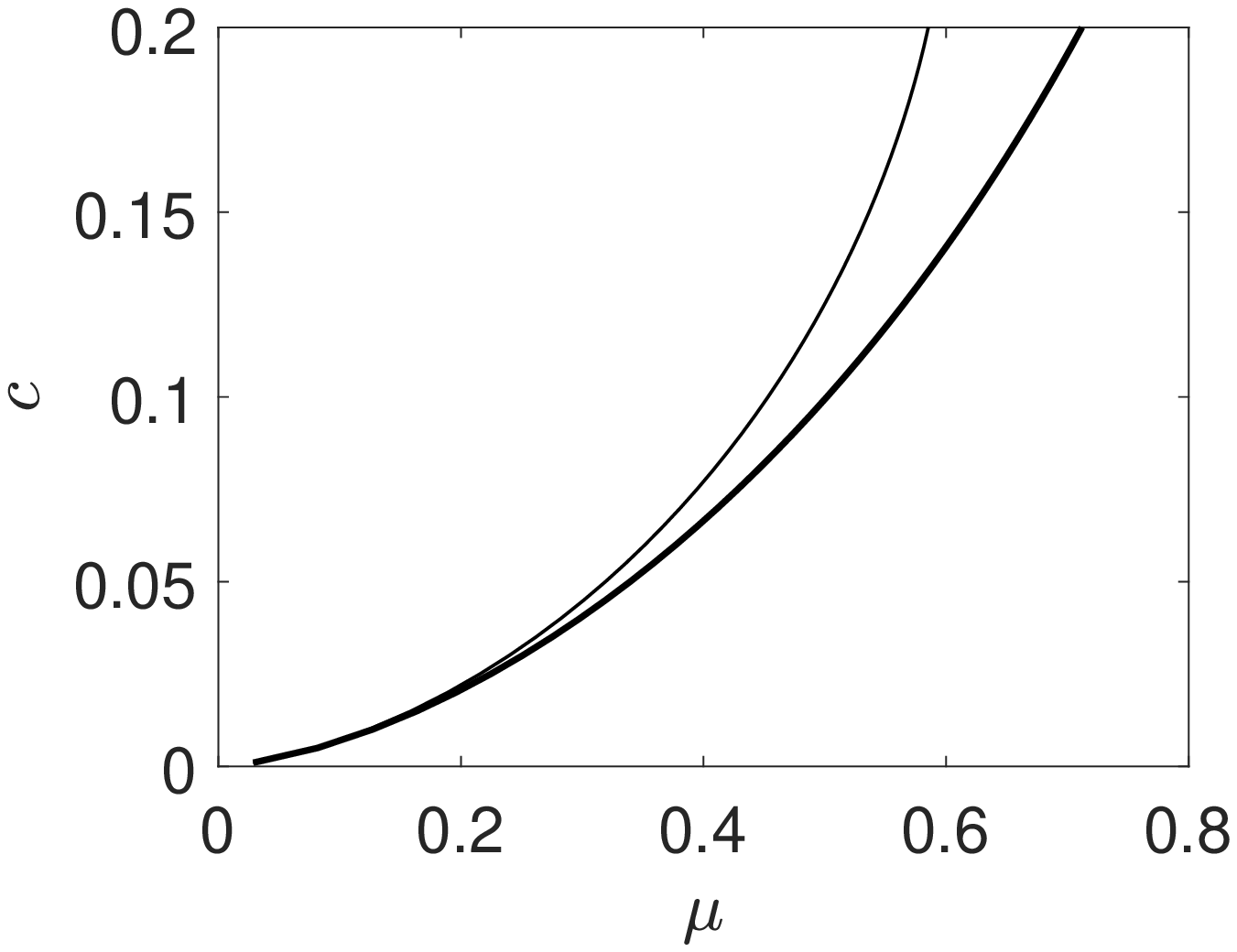}\label{subfig:pinning_type_6}}
		\caption{{Location, i.e., parameter value $\mu$, of the different types of turning points for varying coupling constant $c$. 
				Solid thick line is the actual value obtained from solving Eq.\ \eqref{eq:ac_ti} numerically. 
			}
		}
		\label{fig:pinning_type}
	\end{figure}
	\begin{figure}[h!]
		\centering
		\subfloat[{Type 1}]{\includegraphics[scale=0.27]{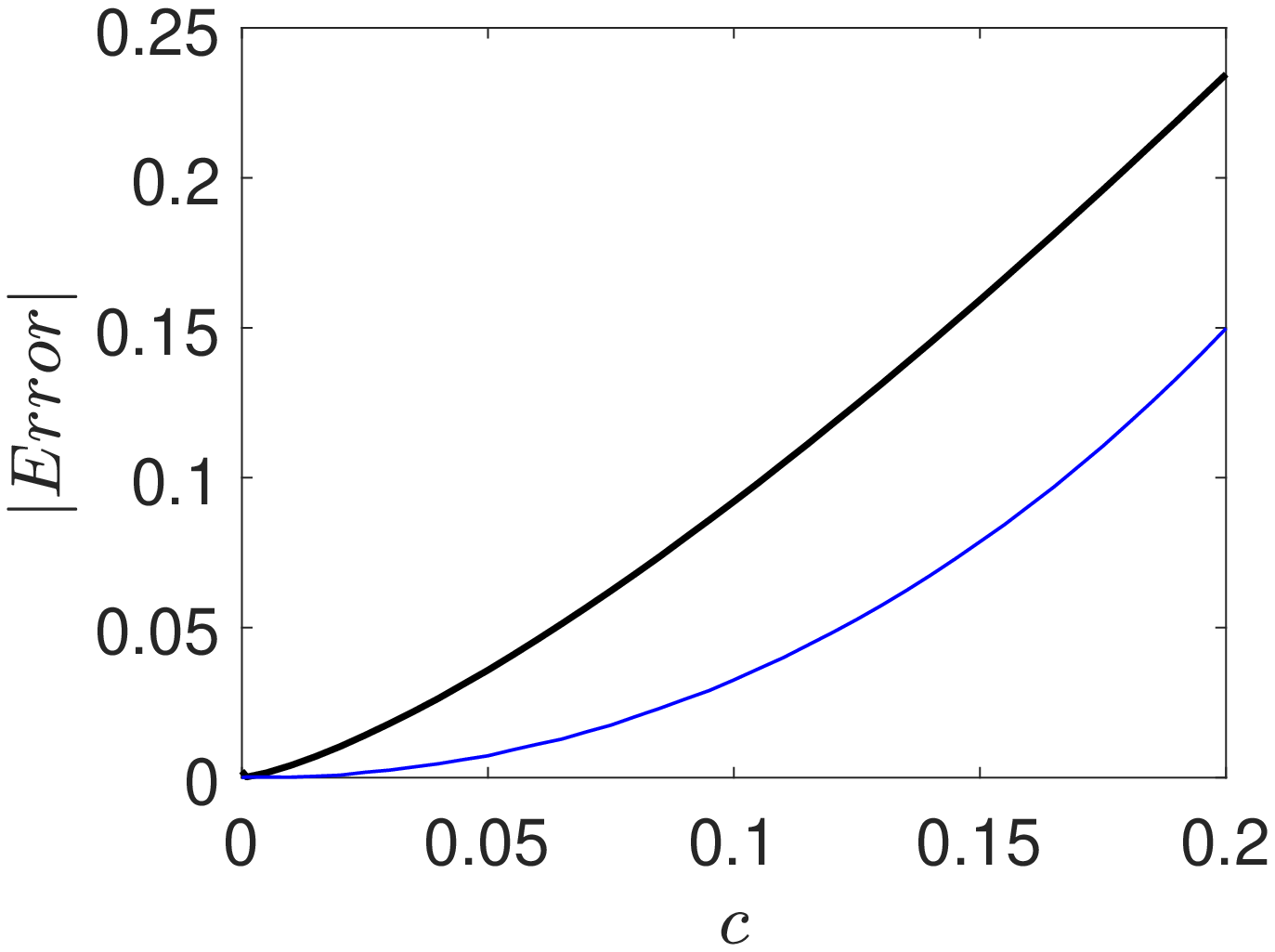}\label{subfig:width_type_1}}\,\,
		\subfloat[{Type 2}]{\includegraphics[scale=0.27]{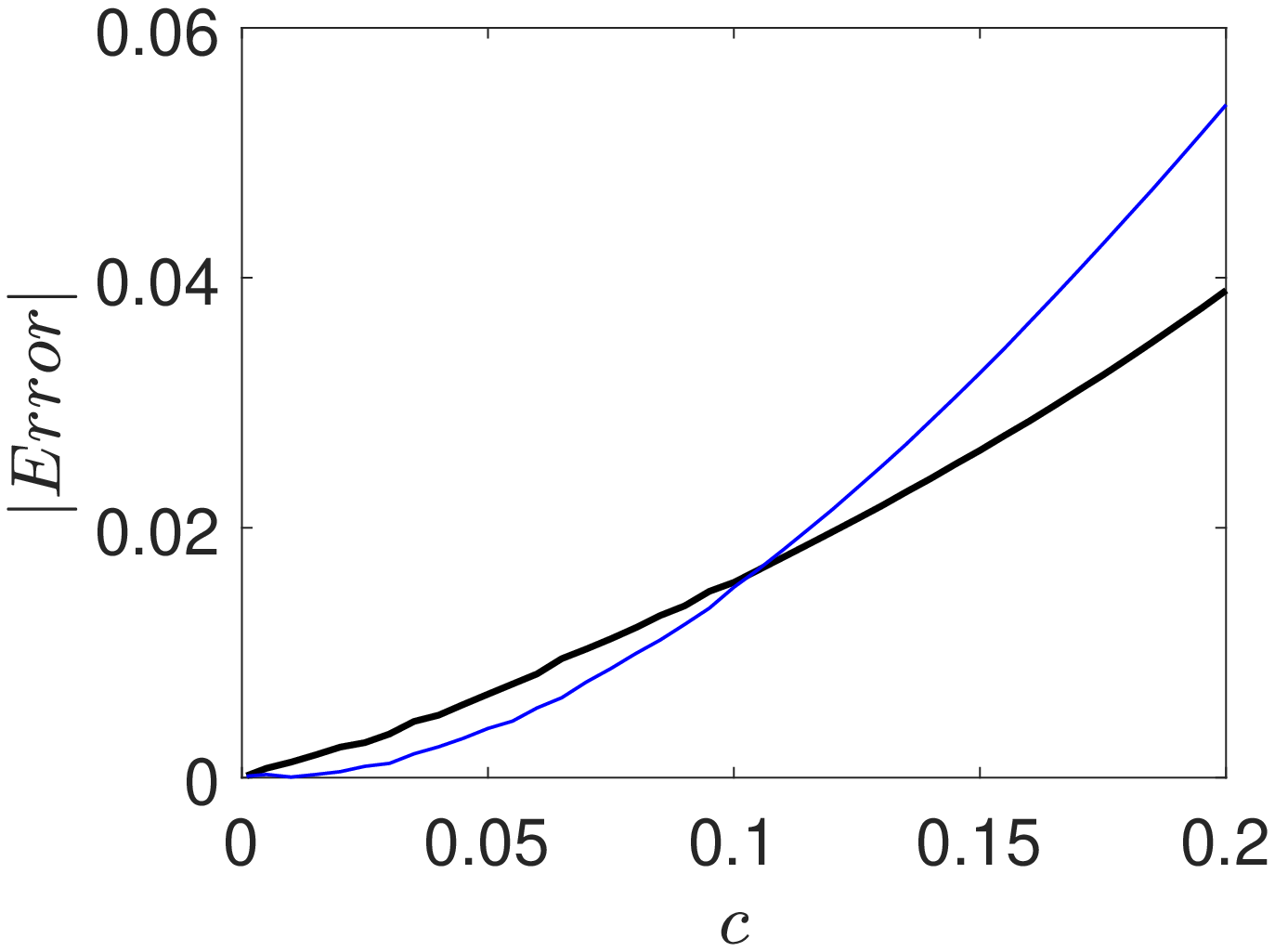}\label{subfig:width_type_2}}\,\,
		\subfloat[{Type 3}]{\includegraphics[scale=0.27]{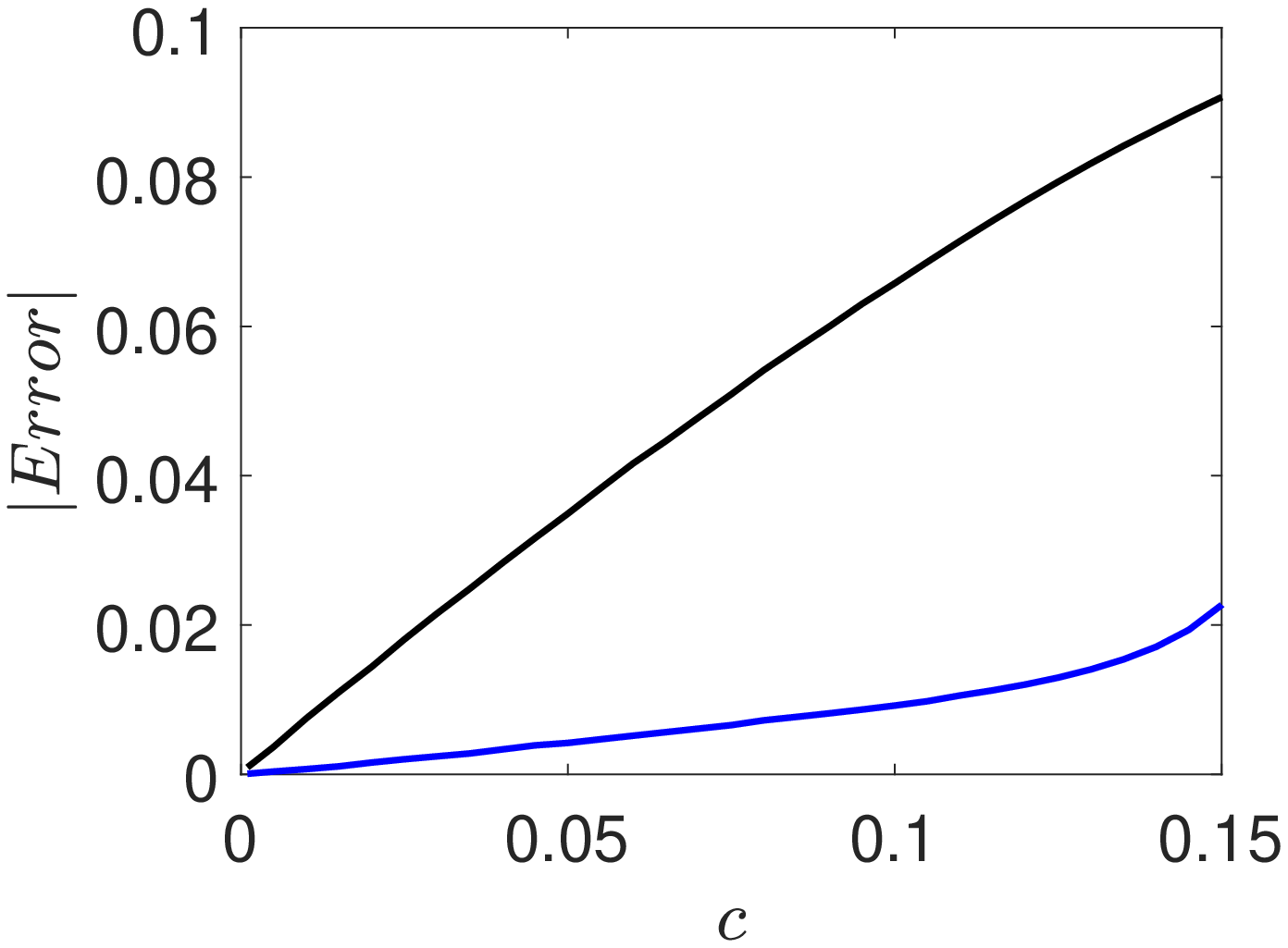}\label{subfig:width_type_3}}\\
		\subfloat[{Type 4}]{\includegraphics[scale=0.27]{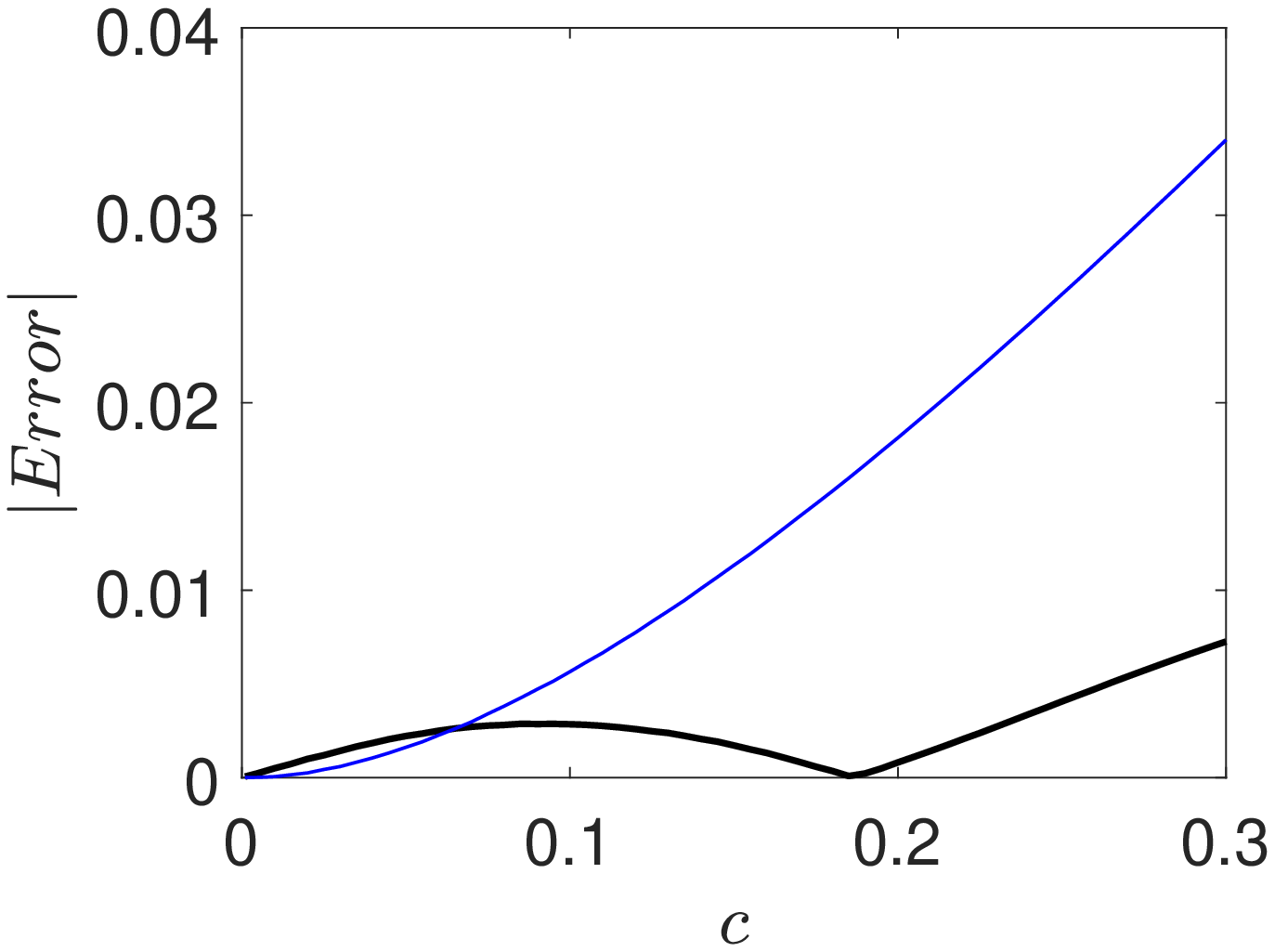}\label{subfig:width_type_4}}\,\,
		\subfloat[{Type 5}]{\includegraphics[scale=0.27]{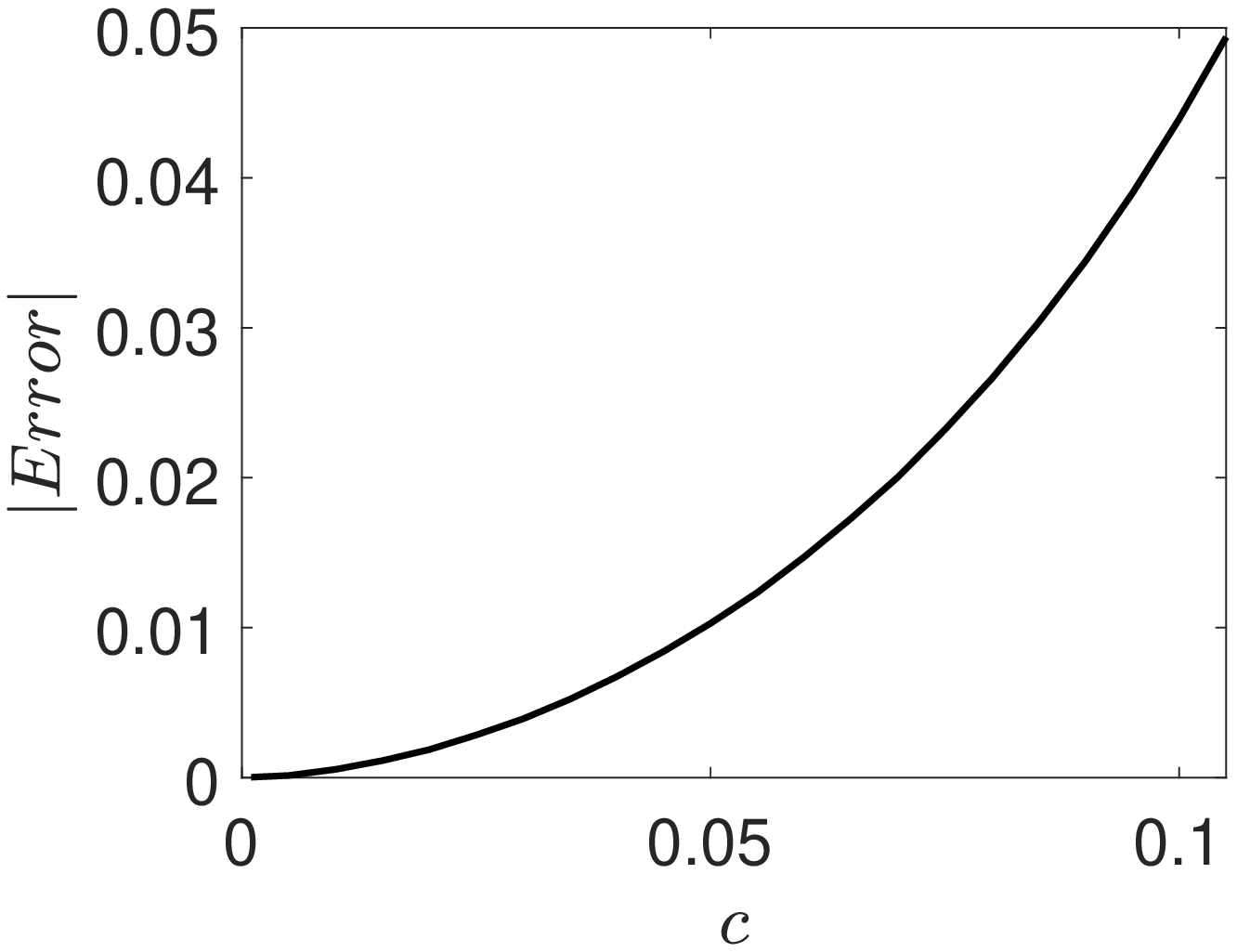}\label{subfig:width_type_5}}\,\,
		\subfloat[{Type 6}]{\includegraphics[scale=0.27]{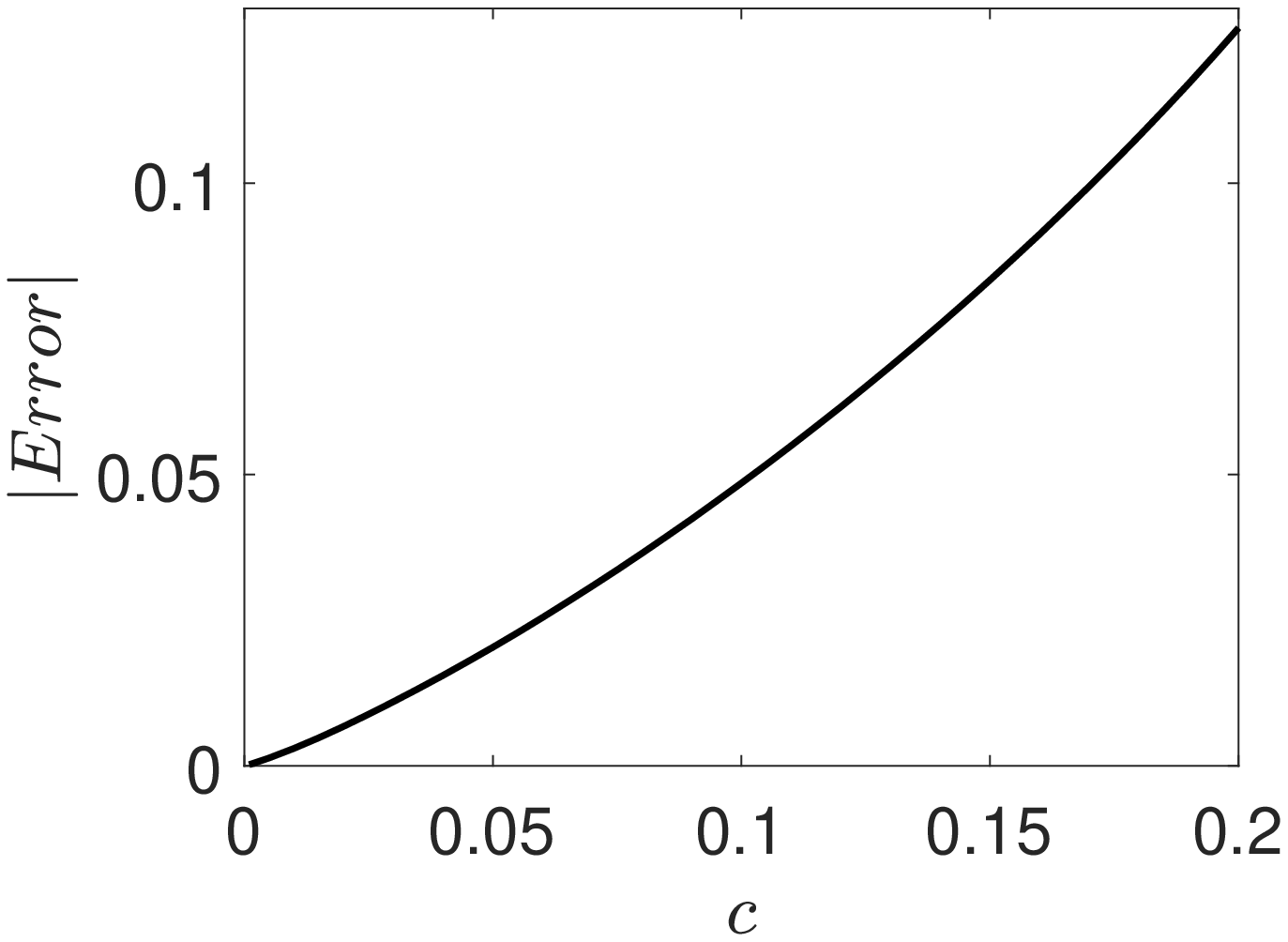}\label{subfig:width_type_6}}
		\caption{{Difference between the actual value and approximation of turning points in Fig.\ \ref{fig:pinning_type}. Shown is the absolute value.  
				Thick black and thin blue lines correspond to the difference along the left and right curves in Fig.\ \ref{fig:pinning_type}, 
				respectively.
			}
		}
		\label{fig:width_type}
	\end{figure}

	Figure \ref{subfig:snake_site_c_0_10} shows the bifurcation diagram of the site-centred solutions at $c = 0.1$ and our approximations of the saddle-node bifurcations. By comparing between $c = 0.05$ and $0.1$, we can see that the active-cell approximations give better results at small coupling strength.
	In this case, the active-cell approximations fail to approximate points (j), (k), and (n), while (h), (i), (l), and (m) are still relatively well approximated.
	
	\begin{figure}[tbp]
		\centering
		\subfloat[]{\includegraphics[scale=0.4]{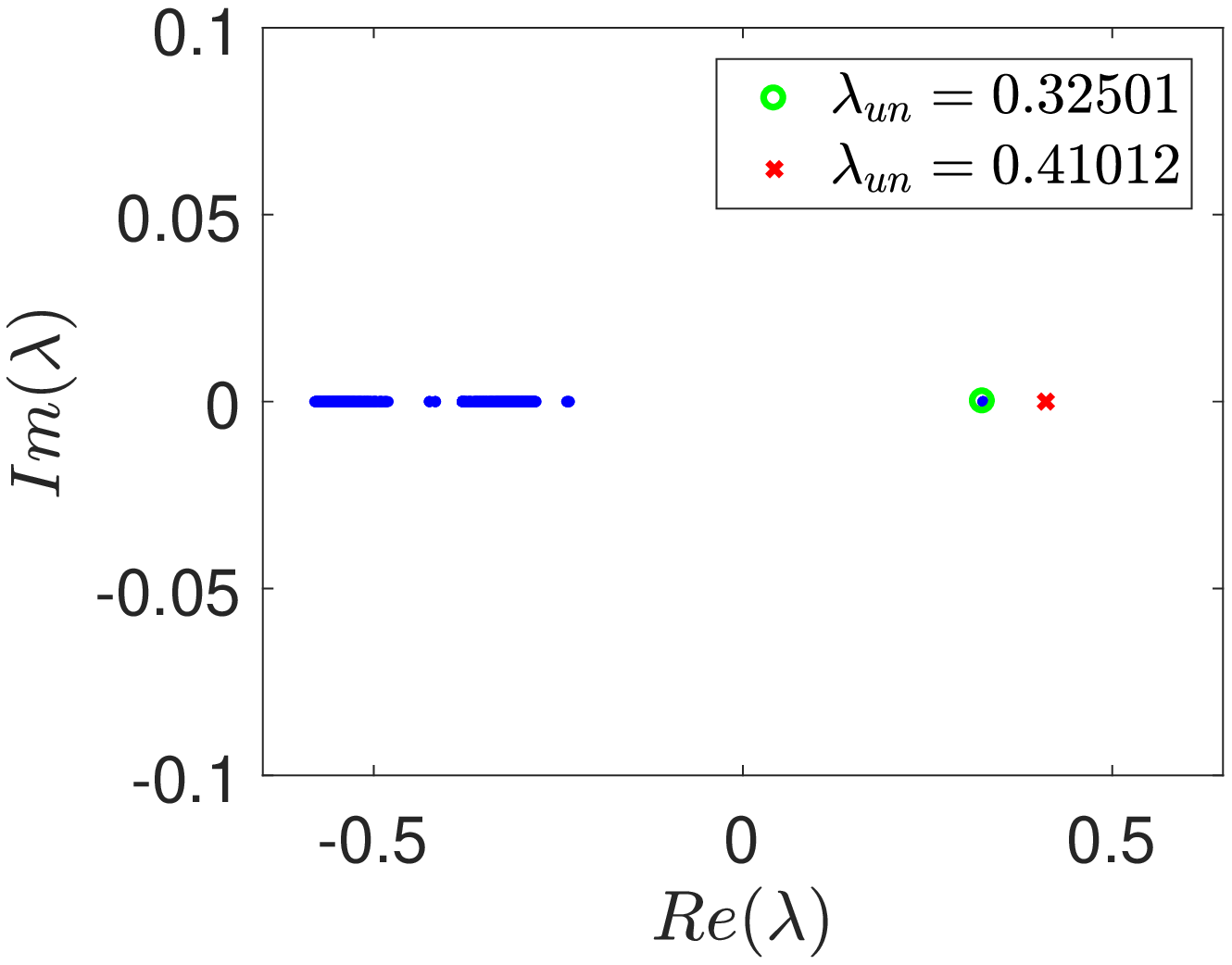}\label{subfig:eig_site_c_0_05_1}}
		\subfloat[]{\includegraphics[scale=0.4]{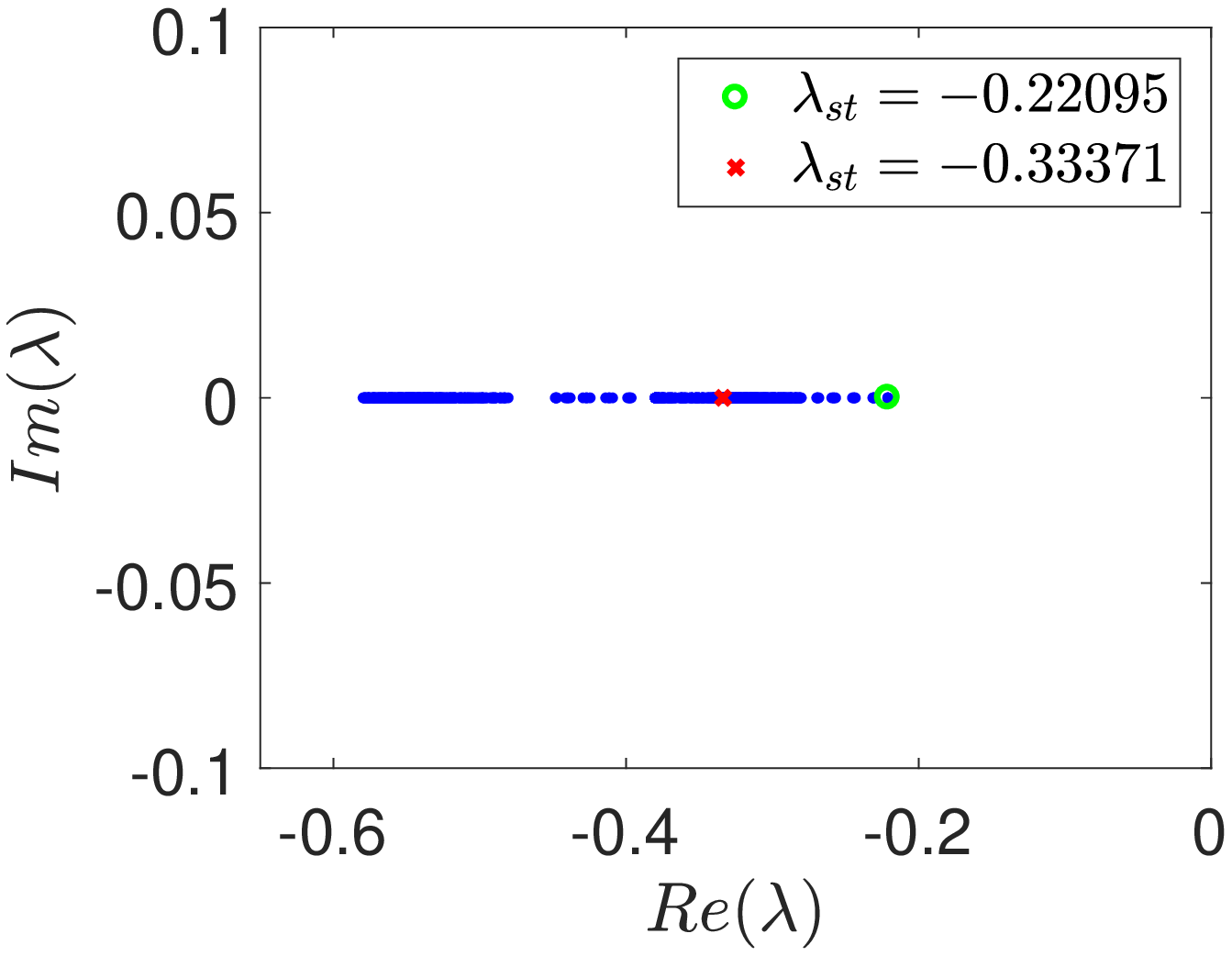}\label{subfig:eig_site_c_0_05_2}}
		\caption{Plot of numerical eigenvalues (blue dots and green circle) and our approximation (red cross) for site-centred solutions at points (1) and (2) in the bifurcation diagram in Fig.\ \ref{fig:bifur_site}, i.e.,\ see also Figs.\ \ref{fig:prof_site}(1) and \ref{fig:prof_site}(2).}
		\label{fig:eig_approx}
	\end{figure}

	\begin{figure}[htbp]
		\centering
		\subfloat[{Type 1}]{\includegraphics[scale=0.27]{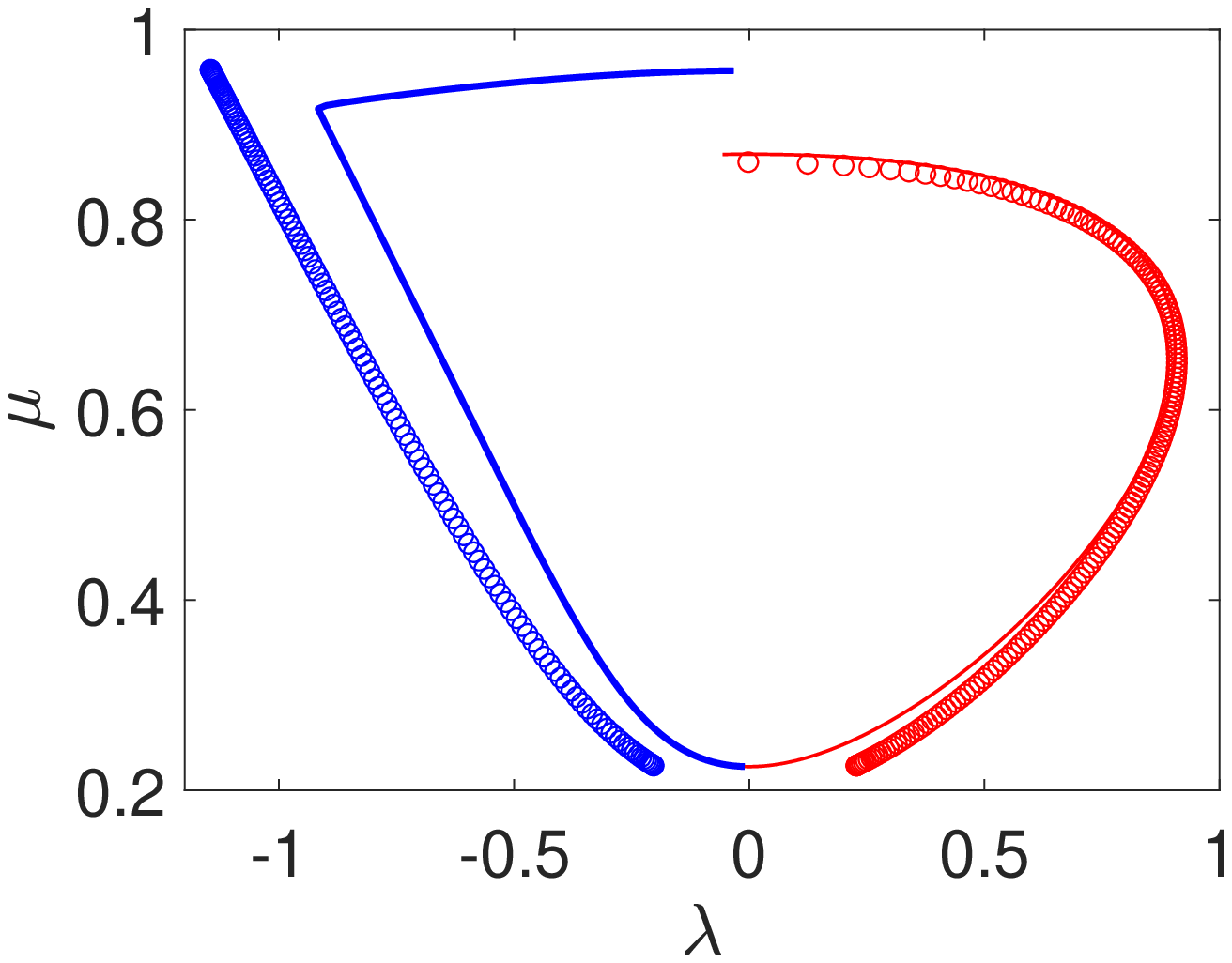}\label{subfig:eig_approx_type_1}}\,\,
		\subfloat[{Type 2}]{\includegraphics[scale=0.27]{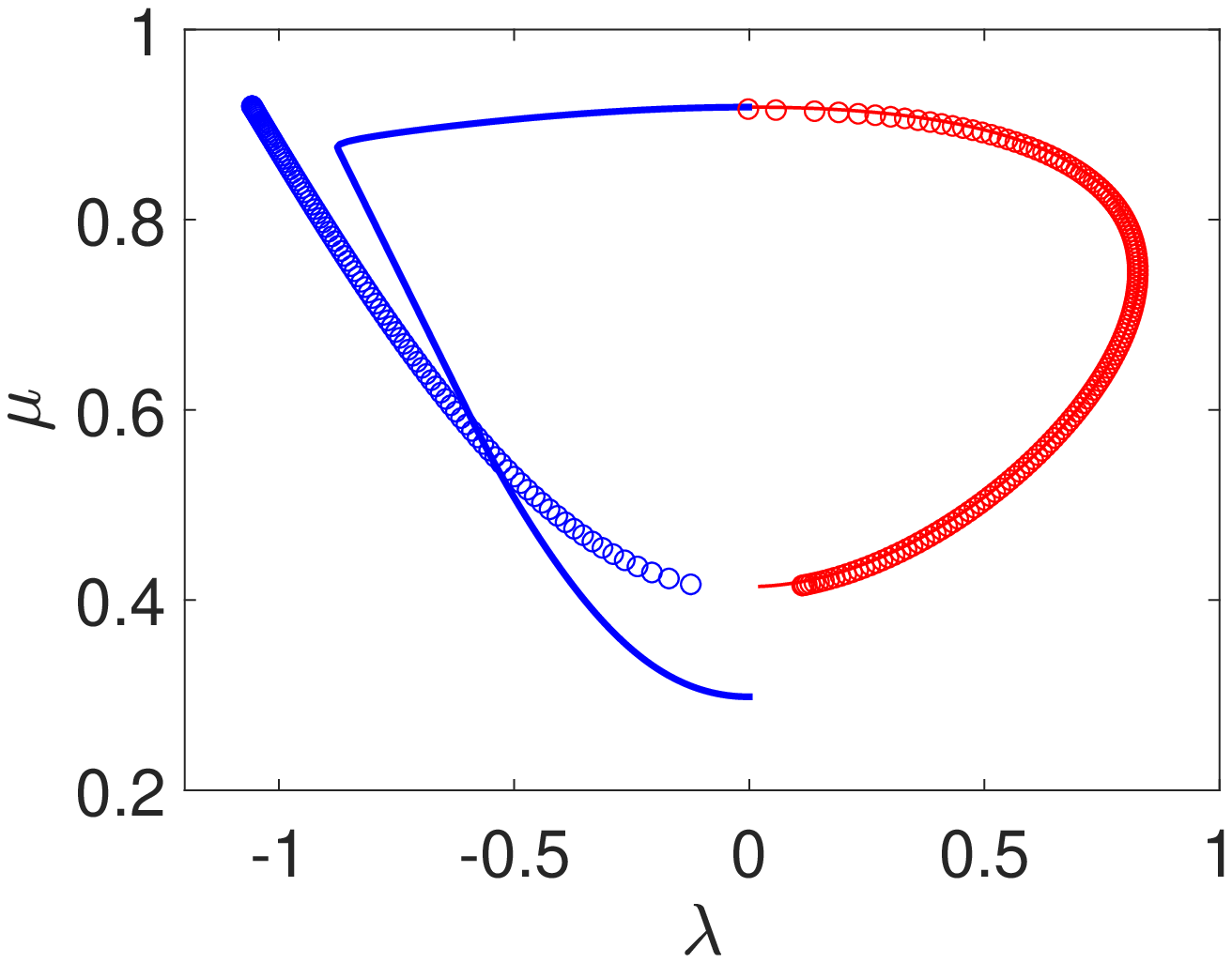}\label{subfig:eig_approx_type_2}}\,\,
		\subfloat[{Type 3}]{\includegraphics[scale=0.27]{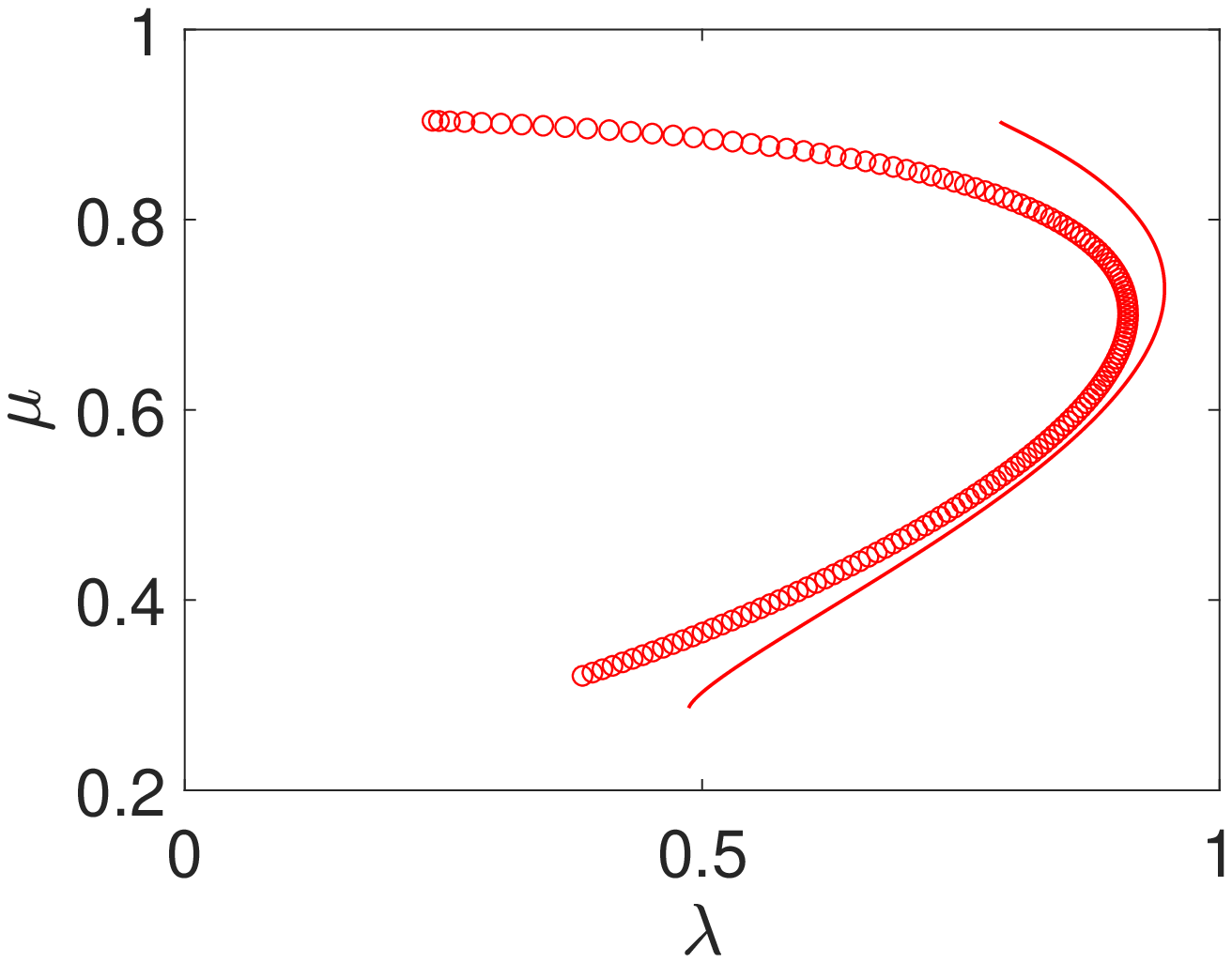}\label{subfig:eig_approx_type_3}}\\
		\subfloat[{Type 4}]{\includegraphics[scale=0.27]{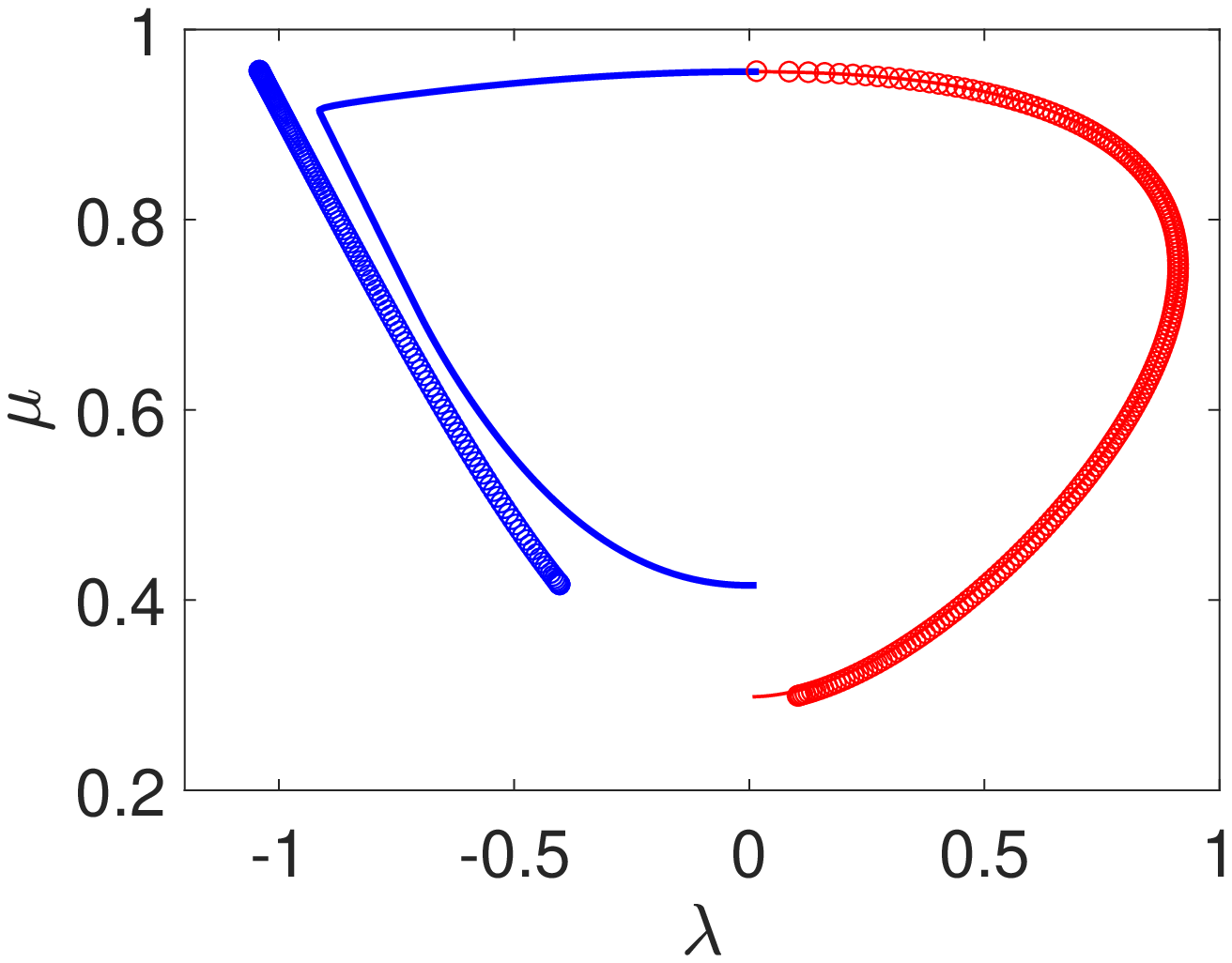}\label{subfig:eig_approx_type_4}}\,\,
		\subfloat[{Type 5}]{\includegraphics[scale=0.27]{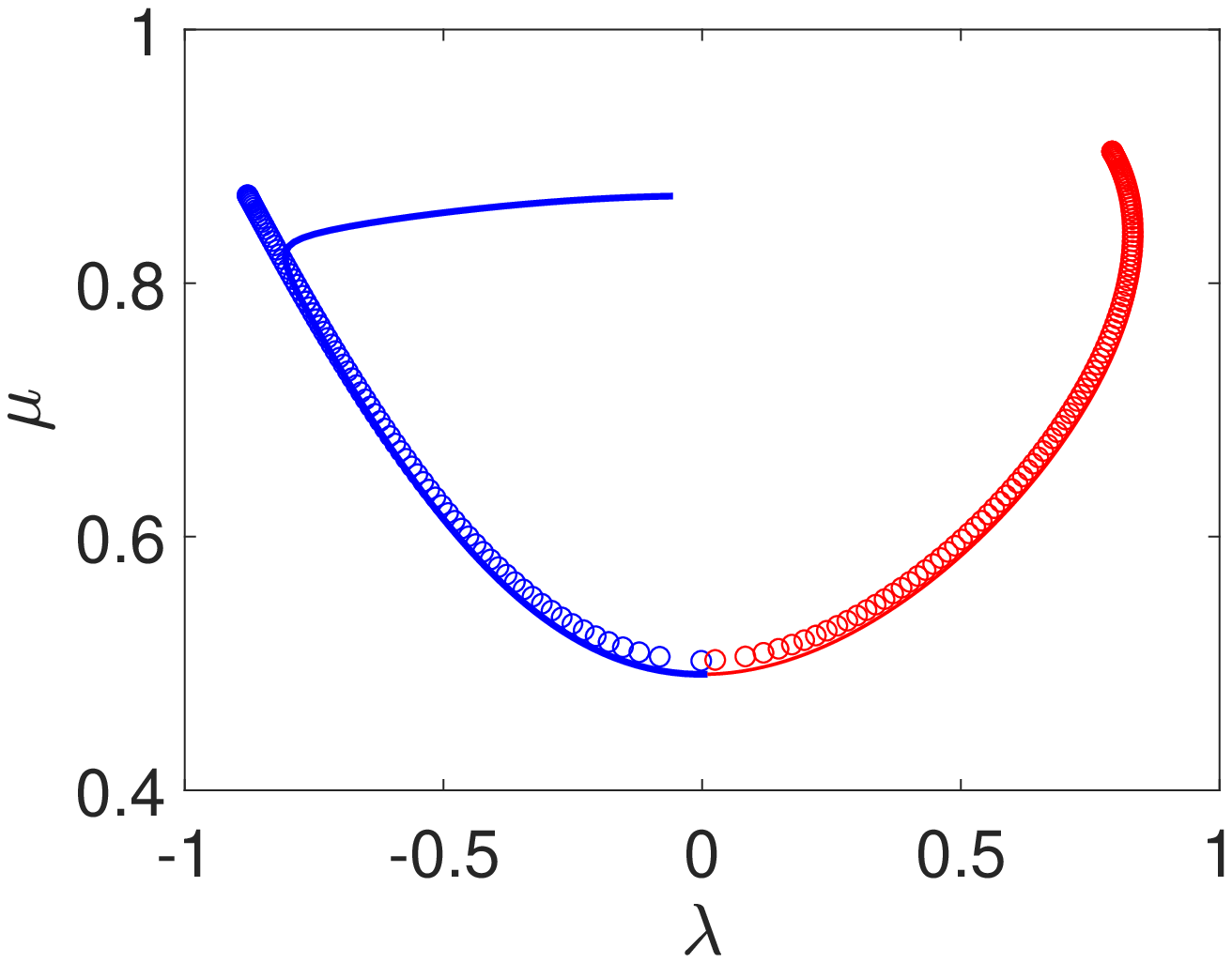}\label{subfig:eig_approx_type_5}}\,\,
		\subfloat[{Type 6}]{\includegraphics[scale=0.27]{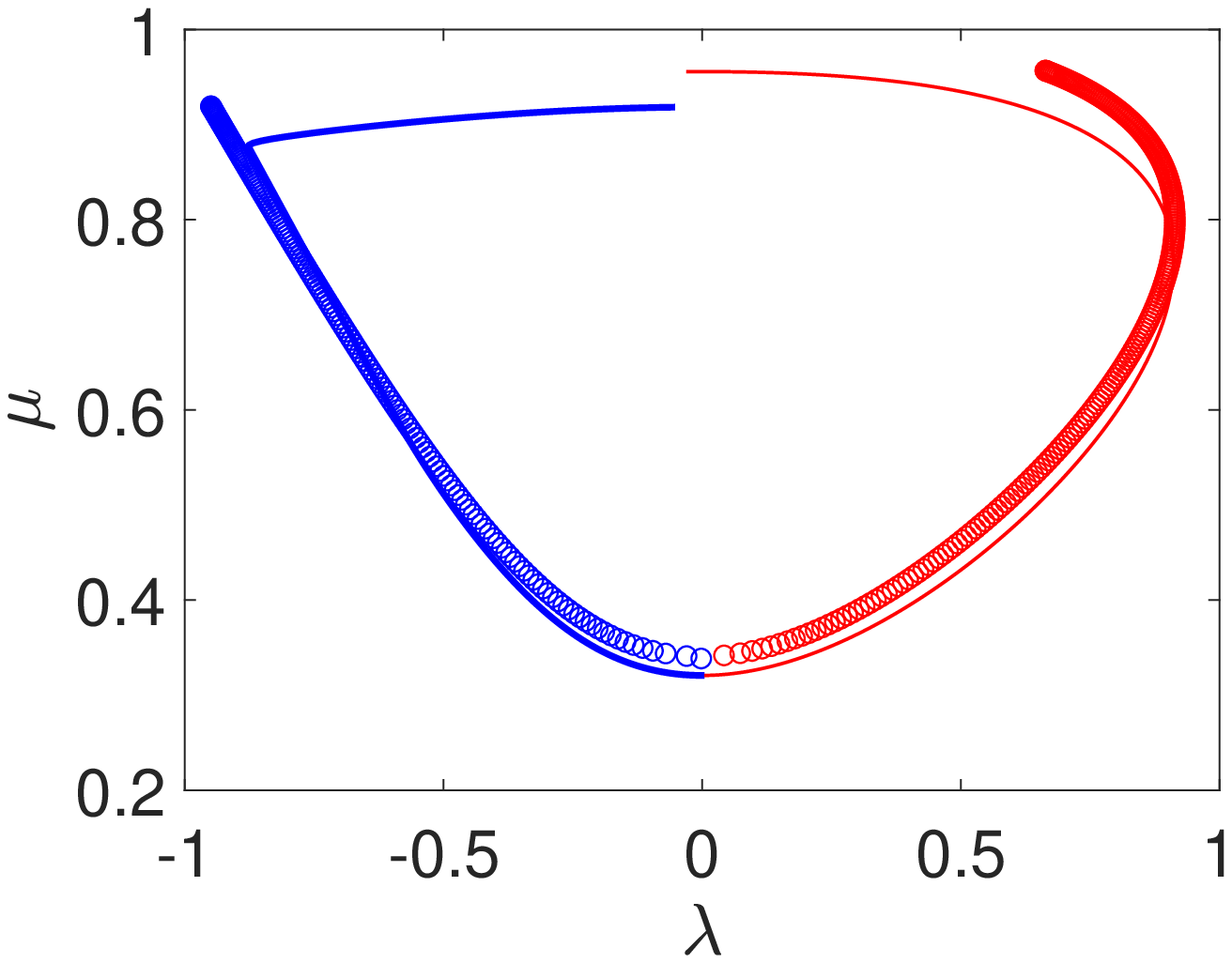}\label{subfig:eig_approx_type_6}}
		\caption{The critical eigenvalue of solutions in the neighborhood of some of the turning points in Fig.\ \ref{fig:bifur_bond} for $c=0.05$. The solutions correspond to the different types of saddle-node bifurcations. The solid lines indicate the critical eigenvalue from numerically solving the eigenvalue problem \eqref{ei}, while the circles are our approximation \eqref{ei2}.
		}
		\label{fig:eig_approx_type}
	\end{figure}
	
	\begin{figure}[htbp!]
		\centering
		\subfloat[{Type 1}]{\includegraphics[scale=0.27]{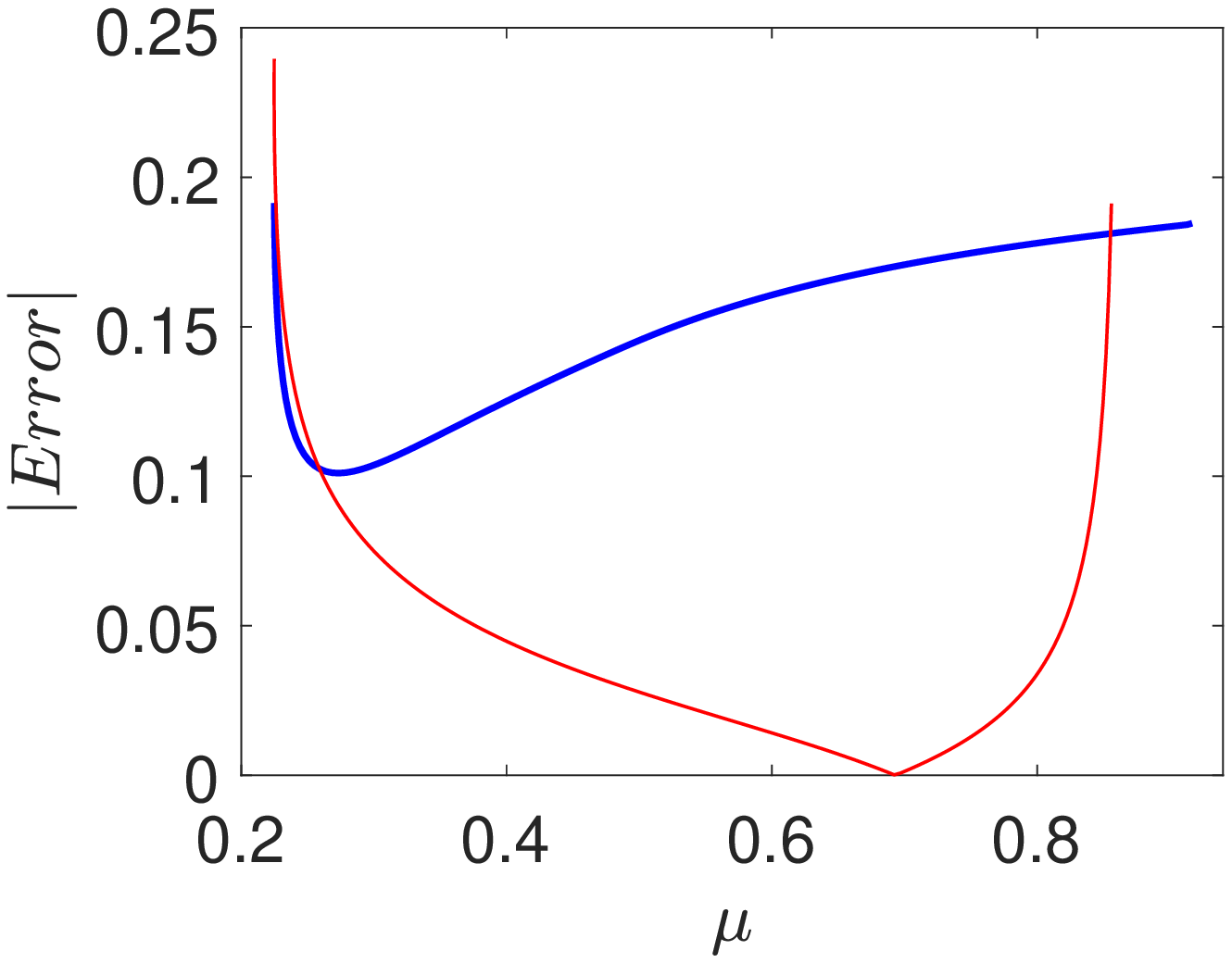}\label{subfig:eig_error_type_1}}\,\,
		\subfloat[{Type 2}]{\includegraphics[scale=0.27]{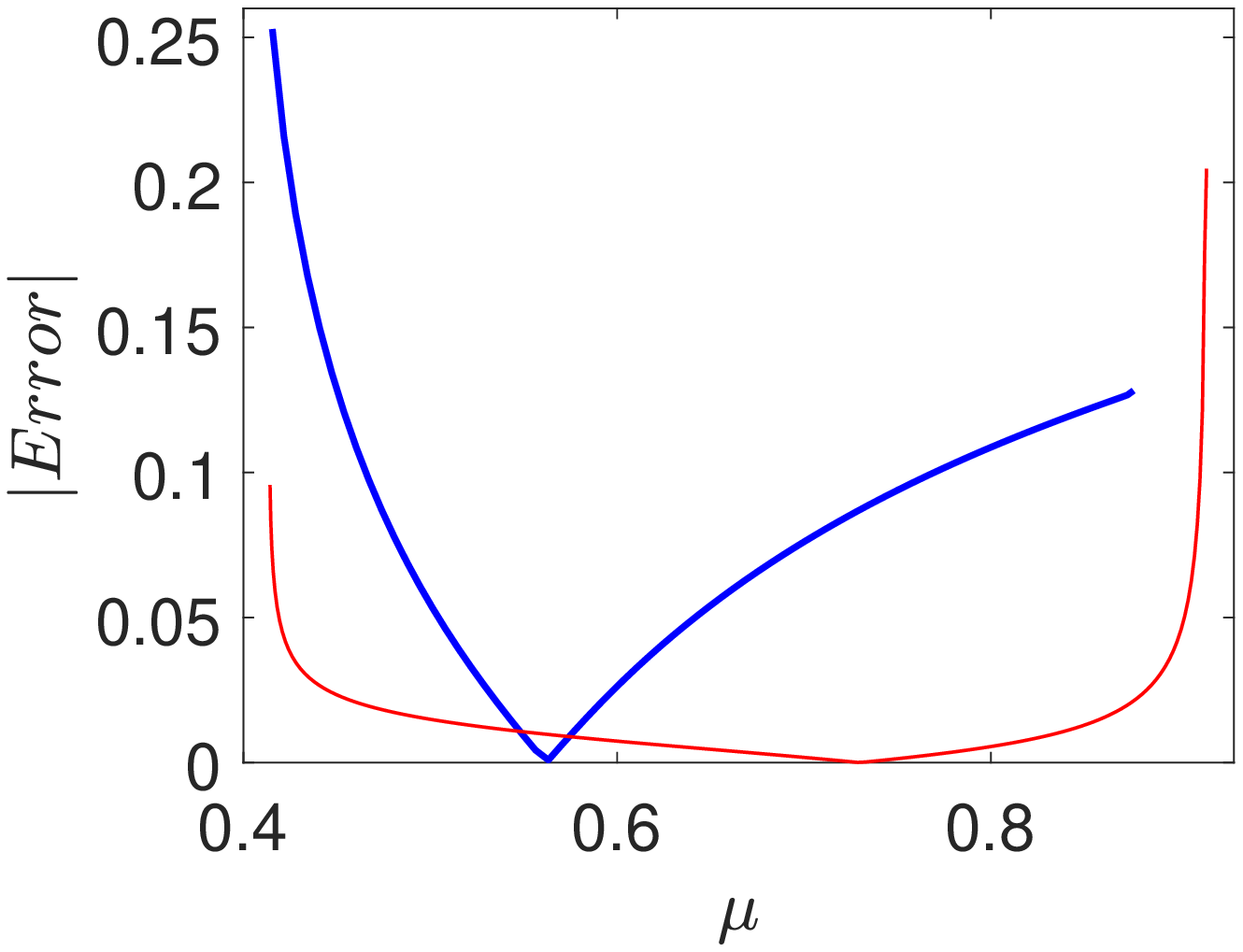}\label{subfig:eig_error_type_2}}\,\,
		\subfloat[{Type 3}]{\includegraphics[scale=0.27]{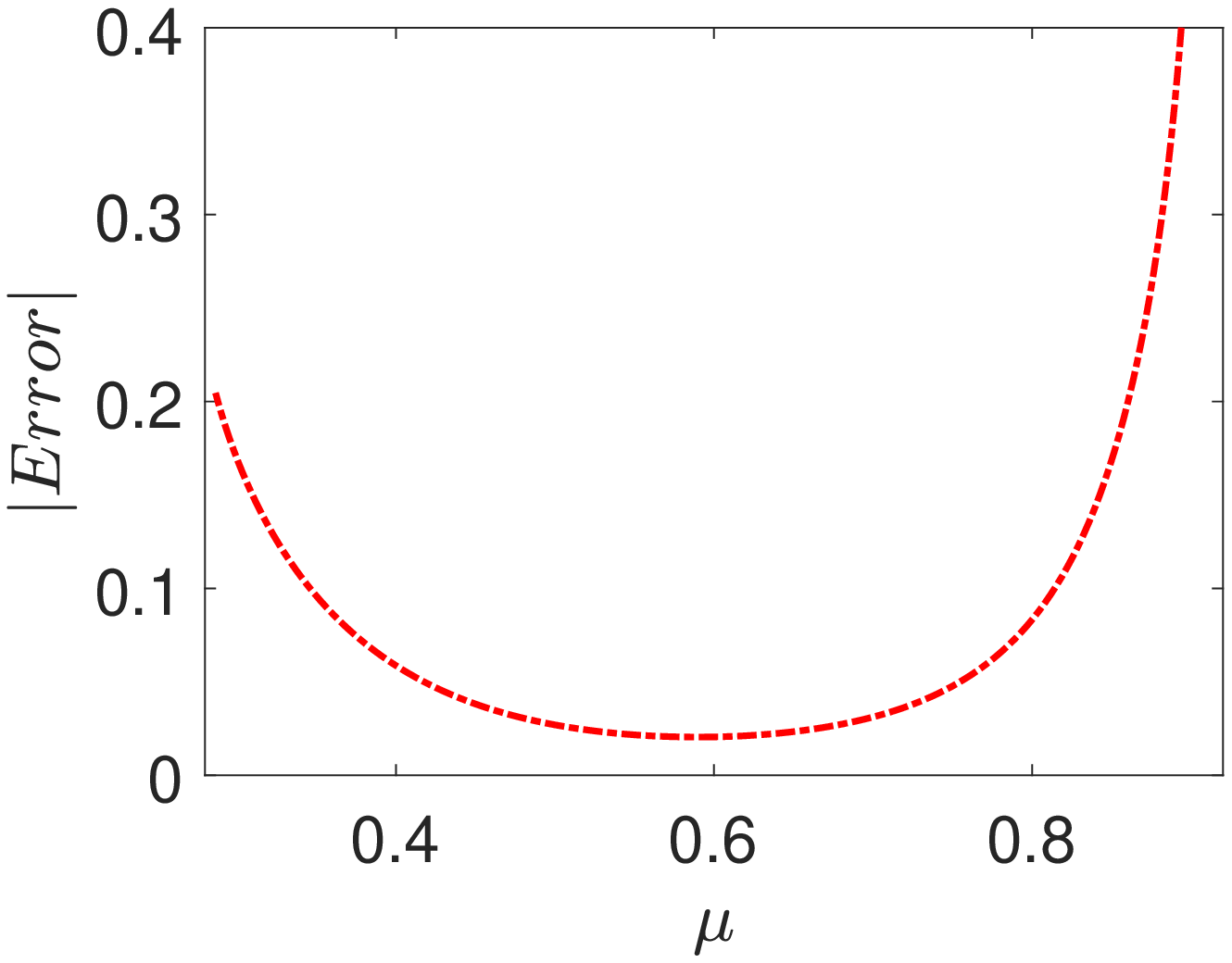}\label{subfig:eig_error_type_3}}\\
		\subfloat[{Type 4}]{\includegraphics[scale=0.27]{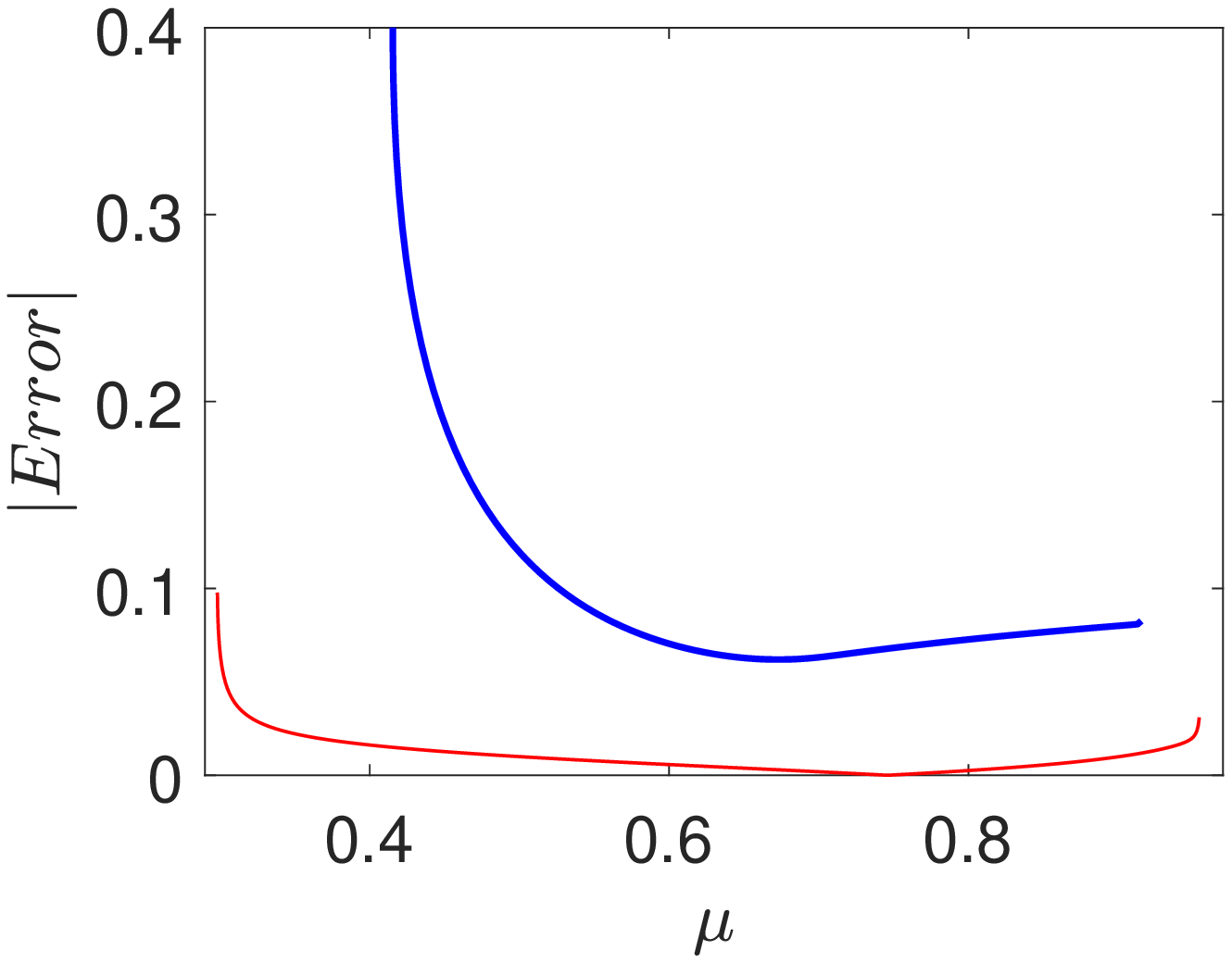}\label{subfig:eig_error_type_4}}\,\,
		\subfloat[{Type 5}]{\includegraphics[scale=0.27]{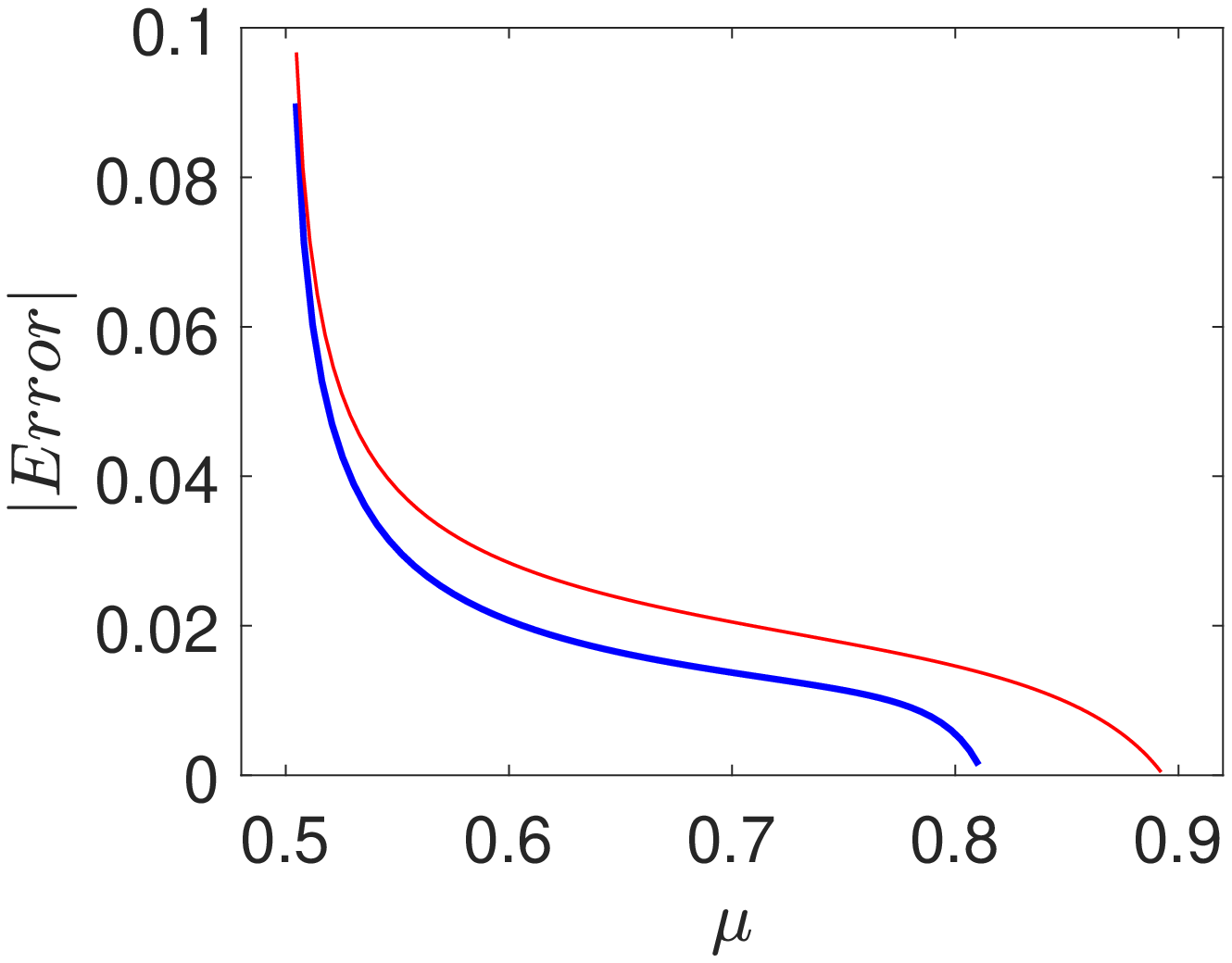}\label{subfig:eig_error_type_5}}\,\,
		\subfloat[{Type 6}]{\includegraphics[scale=0.27]{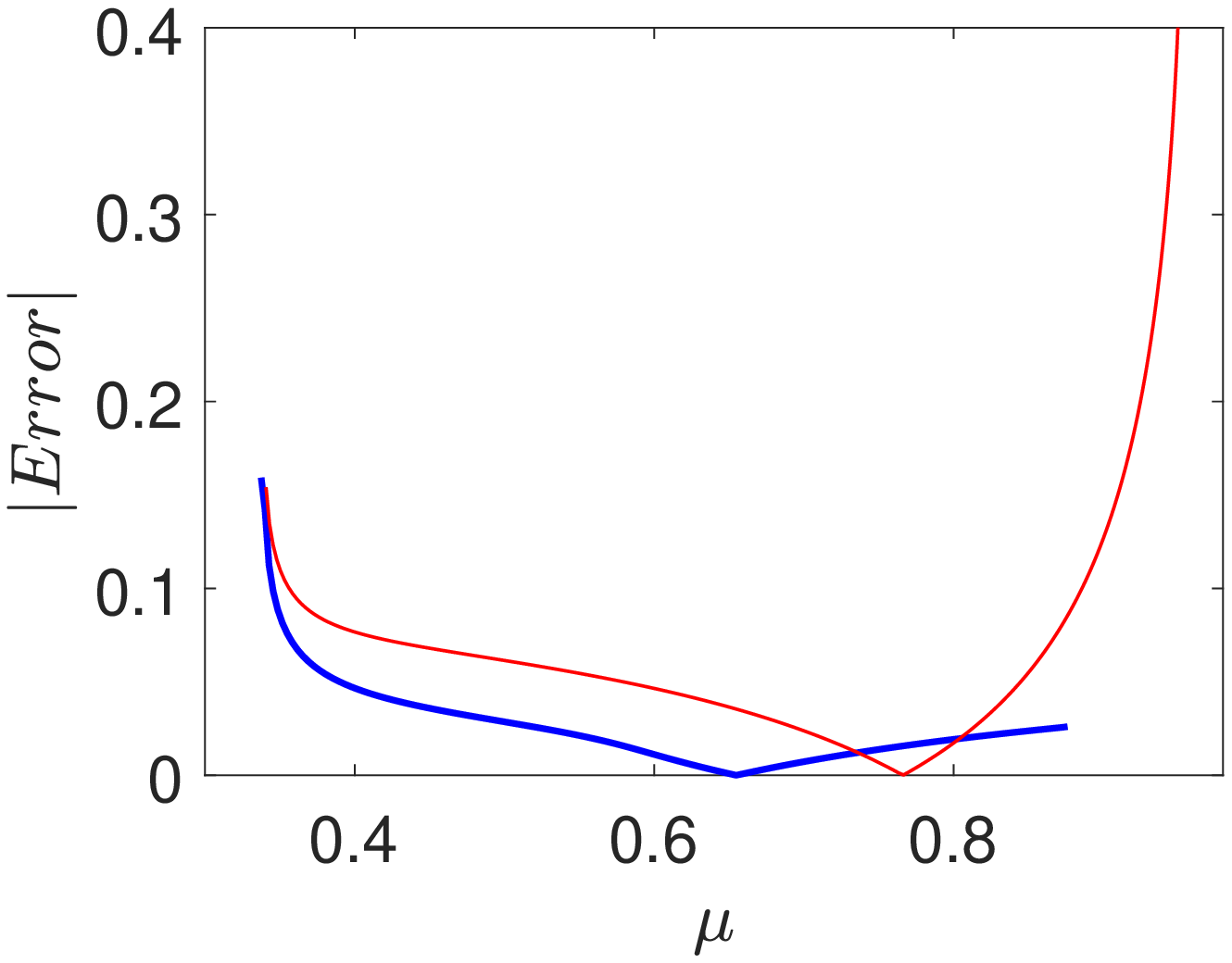}\label{subfig:eig_error_type_6}}
		\caption{{Error made by our approximation in predicting a critical eigenvalue, i.e., difference between curves depicted in Fig.\ \ref{fig:eig_approx_type}. Thick blue and thin red lines show the error along stable and unstable branches, respectively. Shown is the absolute value of the difference. 
			}
		}
		\label{fig:eig_error_type}
	\end{figure}
	
	Figure \ref{subfig:snake_bond_c_0_05} shows the bifurcation diagram of the bond-centred solutions at $c = 0.05$ and our approximations to the saddle-node bifurcations. Turning points of all types appear in bond-centred solutions. {Similarly to the site-centred solution, the approximations give good agreement for both the `lower' and `upper' saddle-node bifurcations. }
	
	Figure \ref{subfig:snake_bond_c_0_10} shows the case for the bond-centred solutions at $c = 0.1$.
	We also compare the diagrams with $c = 0.05$ and $0.1$. One can see that the active-cell approximations also give better results at smaller coupling strength. {
	One can see that the active-cell approximation fails to approximate points (o)-(t). In contrast, points (k)-(n) are still relatively well approximated. }
	
{
We provide in Figs.\ \ref{fig:pinning_type} a clear comparison of the different types of turning points from the original system Eq.\ \eqref{eq:ac_ti} (obtained numerically) and our asymptotic approximation Eq.\ \eqref{eq:active_cell}. We plot the location $\mu$ of the turning points 
as a function of coupling strength $c$. In general, the active-cell approximation gives an excellent result for relatively small values of $c$ (weakly coupled condition) and deviates from the actual value as $c$ increases. This is because the underlying assumption of the asymptotic analysis is no longer satisfied in the latter case as more cells become ``active'' with the increment of the coupling strength, see the top-view profile solutions in Figs.\ \ref{fig:prof_site} and \ref{fig:prof_bond}. Deviation of our approximations from the actual values is presented in Fig.\ \ref{fig:width_type} where types 2, 4 and 5 happen to have the smallest error.
}
	

	\subsection{Critical eigenvalue approximation}
	The active-cell approximation also can be used to approximate the critical eigenvalue of the localized solutions. By considering our assumption in Eq.\ \eqref{eq:active_cell}, it is straightforward that from the eigenvalue problem \eqref{ei}, one can obtain the approximation
	\begin{equation}
		\lambda\zeta=\left.\frac{d}{d\zeta}F(\zeta)\right|_{\zeta=\zeta_{st,un}}\zeta.
	\end{equation}
	Hence, $\lambda$ satisfies
	\begin{equation}
		\lambda(\mu)=-\mu+6\zeta_{st,un}^2-5\zeta_{st,un}^4+\frac{\partial Z}{\partial \zeta_{st,un}}.
		\label{ei2}
	\end{equation}
	The active-cell approximation of the critical eigenvalue at points (1) and (2) approximated by using type 1 indicated in Figs.\ \ref{fig:prof_site}(1) and \ref{fig:prof_site}(2) are shown in Fig.\ \ref{fig:eig_approx}, where good agreement is obtained when the coupling is weak.
	We also compare in Fig.\ \ref{fig:eig_approx_type} the critical eigenvalue along several branches containing the six different types of turning point and its active-cell approximation. 
	{Furthermore, we plot their difference 
		in Fig.\ \ref{fig:eig_error_type}. The results show that our analytical approximation is good enough in determining the stability of solutions, particularly when they contain a turning point of type 5
	.}

	\begin{figure*}[h!]
		\centering
		\subfloat[]{\includegraphics[scale=0.45]{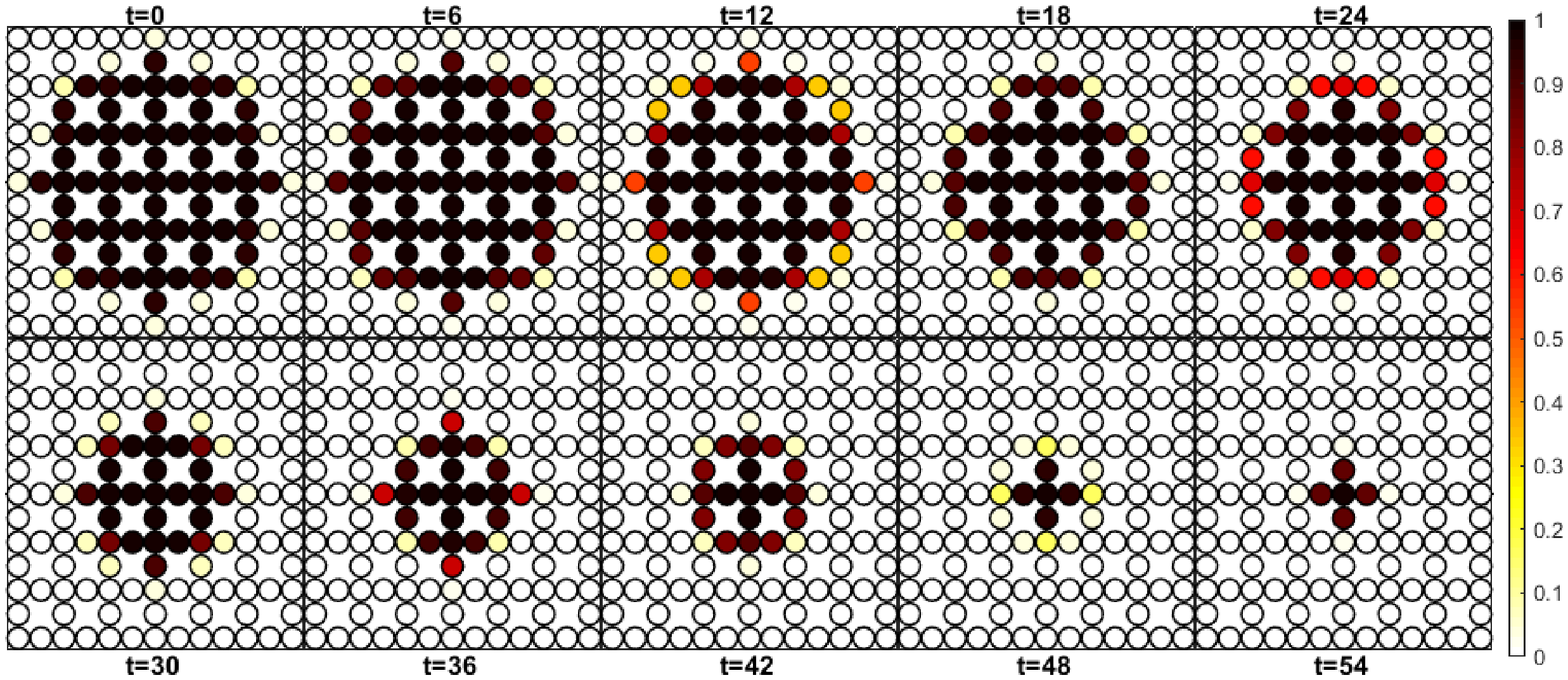}\label{subfig:prof_time_site_c_0_05_lower2}}\\
		\subfloat[]{\includegraphics[scale=0.45]{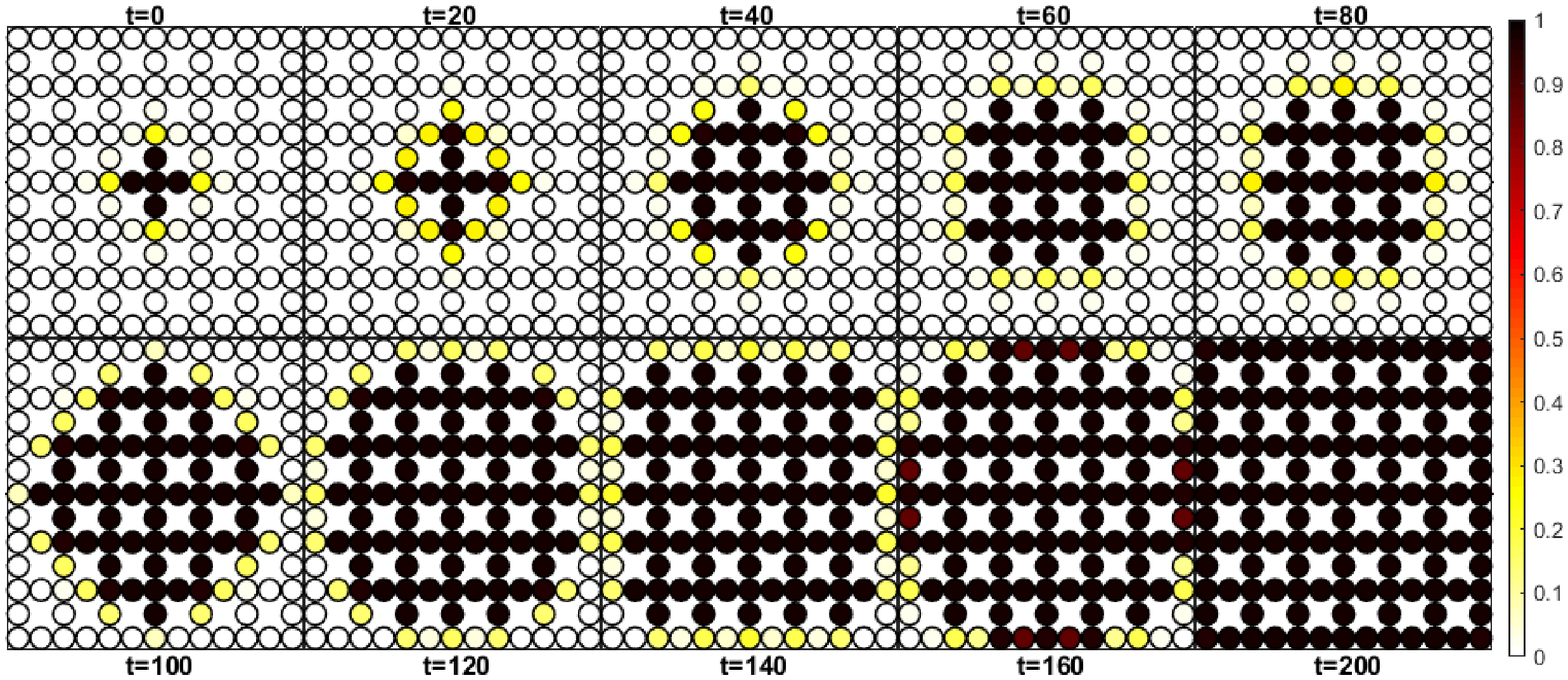}\label{subfig:prof_time_site_c_0_05_upper2}}
		\caption{Time evolution of site-centred solutions outside the snaking region, i.e., $\mu=0.97$ (a) and $\mu=0.22$ (b) for $c=0.05$. Shown are top-view snapshots at particular times.}
		\label{fig:time_site}
	\end{figure*}
	
	\begin{figure*}[h!]
		\subfloat[]{\includegraphics[scale=0.45]{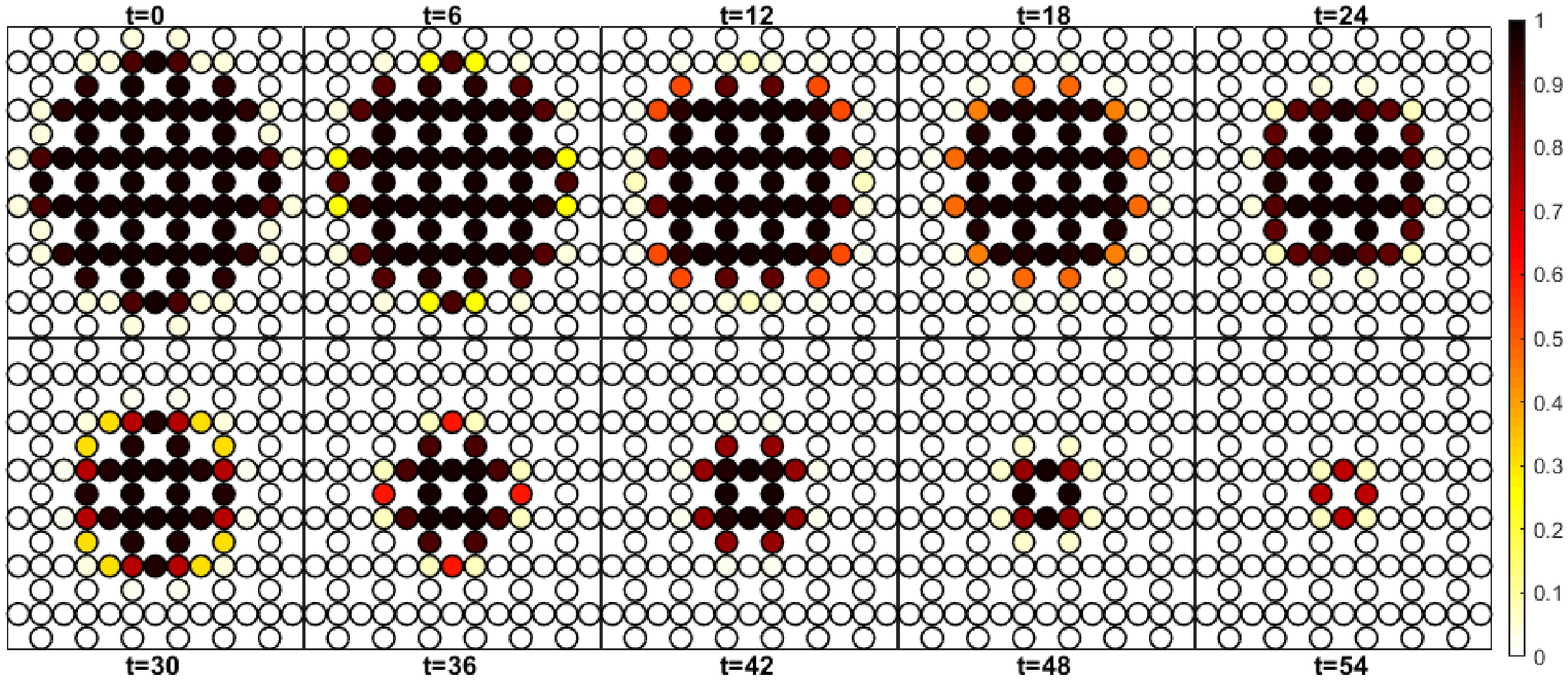}\label{subfig:prof_time_bond_c_0_05_lower2}}\\
		\subfloat[]{\includegraphics[scale=0.45]{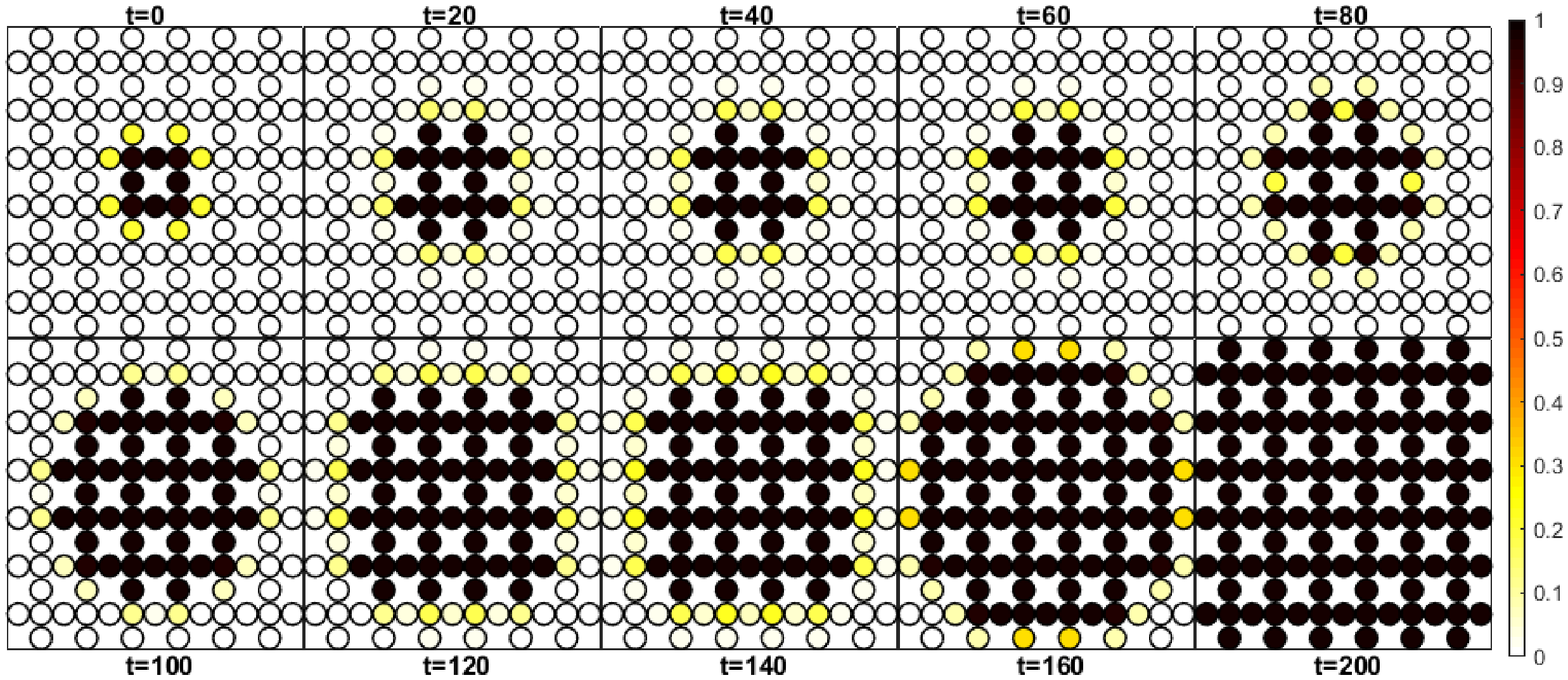}\label{subfig:prof_time_bond_c_0_05_upper2}}
		\caption{The same as Fig.\ \ref{fig:time_site} for  bond-centred solutions outside the snaking region.}
		\label{fig:time_bond}
	\end{figure*}
	
	\section{Dynamics inside and outside the pinning region}\label{sec:time}
	{
	When a solution is unstable, it is natural to question its dynamics in time. For an unstable solution that lies within the pinning region, it will evolve in time into a neighbouring stable solution, i.e., a `nearby' profile structure with a lower energy level $\mathcal{H}$. We integrate the governing equation \eqref{eq:ac_ori} in time where we obtain that, e.g., unstable solutions shown as point (1) and (3) in Fig.\ \ref{subfig:snake_site_c_0_05} will go to stable solutions indicated as point (2) and (4), respectively. This is the typical dynamics of a dissipative system inside a pinning region (see, e.g., \cite{Kusdiantara2017}). }

{	
We also consider dynamics of the system with parameter values outside the snaking region. While it may be obvious that there shall be no static localised solutions, the goal of the consideration is two fold: studying the difference between dynamics above and below the pinning region, and characterising time-dependent dynamics of the system \eqref{eq:ac_ori} from its static solutions along their bifurcation diagram. 
As a test case, we show in Figs.\ \ref{fig:time_site} and \ref{fig:time_bond} an example of such time dynamics, 
depicting time evolution of site and bond-centred solutions outside the snaking region respectively, for $\mu=0.97$ and $\mu=0.22$ with $c=0.05$.  Initial conditions for the simulations in Figs.\ \ref{subfig:prof_time_site_c_0_05_lower2} and \ref{subfig:prof_time_bond_c_0_05_lower2} are a solution up in the snaking of Fig.\ \ref{subfig:snake_site_c_0_05} with $M^2\approx63.7783$ and Fig.\ \ref{subfig:snake_bond_c_0_05} with $M^2\approx66.2366$, respectively, while 
those for Figs.\ \ref{subfig:prof_time_site_c_0_05_upper2} and \ref{subfig:prof_time_bond_c_0_05_upper2} are solutions depicted in Figs.\ \ref{subfig:prof_site_c_0_05_c} and \ref{subfig:prof_bond_c_0_05_c}, respectively.}

{
Above the pinning region, i.e., Figs.\ \ref{subfig:prof_time_site_c_0_05_lower2} and \ref{subfig:prof_time_bond_c_0_05_lower2}, we obtain that as $t\to\infty$, the solution goes to the trivial state. On the other hand, time evolution of the system below the pinning region as depicted in Figs.\ \ref{subfig:prof_time_site_c_0_05_upper2} and \ref{subfig:prof_time_bond_c_0_05_upper2}, shows that the solution tends to the non-zero uniform state. Generally, this occurs because the system prefers a lower energy state. We can explain the dynamics here from the energy $\mathcal{H}$ of the solutions along the bifurcation diagrams in Figs.\ \ref{subfig:snake_site_c_0_05} and \ref{subfig:snake_bond_c_0_05}.
Tables \ref{tab:site_energy} and \ref{tab:bond_energy} show the energy value for each solution indicated in Figs.\ \ref{subfig:snake_site_c_0_05} and \ref{subfig:snake_bond_c_0_05} (their profiles are depicted in Figs.\ \ref{fig:prof_site} and \ref{fig:prof_bond}).
The solution energy around the ``upper'' turning points has positive values and becomes larger as the norm $M^2$ increases. Because of the gradient property of the energy \eqref{dH}, above the pinning region, the system will therefore evolve in time to the trivial solution. In contrast, the energy around the `lower' saddle-node bifurcations has negative values and becomes smaller as the norm $M^2$ increases. It is the reason below the lower boundary, the system will tend to evolve into the non-zero uniform solution.}

\begin{figure*}[h!]
	\subfloat[Site-centred $\mu=0.97$]{\includegraphics[scale=0.45]{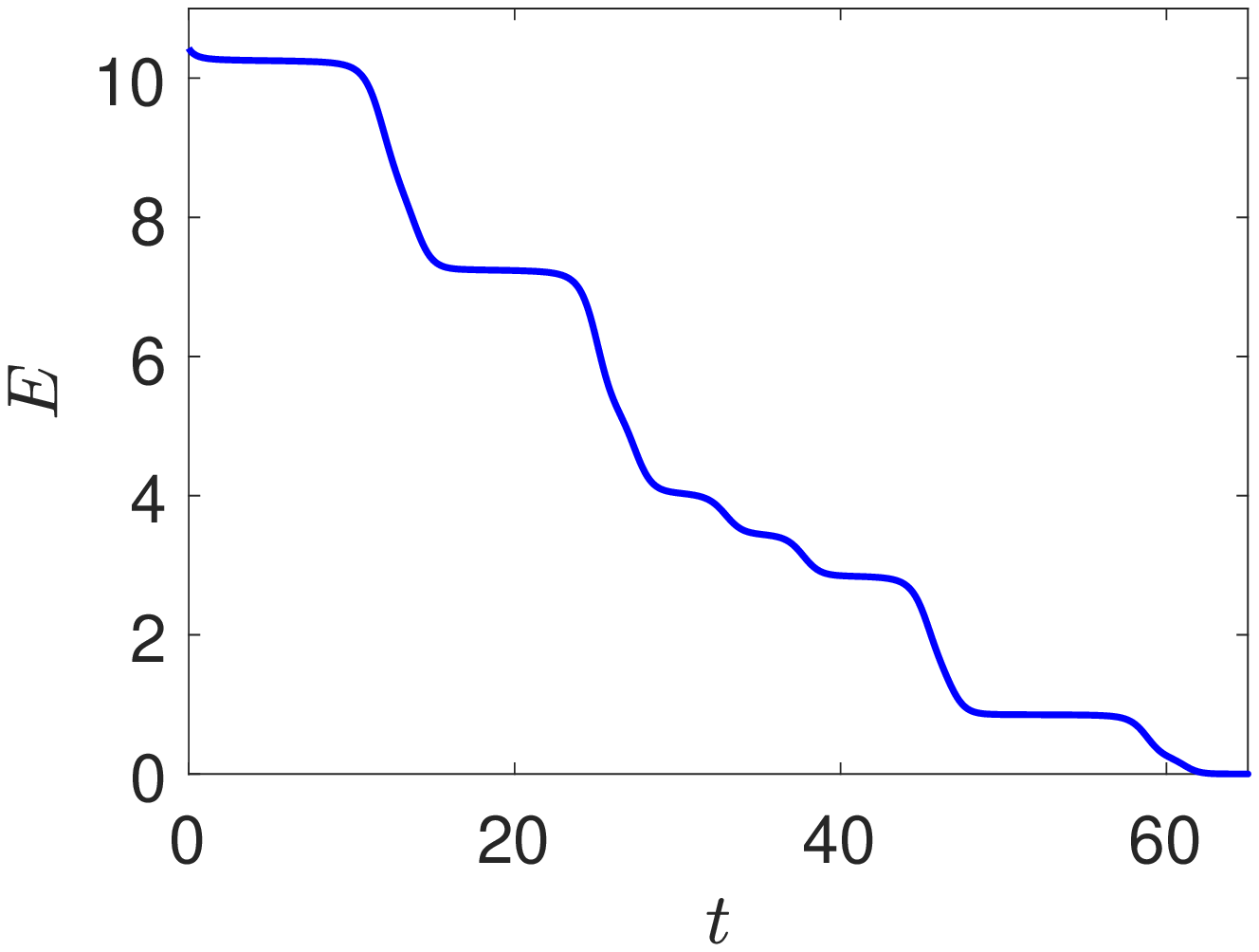}\label{subfig:Ha}}
	\subfloat[Site-centred $\mu=0.22$]{\includegraphics[scale=0.45]{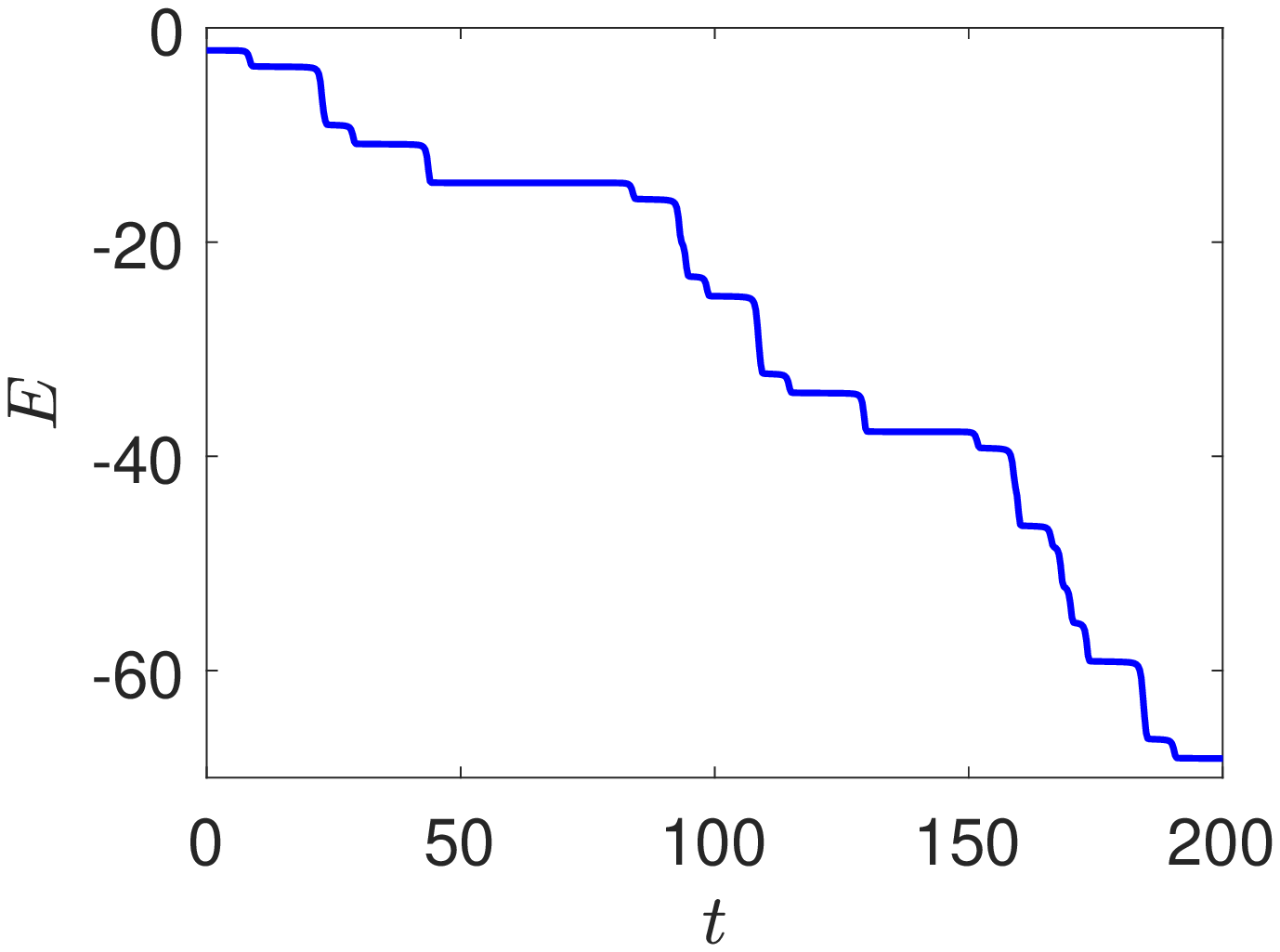}\label{subfig:Hb}}\\
	\subfloat[Bond-centred $\mu=0.97$]{\includegraphics[scale=0.45]{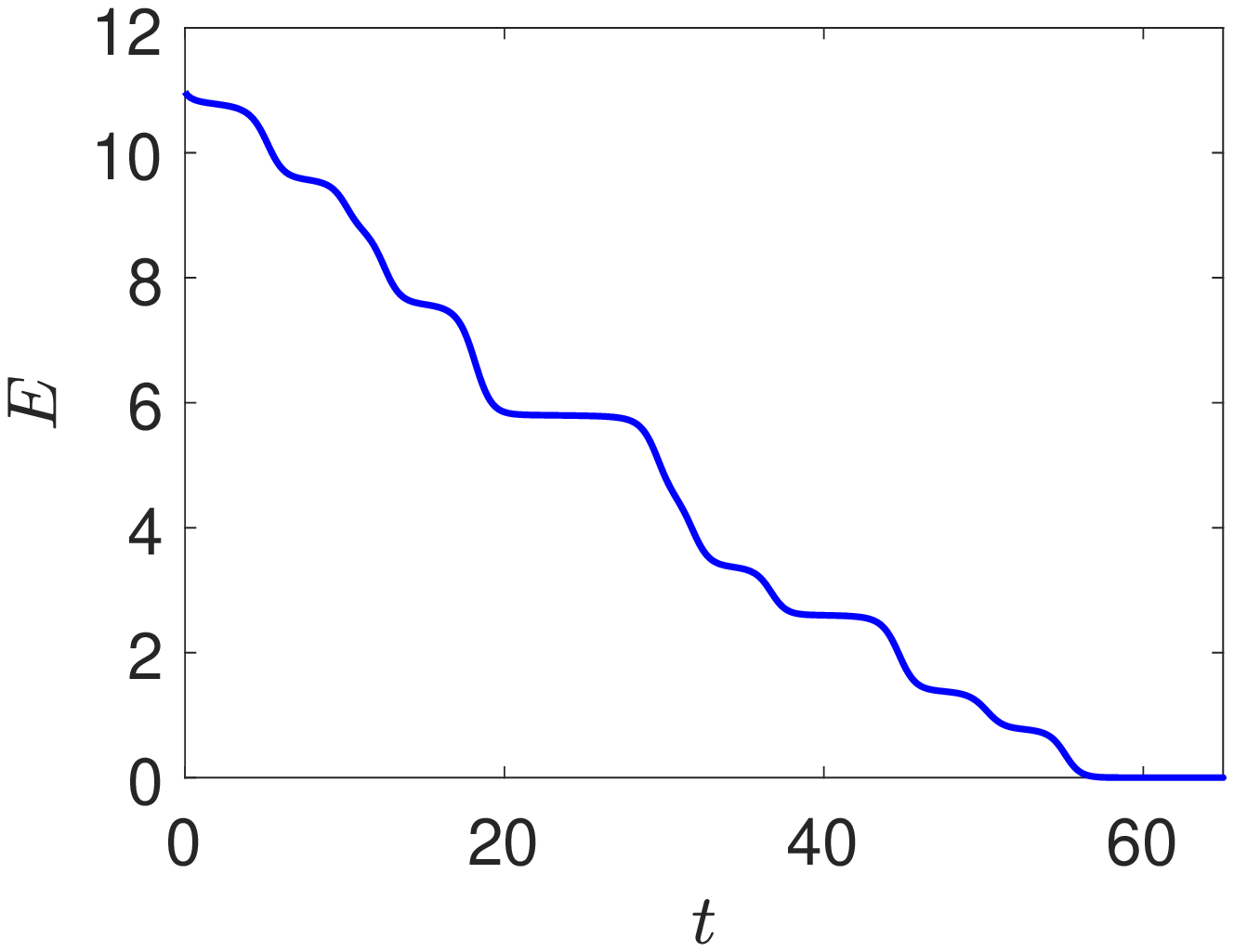}\label{subfig:Hc}}
	\subfloat[Bond-centred $\mu=0.22$]{\includegraphics[scale=0.45]{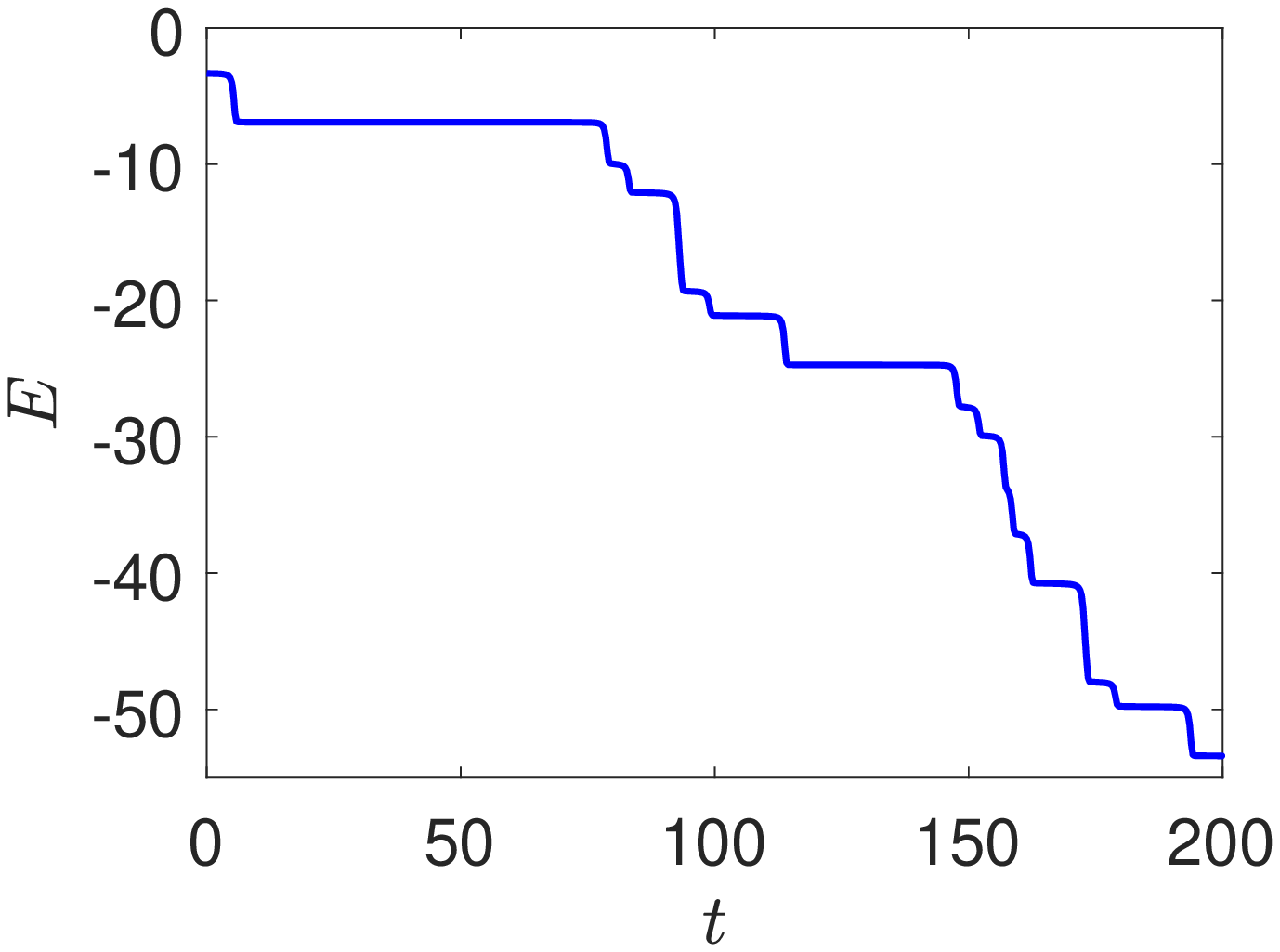}\label{subfig:Hd}}
	\caption{Energy evolution $\mathcal{H}(t)$ (cf.\ \eqref{eq:energy}) of the dynamics in Figs.\ \ref{fig:time_site} and \ref{fig:time_bond}.}
	\label{fig:H}
\end{figure*}

{
Moreover, we plot in Fig.\ \ref{fig:H} the corresponding energy of the solution dynamics depicted in Figs.\ \ref{fig:time_site} and \ref{fig:time_bond}. In agreement with the inequality \eqref{dH}, the energy flows down along a negative slope. Moreover, the curves have clear plateaus indicating that the system transits for some time at `almost' static state. Studying the profiles of those states (see Figs.\ \ref{fig:time_site} and \ref{fig:time_bond}), we obtain that they look like time-independent solutions at turning points.} 
	
	\begin{table}[htbp]
		\centering
		\caption{The energy \eqref{eq:energy} of the site (a) and bond-centred (b) solution profiles shown in Figs.\ \ref{fig:prof_site} and \ref{fig:prof_bond}.}
		\subfloat[]{
			\begin{tabular}{|c|r|l|c|r|}
				\cline{1-2} \cline{4-5}
				Point &\multicolumn{1}{c|}{ $\mathcal{H}$ }&\quad & Point & \multicolumn{1}{c|}{ $\mathcal{H}$ } \\ \cline{1-2} \cline{4-5} 
				(a)   & -0.235       &  & (j)   & -1.684       \\ 
				(b)   & 0.817        &  & (k)   & 0.741        \\ 
				(c)   & -2.115       &  & (l)   & -1.868       \\ 
				(d)   & 1.103        &  & (m)   & 2.285        \\ 
				(e)   & -4.223       &  & (n)   & -7.553       \\ 
				(f)   & 2.822        &  & (1)   & {{-11.938}}       \\ 
				(g)   & -12.580      &  & (2)   & {{-32.934}}      \\ 
				(h)   & -0.023       &  & (3)   & 
				{{5.525}}        \\ 
				(i)   & 0.808         &  & (4)   & {{3.867}}        \\ \cline{1-2} \cline{4-5} 
			\end{tabular}
			\label{tab:site_energy}}
		\hspace{1cm}
		\subfloat[]{
			\begin{tabular}{|c|r|l|c|r|}
				\cline{1-2} \cline{4-5}
				Point & \multicolumn{1}{c|}{ $\mathcal{H}$ } &\quad  & Point & \multicolumn{1}{c|}{ $\mathcal{H}$ } \\ \cline{1-2} \cline{4-5} 
				(a)   & -0.825       &  & (k)   & -0.165       \\ 
				(b)   & 1.168        &  & (l)   & 1.048        \\ 
				(c)   & -2.766       &  & (m)   & -1.672       \\ 
				(d)   & 2.490        &  & (n)   & 2.342        \\ 
				(e)   & -6.853        &  & (o)   & -5.488       \\ 
				(f)   & 2.734        &  & (p)   & 1.642        \\ 
				(g)   & -4.118       &  & (q)   & -0.060       \\ 
				(h)   & -21.411      &  & (r)   & -15.301      \\ 
				(i)   & 8.148        &  & (s)   & 4.790        \\ 
				(j)   & -23.504      &  & (t)   & -14.400      \\ \cline{1-2} \cline{4-5} 
			\end{tabular}
			\label{tab:bond_energy}}
	\end{table}

	\section{Conclusions}\label{sec:conclusion}
	We have considered a two-dimensional discrete Allen-Cahn equation with cubic and quintic nonlinearities in the domain of Lieb lattice. We have studied numerically and analytically time-independent solutions in the form of uniform and localized solutions and their stability. 
	
	
	We have shown that the localized solutions form a snaking structure in their bifurcation diagram. However, different from the previously reported cases, the snaking has more than one type of turning points. 
	While such a point is usually associated with a change of stability of the corresponding solution, in here we also obtained a {`switchback'} phenomenon in the bifurcation diagram, where no stability change occurs following a turning point.
	
	We have developed an active-cell approximation to estimate the saddle-node bifurcations. We showed that our analytical approximation gives good agreement with the numerical results for small coupling strength (relative to the parameter $\mu$, i.e., $|c|\ll\mu$). 
	{Furthermore, we also showed that the approximation can be used to determine the critical eigenvalue (stability) of localized solutions for small coupling constant. In particular, we obtained that on the Lieb lattice,  we can approximate the first saddle node well for bond-centred snaking (see Fig.\ \ref{fig:bifur_bond}) which was an issue in our previous work on regular (i.e., square, honeycomb, and triangular) lattices \cite{Kusdiantara2019}.}
	
	{
		We have derived the energy (potential) $\mathcal{H}$ of the Allen-Cahn equation on the Lieb lattice and performed time integration of unstable localized solutions. Our simulations showed that they {tend} to stable states that are `close' in structure with a lower energy level.}

{As we reported herein, the flat band that appears in the dispersion relation of the Lieb lattice has no effect on the structure of the homoclinic snaking. It is because the band is located in the opposite side of the existence domain of the localized solutions. As a follow up of the present work, we will 
seek for particular lattices with flat band that lies in the same parameter region as the localized solutions and study its effect on the snaking structure. {We also propose to consider the presence of anisotropy in the coupling strength between sites. Even in a simple square lattice configuration, such a non-uniformity can create nontrivial effects to a bifurcation diagram \cite{Taylor2010}. 
It will also be interesting to apply the active-cell approximation to snaking in even higher dimensional problems.}

	\begin{acknowledgements}
		RK gratefully acknowledged financial support from WCU ITB 2019 ``In House Post Doctoral Program''. 
		FTA gratefully acknowledged financial support from Riset ITB 2021.
		The work of BEG is partly supported by PPD RistekBRIN 2020-2021. HS is supported by Khalifa University through a Faculty Start-Up Grant (No.\ 8474000351-FSU-2021-011).
		RK, FTA, NN, and BEG gratefully acknowledged financial support from P2MI FMIPA ITB 2022.
	\end{acknowledgements}

	%
	\section*{Conflict of interest}
	
	The authors declare that they have no conflict of interest.


\begin{thebibliography}{100}
		


%

%
		\bibitem{Avitabile2010}
		{ Avitabile, D., Lloyd, D. J.~B., Burke, J., Knobloch, E., and Sandstede,
			B.}:
		\newblock {To Snake or Not to Snake in the Planar Swift–Hohenberg Equation}.
		\newblock {SIAM J. Appl. Dyn. Syst. \textbf {9}}(3), 704--733 (2010)
		
		\bibitem{Barbay2008}
		{ Barbay, S., Hachair, X., Elsass, T., Sagnes, I., and Kuszelewicz, R.}:
		\newblock {Homoclinic snaking in a semiconductor-based optical system}.
		\newblock {Phys. Rev. Lett.\textbf  {101}}(25), 253902 (2008)
%
		\bibitem{Beaume2011}
		{ Beaume, C., Bergeon, A., and Knobloch, E.}:
		\newblock {Homoclinic snaking of localized states in doubly diffusive
			convection}.
		\newblock {Phys. Fluids \textbf {23}}(9), 094102 (2011)
%
		\bibitem{Beaume2013}
		{ Beaume, C., Bergeon, A., and Knobloch, E.}:
		\newblock {Convectons and secondary snaking in three-dimensional natural doubly
			diffusive convection}.
		\newblock {Phys. Fluids \textbf {25}}(2), 024105 (2013)
%
		\bibitem{Bensimon1988}
		{ Bensimon, D., Shraiman, B.~I., and Croquette, V.}:
		\newblock {Nonadiabatic effects in convection}.
		\newblock {Phys. Rev. A \textbf {38}}(10), 5461(R) (1988)
%
%
		
		\bibitem{Bortolozzo2009}
		{ Bortolozzo, U., Clerc, M.~G., and Residori, S.}:
		\newblock {Solitary localized structures in a liquid crystal light-valve
			experiment}.
		\newblock {New J. Phys. \textbf {11}}, 093037 (2009)
%
		\bibitem{Bortolozzo2009a}
		{ Bortolozzo, U., Clerc, M.~G., Haudin, F., Rojas, R. G., and Residori, S.}:
		\newblock {Localized states in bi-pattern systems}.
		\newblock {Advances in Nonlinear Optics \textbf {2009}}, 926810 (2009)
		
		\bibitem{Boudebs2003}
		{Boudebs, G., Cherukulappurath, S., Leblond, H., Troles, J., Smektala, F., and Sanchez, F.}:
		\newblock {Experimental and theoretical study of higher-order nonlinearities in chalcogenide glasses}.
		\newblock {Opt. Commun. \textbf {219}}(1-6), 427--433 (2003)
%
		
		\bibitem{Bramburger2020a}
		{ Bramburger, J.~J., and Sandstede, B.}:
		\newblock Localized patterns in planar bistable weakly coupled lattice systems.
		\newblock {Nonlinearity \textbf {33}}(7), 3500 (2020)
%
		\bibitem{Braun2004}
		{Braun, O. M., Kivshar, Y., and Kivshar, Y. S.}:
		\newblock {The Frenkel-Kontorova model: concepts, methods, and applications}.
		\newblock {Springer Science \& Business Media}, (2004)
%
		
		
		\bibitem{Burke2012}
		{ Burke, J., and Dawes, J. H.~P.}:
		\newblock {Localised states in an extended Swift-Hohenberg equation}.
		\newblock {SIAM J. Appl. Dyn. Syst. \textbf {11}}(1), 261--284 (2012)
%
		\bibitem{Burke2007}
		{ Burke, J., and Knobloch, E.}:
		\newblock {Homoclinic snaking: Structure and stability}.
		\newblock {Chaos \textbf {17}}, 037102 (2007)
%
		\bibitem{Burke2007a}
		{ Burke, J., and Knobloch, E.}:
		\newblock {Snakes and ladders: Localized states in the Swift-Hohenberg
			equation}.
		\newblock {Phys. Lett. A \textbf {360}}(6), 681--688 (2007)
		
		\bibitem{Carretero-Gonzalez2006}
		{Carretero-Gonz\'alez, R., Talley, J. D., Chong, C., and Malomed, B. A.}:
		\newblock { Multistable solitons in the cubic–quintic discrete nonlinear Schrödinger equation}.
		\newblock {Physica D \textbf {216}}(1), 77--89 (2006)
%
		\bibitem{Chen2019}
		{ Chen, Y., and Chen, H.}:
		\newblock {Photonic zero-energy modes in a metal-based Lieb lattice}.
		\newblock {New Journal of Physics \textbf {21}}(11), (2019)
		
%
		\bibitem{Chong2009}
		{ Chong, C., Carretero-Gonz{\'{a}}lez, R., Malomed, B.~A., and Kevrekidis,
			P.~G.}:
		\newblock {Multistable solitons in higher-dimensional cubic-quintic nonlinear
			Schr{\"{o}}dinger lattices}.
		\newblock {Physica D \textbf {238}}(2), 126--136 (2009)
		
		\bibitem{Chong2011}
		{Chong, C., and Pelinovsky, D. E.}:
		\newblock { Variational approximations of bifurcations of asymmetric solitons in cubic-quintic nonlinear Schr\"odinger lattices}.
		\newblock {Disc. Cont. Dyn. Sys. S \textbf {4}}(1), 1019--1032 (2011)
%
		\bibitem{Cisternas2020}
		{ Cisternas, J., Escaff, D., Clerc, M.~G., Lefever, R., and Tlidi, M.}:
		\newblock {Gapped vegetation patterns: Crown/root allometry and snaking
			bifurcation}.
		\newblock {Chaos, Solitons and Fractals \textbf {133}} (2020).
%
		\bibitem{Clerc2017}
		{ Clerc, M. G., Ferr\'e, M. A., Coulibaly, S., Rojas, R. G., and Tlidi, M. }:
		\newblock {Chimera-like states in an array of coupled-waveguide resonators}.
		\newblock {Opt. Lett. \textbf {42}}(15), 2906-2909 (2017).
		
		\bibitem{Clerc2020}
		{ Clerc, M. G., Coulibaly, S., Ferré, M. A., and Tlidi, M.}:
		\newblock {Two-dimensional optical chimera states in an array of coupled waveguide resonators}.
		\newblock {Chaos \textbf {30}} 043107 (2020).
%
%
		\bibitem{Coullet2000}
		{ Coullet, P., Riera, C., and Tresser, C.}:
		\newblock {Stable static localized structures in one dimension}.
		\newblock {Phys. Rev. Lett. \textbf {84}}(14), 3069 (2000)
%
		\bibitem{Cui2020}
		{ Cui, B., Zheng, X., Wang, J., Liu, D., Xie, S., and Huang, B.}:
		\newblock {Realization of Lieb lattice in covalent-organic frameworks with
			tunable topology and magnetism}.
		\newblock {Nature Communications \textbf {11}}(1), (2020)
%
%
		\bibitem{Dean2015}
		{Dean, A. D., Matthews, P. C., Cox, S. M., and King, J. R.}:
		\newblock {Orientation-dependent pinning and homoclinic snaking on a planar lattice}.
		\newblock {SIAM J. Appl. Dyn. Syst., \textbf {14}}(1), 481–521 (2015)
		
		\bibitem{DeWitt2019}
		{ {De Witt}, H.}:
		\newblock {Beyond all order asymptotics for homoclinic snaking in a
			Schnakenberg system}.
		\newblock {Nonlinearity \textbf {32}}(7), 2667--2693 (2019)
%
%
		\bibitem{Dionne1997}
		{ Dionne, B., Silber, M., and Skeldon, A.~C.}:
		\newblock {Stability results for steady, spatially periodic planforms}.
		\newblock {Nonlinearity \textbf {10}}(2), 321 (1997)
		
		\bibitem{Drost2017}
		{ Drost, R., Ojanen, T., Harju, A., and Liljeroth, P.}:
		\newblock {Topological states in engineered atomic lattices}.
		\newblock {Nature Physics \textbf {13}}(7), 668--671 (2017)
%
%
		\bibitem{Egorov2013}
		{ Egorov, O. A., and Lederer, F.}:
		\newblock {Spontaneously walking discrete cavity solitons}.
		\newblock {Opt. Lett. \textbf {38}}(7), 1010 (2013)
%
		\bibitem{Feng2020}
		{ Feng, H., Liu, C., Zhou, S., Gao, N., Gao, Q., Zhuang, J., Xu, X., Hu, Z.,
			Wang, J., Chen, L., Zhao, J., Dou, S.~X., and Du, Y.}:
		\newblock {Experimental Realization of Two-Dimensional Buckled Lieb Lattice}.
		\newblock {Nano Letters \textbf {20}}(4), 2537--2543 (2020)
%
		\bibitem{Firth2007}
		{ Firth, W.~J., Columbo, L., and Maggipinto, T.}:
		\newblock {On homoclinic snaking in optical systems}.
		\newblock {Chaos \textbf {17}}(3), 037115 (2007)
%
%
		\bibitem{Haudin2011}
		{ Haudin, F., Rojas, R.~G., Bortolozzo, U., Residori, S., and Clerc, M.~G.}:
		\newblock {Homoclinic snaking of localized patterns in a spatially forced
			system}.
		\newblock {Phys. Rev. Lett. \textbf{107}}(26), 264101 (2011)
%
		\bibitem{Hilali1995}
		{ Hilali, M.~F., M{\'{e}}tens, S., Borckmans, P., and Dewel, G.}:
		\newblock {Pattern selection in the generalized Swift-Hohenberg model}.
		\newblock {Phys. Rev. E \textbf {51}}(3), 2046 (1995)
%
		\bibitem{Hunt2000}
		{ Hunt, G.~W., Peletier, M.~A., Champneys, A.~R., Woods, P.~D., Wadee,
			M.~A., Budd, C.~J., and Lord, G.~J.}:
		\newblock {Cellular buckling in long structures}.
		\newblock {Nonlinear Dynamics \textbf {21}}(1), 3--29 (2000)
%
		\bibitem{Jiang2019}
		{ Jiang, W., Huang, H., and Liu, F.}:
		\newblock {A Lieb-like lattice in a covalent-organic framework and its Stoner
			ferromagnetism}.
		\newblock {Nature Communications \textbf{ 10}}(1), 1--7 (2019)
		\bibitem{Jiang2020}
		{ Jiang, W., Zhang, S., Wang, Z., Liu, F., and Low, T.}:
		\newblock {Topological Band Engineering of Lieb Lattice in Phthalocyanine-Based
			Metal-Organic Frameworks}.
		\newblock {Nano Letters \textbf {20}}(3), 1959--1966 (2020)
%
		\bibitem{Judd2000}
		{ Judd, S.~L., and Silber, M.}:
		\newblock {Simple and superlattice Turing patterns in reaction–diffusion
			systems: bifurcation, bistability, and parameter collapse}.
		\newblock {Physica D \textbf {136}}(1-2), 45--65  (2000)

\bibitem{julku} Julku, A., Peotta, S., Vanhala, T. I., Kim, D.-H. and T\"orm\"a, P. Geometric origin of superfluidity in the Lieb-lattice flat band. Phys. Rev. Lett. 117, 045303 (2016).

%
		\bibitem{Knobloch2019}
		{ Knobloch, E., Uecker, H., and Wetzel, D.}:
		\newblock {Defectlike structures and localized patterns in the
			cubic-quintic-septic Swift-Hohenberg equation}.
		\newblock {Phys. Rev. E \textbf {100}}(1), 12204 (2019)
%
		\bibitem{Kozyreff2006}
		{ Kozyreff, G., and Chapman, S.~J.}:
		\newblock {Asymptotics of large bound states of localized structures}.
		\newblock {Phys. Rev. Lett. \textbf {97}}(4), 044502 (2006)
%
		\bibitem{Kusdiantara2017}
		{ Kusdiantara, R., and Susanto, H.}:
		\newblock {Homoclinic snaking in the discrete Swift-Hohenberg equation}.
		\newblock {Phys. Rev. E \textbf {96}}(6), 062214 (2017)
		
		\bibitem{Kusdiantara2019}
		{ Kusdiantara, R., and Susanto, H.}:
		\newblock {Snakes in square, honeycomb and triangular lattices}.
		\newblock {Nonlinearity \textbf {32}},(12), 5170--5190 (2019)
%
		\bibitem{Laing2001}
		{ Laing, C.~R., Troy, W.~C., Gutkin, B., and Ermentrout, G.~B.}:
		\newblock {Multiple bumps in a neuronal model of working memory}.
		\newblock {SIAM J. Appl. Math. \textbf {63}}(1), 62--97 (2001)
%
		\bibitem{Le2019}
		{ Le, P.~T., and Yarmohammadi, M.}:
		\newblock {Impurity-tuning of phase transition and mid-state in 2D spin Lieb
			lattice}.
		\newblock {Physica E \textbf {105}}, 56--61 (2019)
%
		\bibitem{Lieb1989}
		{ Lieb, E.~H.}
		\newblock {Two theorems on the Hubbard model}.::
		\newblock { Phys. Rev. Lett. \textbf {62}}(10), 1201--1204 (1989)
%
		\bibitem{Lloyd2009}
		{ Lloyd, D., and Sandstede, B.}
		\newblock {Localized radial solutions of the Swift-Hohenberg equation}.:
		\newblock { Nonlinearity \textbf {22}}(2), 485 (2009)
%
		\bibitem{Lloyd2019}
		{ Lloyd, D.~J.}:
		\newblock {Invasion fronts outside the homoclinic snaking region in the planar
			Swift–Hohenberg equation}.
		\newblock { SIAM J. Appl. Dyn. Syst. \textbf {18}}(4), 1892--1933 (2019)
%
		\bibitem{Lloyd2015}
		{ Lloyd, D.~J., Gollwitzer, C., Rehberg, I., and Richter, R.}:
		\newblock {Homoclinic snaking near the surface instability of a polarisable
			fluid}.
		\newblock { J. Fluid Mech. \textbf {783}}, 283--305 (2015)
%
		\bibitem{Lloyd2008}
		{ Lloyd, D. J.~B., Sandstede, B., Avitabile, D., and Champneys, A.~R.}:
		\newblock {Localized Hexagon Patterns of the Planar Swift–Hohenberg
			Equation}.
		\newblock { SIAM J. Appl. Dyn. Syst. \textbf {7}}(3), 1049--1100 (2008)
%
		\bibitem{Matthews2011}
		{ Matthews, P., and Susanto, H.}:
		\newblock Variational approximations to homoclinic snaking in continuous and
		discrete systems.
		\newblock { Phys. Rev. E \textbf {84}}(6), 066207 (2011)
%
		\bibitem{McCalla2010}
		{ McCalla, S., and Sandstede, B.}:
		\newblock {Snaking of radial solutions of the multi-dimensional Swift-Hohenberg
			equation: A numerical study}.
		\newblock { Physica D \textbf {239}}(16), 1581--1592 (2010)
		
		\bibitem{McCullen2016}
		{McCullen, N., and Wagenknecht, T.}:
		\newblock {Pattern formation on networks: from localised activity to Turing patterns}.
		\newblock {Scientific reports \textbf {6}}(1), 1--8 (2016)
%
		\bibitem{Mukherjee2015}
		{ Mukherjee, S., Spracklen, A., Choudhury, D., Goldman, N., Öhberg, P., Andersson, E. and Thomson, R.R.}:
		\newblock {Observation of a localized flat-band state in a photonic Lieb lattice}.
		\newblock { Phys. Rev. Lett \textbf {114}}(24), 245504 (2015)

%
		\bibitem{Oliveira-Lima2020}
		{ Oliveira-Lima, L., Costa, N.~C., {De Lima}, J.~P., Scalettar, R.~T., and
			Santos, R.~R.}:
		\newblock {Dynamical resilience to disorder: The dilute Hubbard model on the
			Lieb lattice}.
		\newblock { Phys. Rev. B \textbf {101}}(16), 1--9 (2020)
%
		\bibitem{Ozawa2017}
		{ Ozawa, H., Taie, S., Ichinose, T. and Takahashi, Y.}:
		\newblock {Interaction-driven shift and distortion of a flat band in an optical Lieb lattice}.
		\newblock { Phys. Rev. Lett. \textbf {118}}(17), 175301 (2017)

%
		\bibitem{Pomeau1986}
		{ Pomeau, Y.}:
		\newblock {Front motion, metastability and subcritical bifurcations in
			hydrodynamics}.
		\newblock { Physica D \textbf {23}}(1-3), 3--11 (1986)
%
		\bibitem{Sakaguchi1996}
		{ Sakaguchi, H., and Brand, H.~R.}:
		\newblock {Stable localized solutions of arbitrary length for the quintic
			Swift-Hohenberg equation}.
		\newblock { Physica D \textbf {97}}(1-3), 274--285 (1996)
%
		\bibitem{Salewski2019}
		{ Salewski, M., Gibson, J.~F., and Schneider, T.~M.}:
		\newblock {Origin of localized snakes-and-ladders solutions of plane Couette
			flow}.
		\newblock { Phys. Rev. E \textbf {100}}(3), 31102 (2019)
%
		\bibitem{Scafirimuto2021}
		{ Scafirimuto, F., Urbonas, D., Becker, M.A., Scherf, U., Mahrt, R.F. and Stöferle, T.}:
		\newblock {Tunable exciton–polariton condensation in a two-dimensional Lieb lattice at room temperature}.
		\newblock { Communications Physics \textbf {4}}(1), 1--6 (2021)
%

		\bibitem{Schmidt2020}
		{ Schmidt, H., and Avitabile, D.}:
		\newblock {Bumps and oscillons in networks of spiking neurons}.
		\newblock { Chaos \textbf {30}}(3), (2020).
%
		\bibitem{Slot2017}
		{ Slot, M.R., Gardenier, T.S., Jacobse, P.H., van Miert, G.C., Kempkes, S.N., Zevenhuizen, S.J., Smith, C.M., Vanmaekelbergh, D. and Swart, I.}:
		\newblock {Experimental realization and characterization of an electronic Lieb lattice}.
		\newblock { Nature physics \textbf {13}}(7), 672--676 (2017).
%

		\bibitem{Smektala2010}
		{Smektala, F., Quemard, C., Couderc, V., and Barth\'el\'emy, A.}:
		\newblock {Non-linear optical properties of chalcogenide glasses measured by Z-scan}.
		\newblock {J. Non Cryst. Solids \textbf {274}}(1-3), 232--237 (2010)
		
		\bibitem{Susanto2011}
		{ Susanto, H., and Matthews, P.}:
		\newblock Variational approximations to homoclinic snaking.
		\newblock { Phys. Rev. E \textbf {83}}(3), 035201 (2011)
%
%

\bibitem{tamura} Tamura, H., Shiraishi, K., Kimura, T. and Takayanagi, H. Flat-band ferromagnetism in quantum dot superlattices. Phys. Rev. B 65, 085324 (2002).

		\bibitem{Taylor2010}
		{ Taylor, C., and Dawes, J.~H.}:
		\newblock {Snaking and isolas of localised states in bistable discrete
			lattices}.
		\newblock { Phys. Lett. A \textbf {375}}(1), 14--22 (2010)
%
		\bibitem{Thompson2015}
		{ Thompson, J. M.~T.}:
		\newblock {Advances in Shell Buckling: Theory and Experiments}.
		\newblock { Int. J. Bifurcation Chaos \textbf {25}}, 1530001 (2015)
		
		\bibitem{Tian2021}
		{Tian, M., Bramburger, J. J., and Sandstede, B.}:
		\newblock {Snaking bifurcations of localized patterns on ring lattices}.
		\newblock arXiv preprint arXiv:2105.02380 (2021)
%
		\bibitem{Tlidi1994}
		{ Tlidi, M., Mandel, P., and Lefever, A.}:
		\newblock {Localized structures and localized patterns in optical bistability}.
		\newblock {Phys. Rev. Lett. \textbf {73}}(5), 640 (1994)
%
%
		\bibitem{Tlidi2012}
		{ Tlidi, M., Averlant, E., Vladimirov, A., and Panajotov, K.}:
		\newblock {Delay feedback induces a spontaneous motion of two-dimensional
			cavity solitons in driven semiconductor microcavities}.
		\newblock { Phys. Rev. A \textbf {86}}(3), 033822 (2012)
%

		\bibitem{Uecker2014}
		{ Uecker, H., and Wetzel, D.}:
		\newblock {Numerical Results for Snaking of Patterns over Patterns in Some 2D
			Selkov--Schnakenberg Reaction-Diffusion Systems}.
		\newblock { SIAM J. Appl. Dyn. Syst. \textbf {13}}(1), 94--128 (2014)
%
		\bibitem{Uecker2020}
		{ Uecker, H., and Wetzel, D.}:
		\newblock {Snaking branches of planar BCC fronts in the 3D Brusselator}.
		\newblock { Physica D \textbf {406}}, 132383 (2020)
%
		\bibitem{Vladimirov2011}
		{Vladimirov, A.G., Lefever, R. and Tlidi, M.}:
		\newblock {Relative stability of multipeak localized patterns of cavity solitons}.
		\newblock { Phys. Rev. A \textbf {84}}(4), 043848 (2011)


\bibitem{wang} Wang, Y. F., Gu, Z. C., Gong, C., De and Sheng, D. N. Fractional quantum Hall effect of hard-core bosons in topological flat bands. Phys. Rev. Lett. 107, 146803 (2011).

		
		\bibitem{Whittaker2018}
		{ Whittaker, C.E., Cancellieri, E., Walker, P.M., Gulevich, D.R., Schomerus, H., Vaitiekus, D., Royall, B., Whittaker, D.M., Clarke, E., Iorsh, I.V. and Shelykh, I.A.}:
		\newblock {Exciton polaritons in a two-dimensional lieb lattice with spin-orbit coupling}.
		\newblock { Phys. Rev. Lett \textbf {120}}(9), 097401 (2018)

		\bibitem{Woods1999}
		{ Woods, P., and Champneys, A.}:
		\newblock {Heteroclinic tangles and homoclinic snaking in the unfolding of a
			degenerate reversible Hamiltonian Hopf bifurcation}.
		\newblock { Physica D \textbf {129}}(3-4), 147--170 (1999)
%
		\bibitem{Yulin2011}
		{ Yulin, A., and Champneys, A.}:
		\newblock {Snake-to-isola transition and moving solitons via symmetry-breaking
			in discrete optical cavities}.
		\newblock { Discrete {\&} Continuous Dynamical Systems - S \textbf {4}}(5),
		1341--1357 (2011)
%
		\bibitem{Yulin2010}
		{ Yulin, A.~V., and Champneys, A.~R.}:
		\newblock {Discrete snaking: multiple cavity solitons in saturable media}.
		\newblock { SIAM J. Appl. Dyn. Syst. \textbf {9}}(2), 391--431 (2010)
%
		\bibitem{Yulin2008}
		{ Yulin, A.~V., Champneys, A.~R., and Skryabin, D.~V.}:
		\newblock {Discrete cavity solitons due to saturable nonlinearity}.
		\newblock { Phys. Rev. A \textbf {78}}(1), 011804(R) (2008)
		
		\bibitem{Zhan2002}
		{Zhan, C., Zhang, D., Zhu, D., Wang, D., Li, Y., Li, D., Lu, Z., Zhao, L. and Nie, Y.}:
		\newblock {Third-and fifth-order optical nonlinearities in a new stilbazolium derivative}.
		\newblock {J. Opt. Soc. Am. B \textbf {19}}(3), 369--375 (2002)

		
	\end{thebibliography}
	
\end{document}